\documentstyle[aps,epsfig,psfig]{revtex}

\newcommand{\beq}{\begin{equation}}
\newcommand{\eeq}{\end{equation}}
\newcommand{\be}{\begin{eqnarray}}
\newcommand{\ee}{\end{eqnarray}}

\newcommand{\tr}{\rm tr}
\newcommand{\ie}{{\it i.e.}}
\newcommand{\eg}{{\it e.g.}}
\newcommand{\wrt}{{\it w.r.t.}}
\newcommand{\backdel}{\stackrel{\gets}{\delta^2}}
\newcommand{\delslash}{\not\!\partial}
\newcommand{\Dslash}{\not\!\!{D}}

\newcommand{\AmS}{{\protect\the\textfont2
  A\kern-.1667em\lower.5ex\hbox{M}\kern-.125emS}}
\begin{document}
\bibliographystyle{unsrt}
\draft

\preprint{SUNY-NTG-99-04,IAS-SNS-99-40}

\title{High Density QCD and Instantons}

\author{R.~Rapp$^1$, T.~Sch{\"a}fer$^2$, 
        E.~V.~Shuryak$^1$ and M.~Velkovsky$^3$}
\address{
$^1$ Department of Physics and Astronomy, State University of New York, 
     Stony Brook, NY 11794-3800\\
$^2$ School of Natural Sciences, Institute for Advanced Study, 
     Princeton, NJ 08540\\ 
$^3$ Nuclear Theory Group, Brookhaven National Laboratory, Upton, 
     NY 11973-5000}
\date{\today}
\maketitle    

\begin{abstract} 
Instantons generate strong non-perturbative interactions between 
quarks. In the vacuum, these interactions lead to chiral symmetry 
breaking and generate  constituent quark masses on the order of 
300-400~MeV. The observation that the same forces also
provide attraction in the scalar diquark channel leads to the 
prediction that cold quark matter is a color superconductor, with
gaps as large as $\sim$~100~MeV. We provide a systematic treatment
of color superconductivity in the instanton model. We show that
the structure of the superconductor depends on the number of 
flavors. In the case of two flavors, we verify the standard 
scenario and provide an improved calculation of the mass gap. 
For three flavors, we show that the ground state is color-flavor
locked and calculate the chiral condensate in the high density 
phase. We show that as a function of the strange quark mass, 
there is a sharp transition between the two phases. Finally,
we go beyond the mean-field approximation and investigate the role 
of instanton-antiinstanton molecules, which -- besides superconducting
gap formation -- provide a competitive mechanism for chiral restoration
at finite density.
\end{abstract}

%%%%%%%%%%%%%%%%%%%%%%%%%%%%%%%%%%%%%%%%%%%%%%%%%%%%%%%%%%%%%%%%%%%%%%
\section{Introduction}
%%%%%%%%%%%%%%%%%%%%%%%%%%%%%%%%%%%%%%%%%%%%%%%%%%%%%%%%%%%%%%%%%%%%%%

Studying Quantum Chromodynamics (QCD)
at finite baryon density does not require a special 
motivation: this is, after all, the traditional subject of nuclear physics.   
Nevertheless, the investigation of cold quark matter lay dormant  
for some time and was only revived recently when it was realized 
that a number of exciting new phenomena can be predicted with some 
certainty. The first is that we expect the high density phase of 
QCD to be a color superconductor, with sizeable gaps on the order 
of 100~MeV around the phase transition~\cite{ARW_98,RSSV_98}. 
The structure of this phase 
depends sensitively on the number of active flavors. For three
or more flavors, color and flavor quantum numbers are locked
and chiral symmetry is broken even at large chemical potential 
$\mu$~\cite{ARW_98b,SW_98b}. In addition to that, it was argued that 
the second order chiral phase transition at finite temperature $T$
and zero density is likely to turn first order at some critical density. 
This entails the existence of a tricritical point in the $\mu-T$ 
phase diagram, which would persist even if the light quarks are not 
massless~\cite{BR_98,HJS*_98,SRS_98}.

  The idea that asymptotic freedom and the presence of a sharp Fermi 
surface imply that high density QCD should be a color superconductor
goes back to the work of Frautschi, Barrois, Bailin and 
Love~\cite{Frau_78,Bar_77,BL_84}. Color superconductivity has many 
features of the standard model, such as dynamical gauge symmetry 
breaking and the Higgs phenomenon. It is different from electroweak
symmetry breaking in the standard model in the sense that the 
Higgs is composite. And it is different from models with 
compositeness (such as technicolor) in that it does not
require strong interactions. Color superconductivity takes 
place even in weak coupling. This, of course, is a consequence 
of the BCS instability. 

 Detailed numerical calculations of color superconducting gaps were 
carried out by Bailin and Love, who concluded that one-gluon exchange 
(OGE) induces gaps on the order of 1~MeV at several times nuclear 
matter density (at asymptotically large chemical potentials, 
however, magnetic gluon exchanges generate increasingly large 
gaps~\cite{Son_98}). 
The main new feature pointed out in \cite{ARW_98,RSSV_98} is 
that instanton-induced interactions  
can lead to substantially larger gaps, on the order of 100~MeV. 
Furthermore, it was realized that the phase structure of QCD 
at finite baryon density is very rich. Besides the dominant order 
parameter for the superconducting phase transition, which is a 
scalar-isoscalar color antitriplet diquark operator, many other 
forms are possible. 

 Previous work mostly concentrated on two or three massless
flavors and was based on the mean-field approximation (MFA). In the 
present work, we go beyond these approximations in several important 
respects. In the case of two flavors we replace schematic zero 
range interactions with the full momentum dependent instanton-induced 
interaction~\cite{CD_98,CD_99,Vel_98}. We study the three-flavor case 
and show that the ground state exhibits color-flavor
locking. We show that chiral symmetry is broken, calculate 
the chiral condensate, and also assess the effects
of a finite strange quark mass. In order to go beyond the mean 
field approximation we study the role of instanton-antiinstanton
clusters. In particular, we consider the competition between 
random instantons and clusters employing a statistical mechanics
treatment of the partition function
for the instanton liquid. Finally, we consider more speculative 
possibilities such as phases with diquark Bose condensation, and 
give a general discussion of the phase structure of QCD with different
quark masses. 

 Throughout the article we assume that instanton-induced effects 
are the predominant source of strong non-perturbative interactions
in cold quark matter at small and moderate densities. 
This assumption is based on the success of 
the instanton model at zero temperature and zero chemical potential, 
as well as at $\mu=0$ and $T\neq 0$, see \cite{SS_98} for a review. 
The other major type of interaction that has been widely used is the 
(perturbative) OGE. While the latter should be prevailing at high 
densities, it encounters conceptual difficulties  at low and moderate 
densities since the involved momentum transfers at the Fermi surface 
$q^2\le p_F^2$, and the running coupling constant of QCD might 
not be sufficiently small as required for a perturbative treatment. 
%: all energies involved such as the gap $\Delta$ 
%should be large compared to $\Lambda_{QCD}$ (also, the BCS treatment
%requires that the leading log($\Delta/\mu$) term ought to be large
%as implicit in the selection of diagram resummations; this only 
%happens at unrealistically large $\mu\sim TeV$ or so.) 
On the other
hand, the Debye-screening of electric fields suppresses instanton
effects at large densities; however, this suppression seems not  
to be effective below the chiral phase transition, as has 
been explicitely demonstrated in 
 finite-$T$ lattice studies~\cite{CS_95,Alles} 

Further evidence for the importance of instantons in this context 
is related to the
existence of fermion zero modes in the spectrum of the Dirac operator,
which arise as a consequence of the axial anomaly: In 
a topologically non-trivial background field with topological charge
+1 there is -- for each flavor -- a left-handed state that emerges 
out of the Dirac sea,
and a right-handed one that moves from positive to negative
energy. As a result, the axial charge is violated by 2$N_f$ units. For 
$N_f=1$ this immediately implies chiral condensation. For more than
one flavor, chiral condensation is a collective effect. The quark 
condensate is determined by the number density of (almost) zero modes 
of the Dirac operator. These anomalously small eigenmodes can originate 
from the interaction of exact zero modes associated with isolated 
instantons and antiinstantons. The wave function of the condensate is 
the collective state built from instanton and antiinstanton zero modes.
To check this mechanism for chiral symmetry breaking on the lattice
has lately attracted appreciable attention:  the results indeed  
support the suggested picture~\cite{IN_97,TFM_97}.

  At finite baryon chemical potential the axial anomaly is  connected 
with fermion zero modes in exactly the same way as in vacuum. The only
difference is that the zero modes now correspond to extra states appearing 
at the Fermi surface, rather than the surface of the Dirac sea. As  
in the  vacuum the effect of the zero modes can be represented 
as an effective $(2N_f)$-quark interaction that operates near the 
Fermi surface. For two flavors, this interaction directly leads 
to the BCS instability, the formation of Cooper pairs and the 
appearance of a gap. For three and more flavors the instanton 
vertex does not directly support a Cooper pair; some of the 
chiral condensates have to be non-zero to close off external 
quark legs, reducing the $N_f$-body instanton interaction to a 
two-body one. Instanton-antiinstanton molecules, on the other
hand, lead to an effective four quark operator for any number of 
flavors, which, if attractive, will trigger the formation of a gap. 
Nevertheless, direct instantons play an important role even for 
$N_f\ge 3$. In particular, instantons provide a novel mechanism
for chiral condensation: A diquark-driven $\bar q q$ (chiral) 
condensate.

  The investigation of chiral symmetry restoration and color 
superconductivity at finite density should also be placed in a 
broader context, {\it e.g.}, including finite temperatures. 
Moreover, we would like to understand 
the phase structure as a function of parameters that we cannot
control in the real world, such as the number of flavors and 
their masses. After all, the underlying 
mechanisms for the various transitions and the role of non-perturbative
effects (such as instantons) in the different phases have to be 
clarified.  

  At high temperature we expect to find a quark-gluon plasma phase
in which chiral symmetry is restored, \ie, the density of (almost)
zero modes has to vanish. This can be realized if the instanton liquid 
changes from a random ensemble of instantons and antiinstantons to a 
correlated system with finite clusters, \eg,  instanton-antiinstanton 
($I$-$A$) molecules. The formation of molecules and other correlated 
clusters was observed in numerical simulations of the instanton 
liquid~\cite{SS_96}, where a number of consequences of this scenario were 
explored. On the lattice the disappearance of the quasi-zero modes 
in the vicinity of $T_c$ is well established, and the formation of 
clusters has been observed~\cite{DeForcrand}. Nevertheless, many 
details of the transition remain to be understood. 
In the case of many flavors the instanton calculations \cite{SV_97} 
suggest a chirally restored vacuum state already for a fairly small 
number of flavors, around 5. Again, the transition is associated with 
the formation of correlated clusters.

  In this article we would like to understand the interplay of the 
three major phases that have been considered: (i) the hadronic (H) phase, 
with (strongly) broken chiral symmetry (ii) the color superconductor 
(CSC) phase, with  broken color symmetry, and (iii) the quark-gluon 
plasma (QGP) phase. All three phases are associated with three specific 
instanton-induced
interactions. Chiral symmetry breaking is caused by the strong $\bar qq$
attraction. The binding energy of the lightest baryon, the nucleon,
is mostly associated with the $qq$ interaction. The same interaction is
responsible for  superconductivity at large baryon density. 
Finally,  quark exchanges between instantons and antiinstantons
drive  their pairing, which is expected to become the predominant
feature  in the instanton liquid
as temperature increases (at any chemical potential).

 The structure of our  paper is as follows. The first part, comprised
of sects. II-VII contains a mean-field analysis of color superconductivity
in finite-density QCD with 2 and 3 flavors. We begin with a brief 
introduction to the structure of the effective instanton-induced interaction
in sect.~\ref{sec_L_eff}. In sect.~\ref{sec_diq}, we study the physical 
effects of this interaction in different diquark channels at $\mu=0$.  
In sect.~\ref{sec_int} we discuss the modifications of the
instanton-induced interactions at non-zero chemical potential.
The interplay of chiral symmetry breaking and quark superconductivity
in two-flavor QCD is studied in sect.~\ref{sec_nf2}. This section 
employs the mean-field approximation but is based on the exact form 
of the interaction. Using a simplified version of the form factors  
we then consider different $\langle\bar qq\rangle$ and $\langle qq\rangle$ 
condensates of increasing complexity: the three-flavor problem in the 
chiral limit (sect.~\ref{sec_nf3}) and the effect of flavor symmetry 
breaking due to a finite strange quark mass (sect.~\ref{sec_fsb}). 

 The second part  of the article (sect.~\ref{sec_cocktail} and
\ref{sec_phdia}) addresses effects due to clustering, which are 
beyond the mean-field approximation. In sect.~\ref{sec_cocktail} 
we quantitatively discuss only one type of cluster, the 
instanton-antiinstanton molecule, which we believe is the  
most important cluster in the chirally restored phase. In 
sect.~\ref{sec_phdia}, we also discuss the role of correlations
between quarks -- in particular non-condensed diquarks and (the most 
obvious cluster of all!) nucleons in nuclear matter --, and comment
on possible experimental consequences for heavy-ion reactions and 
neutron stars. We summarize and conclude in sect.~\ref{sec_concl}.

%%%%%%%%%%%%%%%%%%%%%%%%%%%%%%%%%%%%%%%%%%%%%%%%%%%%%%%%%%%%%%%%%%%%%%%%%%%
\section{Effective instanton-induced interactions in vacuum}
\label{sec_L_eff}
%%%%%%%%%%%%%%%%%%%%%%%%%%%%%%%%%%%%%%%%%%%%%%%%%%%%%%%%%%%%%%%%%%%%%%%%%%%

%%%%%%%%%%%%%%%%%%%%%%%%%%%%%%%%%%%%%%%%%%%%%%%%%%%%%%%%
\subsection{Single-Instanton Interactions}
\label{oneinst}
%%%%%%%%%%%%%%%%%%%%%%%%%%%%%%%%%%%%%%%%%%%%%%%%%%%%%%%%%

Our starting point is the euclidean QCD partition function
\beq
 {\cal Z}=\int{\cal D}\psi{\cal D} \psi^\dagger {\cal D}A 
  \exp(-{S_{QCD}})=\int {\cal D}A  \det(\Dslash) \exp(-{S_{gauge}}) \ .
\eeq
The main assumption of the instanton model is that the gauge field
is saturated by classical (anti-)instanton solutions. If the instanton 
ensemble is sufficiently dilute, the gauge field can be approximated 
by a sum of individual instanton gauge potentials
\beq
A=\sum_{k \in I,\bar I} A_k \ . 
\label{sumansatz}
\eeq
Collective effects related to chiral symmetry breaking are generated 
through the low-momentum part of the fermion determinant. In particular, 
we will concentrate on the fermion determinant in a basis spanned by 
the zero modes of the individual instantons. Matrix elements of the
Dirac operator in this basis are given by the overlap integrals
\be
T_{IA}(z,u)=\int d^4x \ \phi_{I}^\dagger(x-z_I) 
\ \Dslash \ \phi_{A}(x-z_A) \  .
\label{Tiavac}
\ee
Here, $z_I$ and $z_A$ denote the positions of the instanton and 
antiinstanton, and $\phi_{I,A}$ the corresponding zero mode wave functions,  
which are solutions of the Dirac equation
\be
\Dslash_{I,A} \ \phi_{I,A}(x)=0 \ ,
\label{dirac}
\ee
where the covariant derivative $\not\!\!\!{D}_{I,A}$ includes the gauge 
potential of the (anti-) instanton $I$ ($A$). Using the Dirac
equation (\ref{dirac}) and the sum ansatz (\ref{sumansatz}), 
Eq.~(\ref{Tiavac}) can be simplified by replacing  the covariant derivative 
by an ordinary one. The overlap matrix element can also be viewed as  
the quark ``hopping'' amplitude from an instanton to an antiinstanton.

 To extract effective $2N_f$-quark interaction vertices, we follow the
approach of Diakonov and Petrov~\cite{DP}, who suggested to 
reintroduce free fermion fields according to  
\be
{\cal Z}=\int d\psi d \psi ^\dagger {\exp \{\int d^4 x
 \psi ^\dagger i \delslash  \psi\} \over
N_+!N_-!}\prod_{I=1}^{N_+} \theta_+
 \prod_{\bar I=1}^{N_-} \theta_-  \ ,
\ee
where in the two-flavor case
\be
\theta_+=\int d\Omega_I \prod_{f=1}^{2}\big[\int
d^4x \psi^
\dagger_f(x) i\not\!\partial \phi_I (x-z_I)
\int d^4y\phi_I^\dagger(y-z_I)i\not\!\partial
 \psi_f (y)\big] \ ,
\ee
and the integrals are over the collective coordinates 
$\Omega_I=\{z_I,\rho_I,u_I\}$ (position, size and color orientation) 
of the instantons. The original zero mode determinant can be recovered
by calculating a Green's function with $N_f (N_++N_-)$ external legs.
In order to perform the integration over the centers of the instantons
it is convenient to proceed to momentum space. This automatically induces 
a four-momentum conserving $\delta$-functions at each vertex.

An effective interaction is most easily derived by  
exponentiating the fermion terms. This is accomplished by applying 
an inverse Laplace transformation which gives the following  
partition function: 
\be
 {\cal Z} &=& const \int d\psi d \psi ^\dagger d\beta_+d\beta_-
\exp\left\{ -(N_++1)\log\left(\frac{\beta_+}{c_{\rho}}\right)
 - (N_-+1)\log\left(\frac{\beta_-}{c_{\rho}}\right) \right. 
  \nonumber \\ 
 & & \hspace{1cm}\left.  + \int d^4 x\, (\psi^\dagger i \delslash
  \psi\ +  \beta_+\theta_++\beta_-\theta_- )\right\}. 
\ee
The integrations over $\beta_\pm$ can be performed by the saddle point
method, which becomes exact in the thermodynamic limit as the 
coefficients in the exponent are extensive quantities ($N_\pm=n_\pm V_4$).
For an equal number of instantons and antiinstantons, one 
may consider $g= \beta_+=\beta_-$ as an effective fermion coupling.  
$\beta_\pm$ are then 
eliminated through the final minimization of the free energy, 
leaving the total instanton density $N/V=n_++n_-$ as the physical 
parameter.

In the remainder of this section we restrict ourselves to two flavors.
In this case, four quarks participate at each vertex, and the pertinent
vertex operator $\theta_\pm$ takes the form
\be
{\cal O}_{\theta_+}=\prod_{f=1}^2 d\Omega_I (\Omega_I \chi_L)\otimes 
(\chi_R^\dagger \Omega_I^\dagger) \ .
\ee
Its non-locality can be expressed through a momentum-dependent 
formfactor ${\cal F}(k)$, which is also a matrix in the Dirac space,
attached to each fermion field. We will analyze the formfactors,
including their dependence on density, in sect.~\ref{sec_formfactors}. 
After color-averaging, one obtains the effective interaction lagrangian
\be
\label{l_nf2}
{\cal L} &=& g\frac{1}{4(N_c^2-1)}
 \Big\{ \frac{2N_c-1}{2N_c}\left[
  (\bar\psi {\cal F}^\dagger\tau_\alpha^{-}{\cal F}\psi)^2 +
  (\bar\psi {\cal F}^\dagger\gamma_5\tau_\alpha^{-}{\cal F}\psi)^2
 \right]\cr
 & &  \hspace{0.5cm}
  + \frac{1}{4N_c}(\bar\psi{\cal F}^\dagger \sigma_{\mu\nu}
 \tau_\alpha^{-}{\cal F}\psi)^2 
 \Big\} \ ,
\ee
where $N_c$ is the number of colors and $\tau^-=(\vec\tau,i)$ is an isospin 
matrix. In the pseudoscalar channel the interaction combines attraction 
for the isospin-1 (pion) channel with repulsion (due to the extra $i$) for
isospin-0 ($\eta'$). Similarly, one finds attraction in the scalar 
isospin-0 ($\sigma$) channel (responsible for spontaneous chiral symmetry 
breaking) together with repulsion in the scalar isospin-1 channel ($a_0$).  

  In practice we will calculate correlation functions and the mean
field effective potential in the Hartree-Fock approximation. For this
purpose, it is convenient to construct an effective $s$-channel kernel
including the exchange term. 
This is made possible by the simple (separable) form of the 
momentum dependence. Using this kernel, one can reproduce the 
result of a Hartee-Fock calculation by evaluating the Hartree
term only. In short-hand notation we will refer to the kernel as the
effective meson or diquark lagrangian. From the Fierz identities
given in appendix \ref{fierz}, we obtain the following kernel 
for color singlet and octet $\bar qq$ states 
\be 
\label{l_mes}
{\cal L}_{mes} &=& {g \over 8 N_c^2} \Big\{ 
 \left[(\bar\psi{\cal F}^\dagger \tau^- {\cal F}\psi)^2+
 (\bar\psi{\cal F}^\dagger \tau^- \gamma_5 {\cal F}\psi)^2 \right]
 \nonumber \\ 
 & &  \mbox{}
+ {N_c-2\over 2(N_c^2-1)}
  \left[(\bar\psi{\cal F}^\dagger \tau^- \lambda^a {\cal F}\psi)^2
 +(\bar\psi{\cal F}^\dagger \tau^- \lambda^a \gamma_5 {\cal F}\psi)^2
 \right]\cr   
 & &   \mbox{}
-{N_c \over 4 (N_c^2-1)}(\bar\psi{\cal F}^\dagger \tau^-  \sigma_{\mu \nu}
\lambda^a {\cal F}\psi)^2 \Big\} \ ,  
\ee
again being attractive in the $\sigma$
and $\pi$ channel, repulsive in the $\eta'$ and $a_0$ channel. 
Analogously, we can construct the effective interaction for 
color-antisymmetric ${\bf \bar 3}$ and -symmetric ${\bf 6}$ 
diquarks. The result is 
\be 
\label{l_diq}
{\cal L}_{diq} &=& 
{g\over 8 N_c^2} \left\{
 -{1\over N_c-1}
 \left[ (\psi^T{\cal F}^T C \tau_2 \lambda_A^a {\cal F}\psi)
        (\bar\psi{\cal F}^\dagger\tau_2 \lambda_A^a C {\cal F}^*\bar\psi^T  
\right. \right.\nonumber\\
 & &  \left.\left. \hspace{1.5cm}\mbox{}
       +(\psi^T{\cal F}^T C \tau_2 \lambda_A^a \gamma_5 {\cal F}\psi)
         (\bar\psi{\cal F}^\dagger \tau_2 \lambda_A^a \gamma_5 C {\cal F}^*
\bar\psi^T) \right]\right. \nonumber\\
 & & \left. \mbox{}
      +{1\over 2(N_c+1)}
      (\psi^T{\cal F}^T C \tau_2 \lambda_S^a \sigma_{\mu \nu} {\cal F}\psi)
      (\bar\psi{\cal F}^\dagger \tau_2 \lambda_S^a \sigma_{\mu \nu} C 
{\cal F}^*\bar\psi^T) 
        \right\}  \  ,
\ee 
where $\tau_2$ is the antisymmetric Pauli matrix, and $\lambda_{A,S}$ 
are the antisymmetric (color ${\bf\bar3}$) and symmetric (color {\bf 6}) 
color generators (normalized in an unconventional way, ${\rm tr}(\lambda^a
\lambda^b)=N_c\delta^{ab}$, in order to facilitate the comparison between 
mesons and diquarks). In the color ${\bf\bar3}$ channel, the interaction
is attractive for scalar $(\psi^TC\gamma_5\psi)$ diquarks, and repulsive
for pseudoscalar $(\psi^TC\psi)$ diquarks. 

   As we have already discussed in our previous paper~\cite{RSSV_98}, 
in the  case of two-color ($N_c$=2) QCD there exists an additional 
Pauli-G{\"u}rsey symmetry (PGSY)~\cite{PG,DFL_96} which mixes quarks 
with antiquarks. It also manifests itself in the lagrangians given 
above, as in this case the coupling constants in $\bar qq$ and $qq$ 
channels are identical, \ie, diquarks (the baryons of the $N_c$=2-theory)  
are degenerate with the corresponding mesons. Chiral symmetry breaking 
then implies that scalar diquarks are also Goldstone bosons, with their 
mass vanishing in the chiral limit (\ie, for current quark masses $m=0$).

%%%%%%%%%%%%%%%%%%%%%%%%%%%%%%%%%%%%%%%%%%%%%%%%%%%%%%%%
\subsection{$I$-$A$-Molecule Induced Interactions}
\label{molint}
%%%%%%%%%%%%%%%%%%%%%%%%%%%%%%%%%%%%%%%%%%%%%%%%%%%%%%%%%

  Using the 't Hooft interaction introduced in the last section
one can calculate correlation functions in a systematic expansion
in multi-instanton interactions, starting from direct instantons
graphs and proceeding to two-instanton or instanton-antiinstanton
graphs, as well as more complicated clusters. In simple cases, like the
set of RPA diagrams discussed in sect.~\ref{sec_diq}, one can sum
a whole series of terms involving infinitely many instantons.
  But if the instanton ensemble is strongly correlated, this method
may become very inefficient. In that case it is more useful to
determine the effective vertex for a given cluster, and fix the
strength of the vertex by calculating the concentration of
clusters from the partition function. The simplest kind of
cluster that can arise in the instanton ensemble are instanton-
antiinstanton molecules. We have observed the formation of these
clusters at high temperature and at large $N_f$ in both 
analytic~\cite{IS_94,VS_97} and numerical simulations of the instanton
ensemble~\cite{SS_96}. In these cases, molecules are intimately
connected with chiral symmetry restoration. An ensemble of molecules
does not have delocalized zero modes or collective eigenstates,
and the chiral condensate is zero.

  In the high density problem, the role of molecules is twofold.
First, the concentration of instanton-antiinstanton molecules
in the ensemble may be dynamically enhanced for similar reasons
as in the case of high temperature or large number of flavors.
We will discuss this problem in detail in sect.~\ref{sec_cocktail}.
Second, the BCS instability is due to quark-quark scattering, or
four-fermion operators. The 't Hooft vertex is a $(2N_f)$-fermion 
operator and does not automatically lead to an instability
for $N_f>2$. However, an instanton-antiinstanton molecule
can always generate an effective four-fermion interaction, with
 the additional $(2N_f-4)$ fermion lines being internal.

  The effective four-fermion vertex induced by instanton-antiinstanton
molecules was evaluated in \cite{SSV_95}. The result is particularly
simple if the relative color orientation is fixed such that the
instanton-antiinstanton interaction is most attractive. In that
case one has 
\be 
{\cal L}_{IA} &=& G_{mol} \left\{
    \frac{1}{N_c^2}\left[
        (\bar\psi\gamma_\mu\psi)^2+
        (\bar\psi\gamma_\mu\gamma_5\psi)^2  \right]
   -\frac{1}{2N_c(N_c-1)} \left[
        (\bar\psi\gamma_\mu\lambda^a\psi)^2+
        (\bar\psi\gamma_\mu\gamma_5\lambda^a\psi)^2 \right]
   \right. \nonumber \\
  & &  \left.\hspace{0.4cm} -\frac{1}{N_c^2}
       \left[ (\bar\psi\gamma_\mu\psi)^2-
        (\bar\psi\gamma_\mu\gamma_5\psi)^2\right]
   -\frac{2N_c-1}{2N_c(N_c^2-1)}
       \left[ (\bar\psi\gamma_\mu\lambda^a\psi)^2-
        (\bar\psi\gamma_\mu\gamma_5\lambda^a\psi)^2\right]
        \right\} \ .
\label{l_mesmol} 
\ee
Similar to the procedure leading to Eq.~(\ref{l_diq}), this interaction
can be rearranged into an effective $qq$ vertex. In the color 
antitriplet channel the result is
\be
{\cal L}_{IA}^{ \underline 3} &=&
 G_{mol} \left\{
  \frac{1}{N_c(N_c-1)}\left[
        (\psi^T C\gamma_5 \tau_2 \lambda_A^a \psi)
        (\bar\psi\gamma_5\tau_2 \lambda_A^a C \bar\psi^T)
       -(\psi^T C \tau_2 \lambda_A^a  \psi)
        (\bar\psi \tau_2 \lambda_A^a C \bar\psi^T) \right] \right.
               \nonumber\\
 & &  \left. \mbox{}
    + \frac{1}{4N_c(N_c-1)}\left[
        (\psi^T C\gamma_\mu\gamma_5 \tau_2\lambda_A^a \psi)
        (\bar\psi\gamma_\mu\gamma_5\tau_2 \lambda_A^a C \bar\psi^T)
       -(\psi^T C \gamma_\mu\tau_2\vec\tau \lambda_A^a  \psi)
        (\bar\psi \gamma_\mu\tau_2\vec\tau \lambda_A^a C \bar\psi^T)
       \right] \right\} \ . 
\ee
Even in the case $N_f=2$, there are two important differences as
compared to the single-instanton vertex. First, since molecules
are topologically neutral, the interaction is $U(1)_A$ invariant.
This implies that it does not distinguish between
scalar and pseudoscalar diquarks. Second, whereas the 't Hooft vertex
only operates in scalar (and tensor) channels, molecules also
provide an interaction in vector-meson and diquark channels.
The coupling constant is related  to the density of molecules
and has to be determined from the partition function of the instanton
liquid. We will study this problem in sect.~\ref{sec_coupl}.

%%%%%%%%%%%%%%%%%%%%%%%%%%%%%%%%%%%%%%%%%%%%%%%%%%%%%%%%%%%%%%%%%%%%%
\section{Diquarks in the Random Phase Approximation}
\label{sec_diq} 
%%%%%%%%%%%%%%%%%%%%%%%%%%%%%%%%%%%%%%%%%%%%%%%%%%%%%%%%%%%%%%%%%%%%%

 Color superconductivity implies that the high density phase is
composed of diquark Cooper pairs. In a weakly coupled BCS system,
the expression  'Cooper pair' should not be taken too literally: 
it is not tightly bound and the range of its wave function 
is large compared to average inter-particle separations ({\it i.e.},
the cube root of the inverse particle density, $d=n^{-3/2}$). In 
QCD this is not necessarily the case. The gap can be quite 
large, and the existence of an intermediate phase of diquarks
with or without Bose condensation is not a priori excluded. 

  For this reason we first study the possibility of diquark 
bound states in vacuum~\cite{BL_89,SSV_94,DFL_96}. 
In QCD with $N_c>2$ there are, of course, no gauge invariant diquark 
states. Instead, one can study correlation functions of 
heavy-light $Qqq$ states, where the heavy quark $Q$ serves
to neutralize color. Effectively, this corresponds to a diquark
correlator in the presence of a Wilson line. The ``gauge 
invariant'' diquark masses extracted from these correlators
then measure the mass of the heavy $Qqq$ state minus the mass
of the heavy quark. In a dense medium, the Wilson line is 
not necessary, and color is neutralized by light quarks
of the third color. 

  In the effective fermionic theory derived in the previous
section, diquarks can appear as physical bound states. These
states should be interpreted as building blocks in the formation
of baryons and dense matter. In practice, we study diquark 
correlation functions and look for poles in the diquark 
propagators at momenta $|p|< 2M$, where $M$ is the constituent 
quark mass. For simplicity, instead of the exact instanton 
formfactors ${\cal F}(p)$, we employ in this section an 
euclidean $O(4)$ symmetric cutoff $\Lambda$,  
 being adjusted to yield a realistic constituent quark 
mass $M$.
%%%%%%%%%%%%%%%%%%%%%%%%%%%%%%%%%%%%%%%%%%%%%%%%%%%%%%
\begin{figure}[tbp]
\begin{center}
\psfig{file=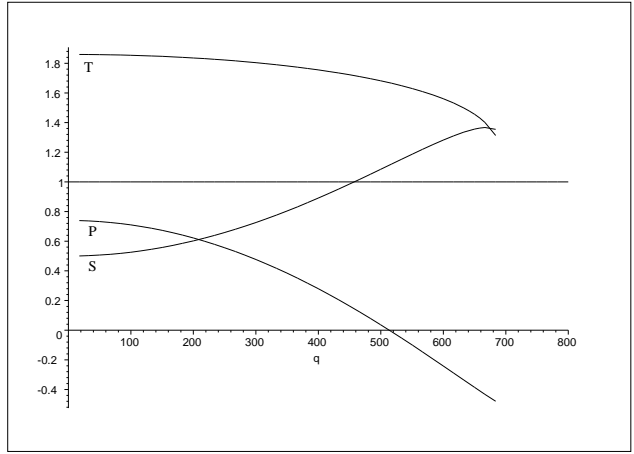,angle=0,height=6. cm}
\end{center}
\vspace{0.2cm}
\caption{The quantity $KJ$ entering the denominators of the $T$ matrix for 
the scalar (S), pseudoscalar (P)
and tensor (T) diquark channels.}
\label{poles}
\end{figure}
%%%%%%%%%%%%%%%%%%%%%%%%%%%%%%%%%%%%%%%%%%%%%%%%%%%%%
The Bethe-Salpeter equation for the two-body ${\cal T}$ matrix in 
a given channel can be written as
\be
{\cal T}(q)&=&\sum_i K_i(C^{-1}{\cal O}^i) \times \left\{ ({\cal O}^i C)
+i \tr 
\int {d^4p\over (2 \pi)^4}S_F(p+q/2)({\cal O}^i C)S_F^T(-p+q/2){\cal
T}(q)  \right\}  \ . 
\label{BS}
\ee
Introducing the notation 
\be
{\cal T}(q)=\sum_{k,k'}(C^{-1}{\cal O}^k)T_{k,k'}(q)
({\cal O}^{k'}C) \ , 
\ee
Eq.~(\ref{BS}) can be expressed schematically as 
\be
T=K(1+JT) \ .
\ee
This equation has the solution 
\be
T=(1-KJ)^{-1}K.
\ee
In both diquark and mesonic channels $J$ takes the form
\be
J_{k,k'}(p)=i \tr
\int {d^4p\over (2 \pi)^4}S_F(p+q/2){\cal O}^k S_F(p-q/2){\cal O}^{k'} \ ,
\ee
since  $C S_F^T(p)C^{-1}=S_F(-p)$. From  
Eq.~(\ref{l_diq}) one has the three different operator structures,  
${\cal O}= \tau_2 \lambda_A^a i \gamma_5$ (scalar channel),
${\cal O}= \tau_2 \lambda_A^a$ (pseudoscalar channel) and  
${\cal O}= \tau_2 \lambda_S^a \sigma_{\mu \nu}$ (tensor channel),  
which lead to 
\be
J_{SS}&=& -2 I_1(M)+2 q^2 I_2(q^2,M)
\nonumber\\
J_{PP}&=& 2I_1(M)+2(4 M^2-q^2)I_2(q^2,M)
\\
J_{TT}&=&-16 M^2 I_2(q^2,M) 
\nonumber
\ee
with the  two standard integrals
\be
I_1(M)&=&8N_c\int_0^\Lambda {d^4p\over (2 \pi)^4}{1\over p^2+M^2}
\\
I_2(-q^2,M)&=&4N_c\int_0^\Lambda {d^4p\over (2 \pi)^4}{1\over
(p+1/2q)^2+M^2}{1\over (p-1/2q)^2+M^2} \ . 
\ee
They are readily evaluated for Euclidean momenta and 
analytically continued to Minkowski space to yield   
\be
I_1(M)&=&{N_c\over 2 \pi^2}\left[ \Lambda^2 +M^2 \ln {M^2\over \Lambda^2
+M^2}\right], \\
I_2(q^2,M)&=&{N_c\over 4 \pi^2}\left[\ln {M^2\over \Lambda^2
+M^2} +2\sqrt{{4 M^2 -q^2\over q^2}}\arctan \sqrt{{q^2\over 4 M^2
-q^2}}\right.\cr
& &\left. -2 \left( 1-{2\Lambda^2 \over4(\Lambda^2+M^2)-q^2}\right)
\sqrt{{4(\Lambda^2+M^2)-q^2\over q^2}}
 \arctan \sqrt{{q^2\over 4(\Lambda^2+M^2)
-q^2}}\right].
\ee
The chiral gap equation in the scalar $\bar q q$ 
channel then becomes  
\be
{g\over 8 N_c^2}I_1(M)=1 \ , 
\ee
which provides a relation between the  coupling $g$ and the constituent
quark mass $M$. 

The conditions for the existence of poles in the corresponding
channels of the diquark ${\cal T}$ matrix are 
\be
-{g\over 8 N_c^2(N_c-1)}(-2 I_1(M)+2 q^2 I_2(q^2,M))=1
\ee
for the scalar channel,
\be
-{g\over 8 N_c^2(N_c-1)}(2I_1(M)+2(4 M^2-q^2)I_2(q^2,M))=1,
\ee
for the pseudoscalar one and
\be
{g\over 16 N_c(N_c+1)}(-16 M^2 I_2(q^2,M))=1
\ee
for the tensor one. If the {\it l.h.s.}~crosses 1, there is a bound state.
This is illustrated in  Fig.~\ref{poles} where we have taken $M=350$~MeV, 
$\Lambda=900$~MeV for definiteness. One finds that only the scalar
diquark is bound, with a binding energy of about 200~MeV.

This result agrees with numerical simulations of the instanton
liquid~\cite{SSV_94}, which included all diagrams in this interaction. 
However, it is at variance with the conclusion drawn in \cite{DFL_96},  
where no bound scalar diquark was found in the same model. We believe
that the discrepancy is due to the fact that the authors of \cite{DFL_96}
only used part of the interaction in the diquark channel. In this work,
we have performed a Hartree-Fock calculation with the full one-instanton
interaction. A lattice calculation of diquark masses was performed
in \cite{HKLW_98}. These authors find a significant scalar-vector
diquark mass splitting, but no scalar diquark bound state. On the 
other hand, they also obtained a too small $N$-$\Delta$ mass splitting,
and a mass ratio $m_N/m_\rho$ that is too large. 

 One can also argue that there should be some continuity when going from
the theory with $N_c=2$ to $N_c=3$. In the former case the scalar diquark 
is the partner of the pion and the vector diquark is the partner of $\rho$.
This implies that the scalar diquark binding is large, $M_{dq,V}-m_{dq,S}=
m_\rho-m_\pi\simeq 600 MeV$. It is then natural that some remnant of 
the binding is left at $N_c=3$, since the coupling constant in the 
scalar diquark channel for $N_c=3$ is only reduced by a factor of 2 
as compared to the $N_c=2$ case. This corresponds to a ``$N_c=2+\epsilon$'' 
picture of the baryon
octet: a tightly bound scalar diquark loosely coupled to the third 
quark. A number of phenomenological observations (reviewed, \eg,  
in~\cite{diquarks}) actually supports the validity of this picture 
for real QCD. The decuplet baryons, on the other hand,  do not contain 
scalar diquarks, and therefore should be generic 3-body objects. 
This picture is quite contrary to another (and much better known) view 
of baryonic structure, the large $N_c$ limit. Here both $N,\Delta$ as 
well as other members of the octet and decuplet are basically the same 
heavy object, slowly rotating with slightly different angular momenta. 
Indeed, as one reads off from the lagrangians given in the previous 
section, in this limit the diquark coupling tends to zero, and the
scalar diquark binding disappears.

%%%%%%%%%%%%%%%%%%%%%%%%%%%%%%%%%%%%%%%%%%%%%%%%%%%%%%%%%%%%%%%%%%%%%%
\section{Instanton-induced interaction in dense matter }
\label{sec_int}
%%%%%%%%%%%%%%%%%%%%%%%%%%%%%%%%%%%%%%%%%%%%%%%%%%%%%%%%%%%%%%%%%%%%%%

%%%%%%%%%%%%%%%%%%%%%%%%%%%%%%%%%%%%%%%%%%%%%%%%%%%%%%%%%%%%%%%%%%%%%%
\subsection{Quark Zero Modes} 
\label{sec_zeromodes} 
%%%%%%%%%%%%%%%%%%%%%%%%%%%%%%%%%%%%%%%%%%%%%%%%%%%%%%%%%%%%%%%%%%%%%%

    In the previous section the instanton-induced interactions between 
quarks were approximated by effective {\em local} 4-fermion vertices. 
In the microscopic treatment of sect.~\ref{sec_L_eff}, 
the external quarks couple to the 
quark zero modes in the instanton field, leading to a  {\em nonlocal} 
profile function for the interaction vertex with a size characterized 
by the typical instanton radius of about $\rho$=0.33~fm. In 
dense matter at zero temperature, the single-instanton solution 
itself is not affected by the surrounding quarks. The zero-mode
wave functions, however, {\em  are} density dependent leading to  
important modifications of the instanton-induced interactions 
in the medium. They can be constructed from the Dirac equation  
at finite (quark-) chemical potential,  
\be
(i\not\!\!D_I-i\mu\gamma_4)\phi_I=0 \ .
\ee 
The correct solution was obtained in \cite{Car_80,Abr_83}: 
\be
\phi_I=i\frac{\rho}{2\pi}{e^{\mu\,t}\over x}\sqrt {\rho^2+{x}^2}\delslash 
\frac{[\cos(\mu r) + {t\over r} \sin(\mu r)] 
e^{-\mu\,t}}{\rho^2+{x}^2 } \ \chi_L \ , 
\label{zm}
\ee
where the spinor $\chi_L$  arises from an antisymmetric (singlet) 
coupling  of spin and color wave functions, as before. Note that 
the solution of the adjoint Dirac equation,
\be
\phi_I^\dagger(x;-\mu) \ (i\not\!\!D_I-i\mu\gamma_4)=0 \ ,
\ee
carries the chemical potential argument with opposite sign. This
is necessary for a consistent definition of expectation values
at finite $\mu$, and in particular renders a finite norm, 
\be
\int d^4x \ \phi_I^\dagger(x;-\mu) \ \phi_I(x;\mu) =1 \ , 
\ee
whereas without the extra sign one has 
\be
\int d^4x \ \phi_I^\dagger(x;\mu) \ \phi_I(x;\mu) =\infty \ .
\ee
This singularity, corresponding to the particle-particle channel, is in 
fact directly related to well-known BCS singularity one encounters 
when resumming an effective (attractive) particle-particle interaction 
around the Fermi surface (see, \eg , ref.~\cite{AGD}).   

%%%%%%%%%%%%%%%%%%%%%%%%%%%%%%%%%%%%%%%%%%%%%%%%%%%%%%%%%%%%%%%%%%%%%%
\subsection{Instanton Form Factors}
\label{sec_formfactors} 
%%%%%%%%%%%%%%%%%%%%%%%%%%%%%%%%%%%%%%%%%%%%%%%%%%%%%%%%%%%%%%%%%%%%%%
  
 Using the explicit form of the zero mode wave function, we 
calculate the form factor from the Fourier transform 
\be
\tilde \phi&=&\int d^4x \phi(x) e^{-i k\cdot x} \nonumber \\
&=& 2i \rho \int_0^{\infty}dR\, R^3 \int_0^{\pi} d\eta \sin^2 \eta
\int_0^{\pi} d \theta \sin \theta {e^{\mu\,t}\over x}
\sqrt{\rho^2+x^2}(\gamma_0 \partial_t+\vec \gamma \cdot \hat k \cos \theta
\partial_r)\nonumber \\
& & \hspace{1cm}{(\cos(\mu r) + {t\over r} \sin(\mu r)) e^{-\mu\,t}
 \over \rho^2 + x^2}
e^{-i(\omega t+ k r \cos \theta)}\chi_L,
\ee
where $x^2=r^2+t^2$ and $\hat k=\vec k / k$. Introducing  
hyper-spherical coordinates for the integration, $r=R\sin\eta$, 
$t=R\cos\eta$, the result can be expressed through two
scalar functions $A(\omega,k,\mu)$ and $B(\omega,k,\mu)$, given in 
appendix~\ref{app_ff}, as   
\be
\tilde \phi=[\gamma_0 B(\omega,k,\mu)+\vec\gamma\cdot \hat k
A(\omega,k,\mu)] \chi_L\equiv \tilde \psi \chi_L \ ,
\ee
The finite density zero mode wave functions $\tilde \psi$ have the 
symmetry properties
\be
\tilde \psi(\omega,\vec k,\mu)=\tilde \psi^*(-\omega,-\vec k,\mu)i
=\tilde \psi^* (\omega,\vec k,-\mu) \ .
\label{sym} 
\ee
The combination ${\cal F}(\omega,\vec k,\mu)= 
\psi^*(\omega,\vec k,\mu)G_0^{-1}(\omega,
\vec k,\mu)$ appears in the effective quark interaction on each 
propagator entering or exiting the instanton-induced vertex. In the mean 
field approach, when two of the propagators participating in the vertex have 
the same momentum, it is useful to 
combine them into two new formfactors. For a propagator entering 
the instanton vertex and another exiting with the same momentum one obtains 
\be
\alpha={\cal F}(-\omega,-\vec k,\mu)^\dagger
{\cal O} {\cal F}(\omega,\vec k,\mu) \ ,
\ee
where $ {\cal O}$ is a matrix with Dirac, color and flavor indices.
For an overall unit matrix, one has  
\be
\label{alpha}
\alpha&=&(\gamma_0 i(\omega-i\mu)+i\vec\gamma\cdot \vec k)
(\gamma_0 B^*(\omega,k,\mu)+\vec\gamma\cdot \hat k
A^*(\omega,k,\mu))^2\cr
& &(\gamma_0 i(\omega-i\mu)+i\vec\gamma\cdot \vec k)\cr
&=&({A^*}^2+{B^*}^2)(k^2+(\omega-i\mu)^2)\cr
&\equiv & \alpha_r+i \alpha_i  \ .
\ee
For a propagator entering the instanton vertex and a transposed one exiting 
with the same momentum one finds  
\be
\beta={\cal O}{\cal F}(-\omega,-\vec k,\mu)^T{\cal O} 
{\cal F}(\omega,\vec k,\mu) \ .
\ee
When the Dirac part of $ {\cal O}$ is $ C \gamma_5$, where $C$ is the charge 
conjugating matrix,
\be
\label{beta}
\beta&=& C \gamma_5 (\gamma_0^T i(-\omega-i\mu)-i\vec\gamma^T\cdot \vec k)
(\gamma_0^T B^*(-\omega,k,\mu)-\vec\gamma^T\cdot \hat k
A^*(-\omega,k,\mu))\cr 
& & C \gamma_5(\gamma_0 B^*(\omega,k,\mu)+\vec\gamma\cdot
\hat k
A^*(\omega,k,\mu)) (\gamma_0 i(\omega-i\mu)+i\vec\gamma\cdot \vec k)\cr &=&
 (\gamma_0(\omega+i\mu)+\vec\gamma\cdot \vec k)
(\gamma_0 B(\omega,k,\mu)+\vec\gamma\cdot \hat k
A(\omega,k,\mu))\cr
& &(\gamma_0 B^*(\omega,k,\mu)+\vec\gamma\cdot \hat k
A^*(\omega,k,\mu)) (\gamma_0 (\omega-i\mu)+\vec\gamma\cdot \vec k)\cr &=&
(\omega^2+k^2+\mu^2)(|A|^2+|B|^2)+2\mu k i(A^*B-AB^*)\cr
& &+i\gamma_0\vec\gamma\cdot
 \hat k[2\mu k(|A|^2+|B|^2)+(\omega^2+k^2+\mu^2) i(A^*B-AB^*)]\cr 
&\equiv & \beta_r+i\gamma_0\vec\gamma\cdot\hat k\beta_i \ .
\ee
We note that $\alpha$ and $\beta$ have the same symmetry as in 
Eq.~(\ref{sym}).
%%%%%%%%%%%%%%%%%%%%%%%%%%%%%%%%%%%%%%%%%%%%%%%%%%%%%%%%%%%%%%%%%%%%%%%%%%%
\begin{figure}
\begin{center}
\psfig{file=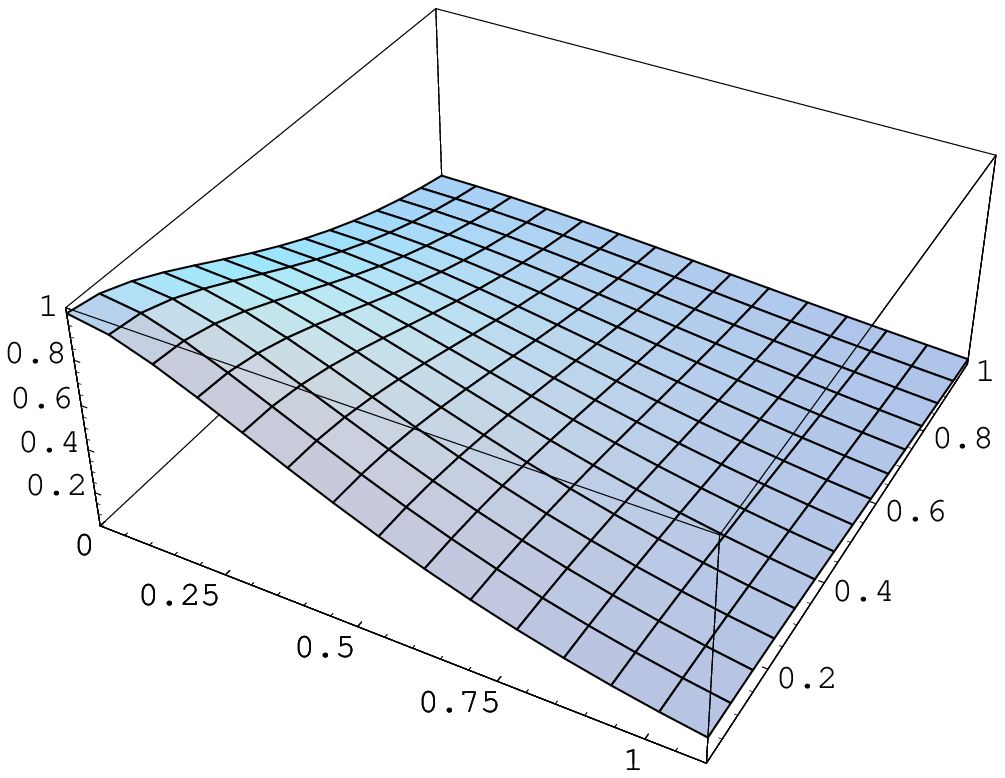,angle=0,height=5 cm}
\hspace{0.5cm}
\psfig{file=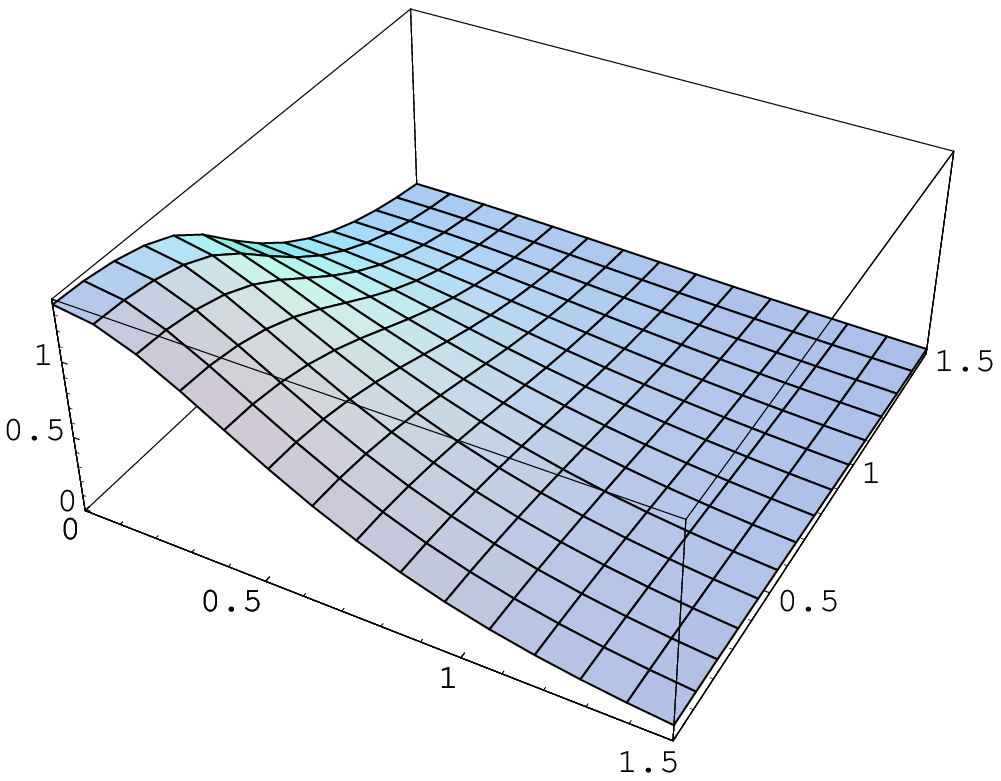,angle=0,height=5 cm}
\psfig{file=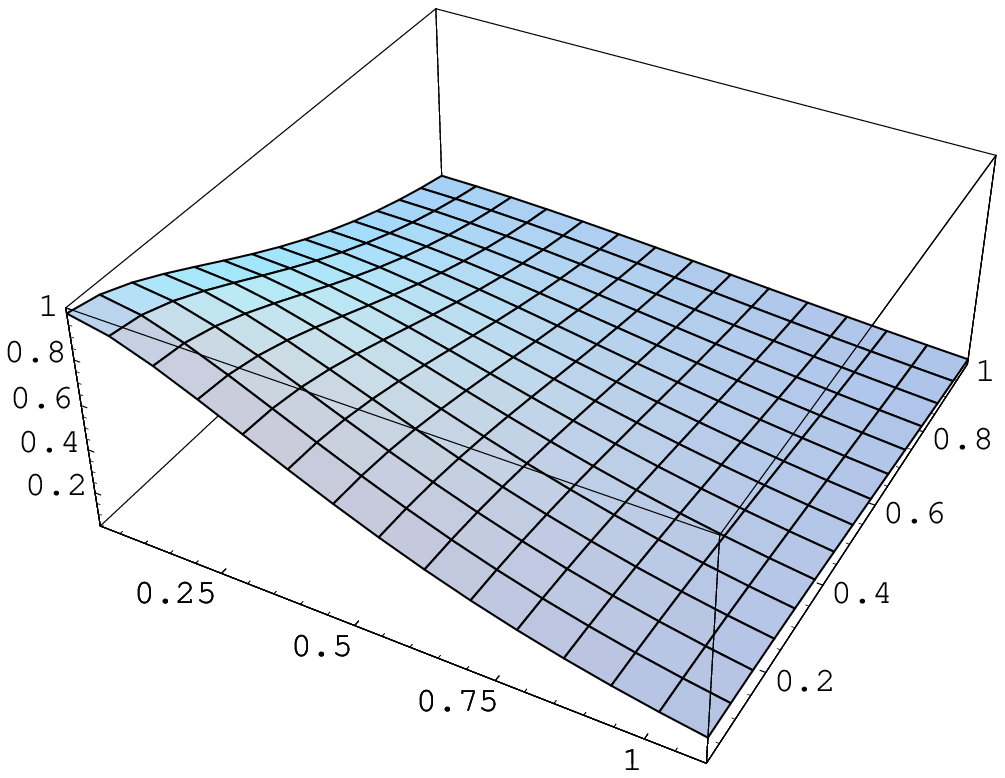,angle=0,height=5 cm}
\hspace{0.5cm}
\psfig{file=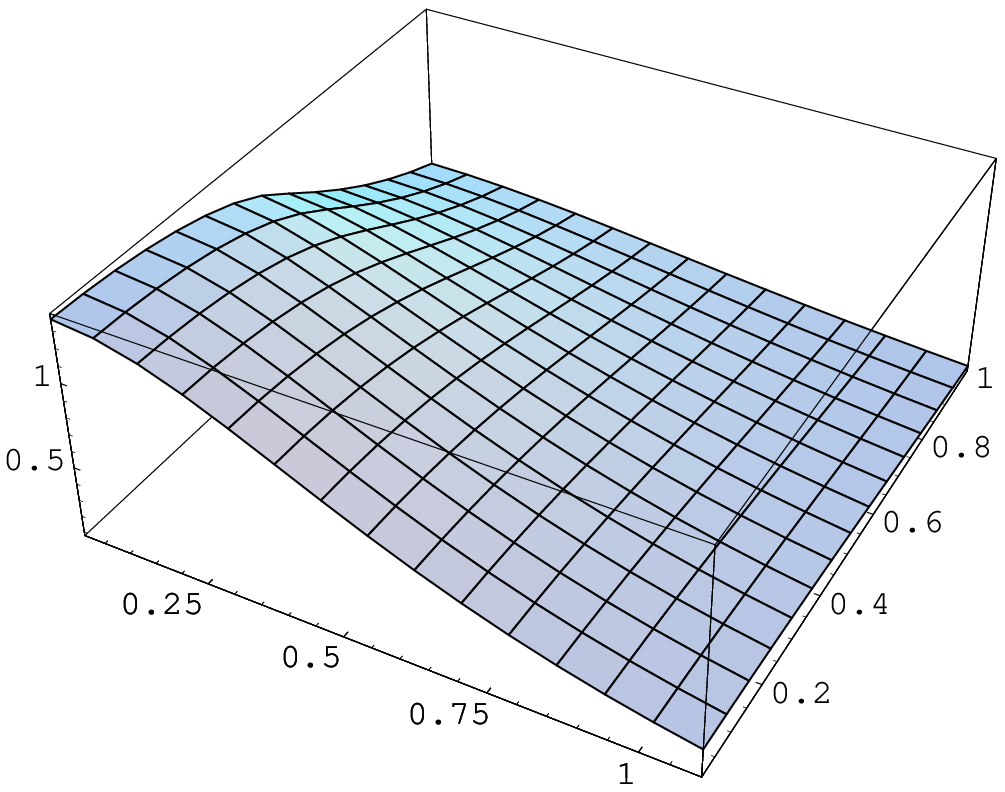,angle=0,height=5 cm}
\end{center}
\vspace{0.2cm}
\caption{The moduli squared of the form factors, 
$|\alpha|^2=\alpha_r^2+\alpha_i^2$  (top panels) and
$|\beta|^2=\beta_r^2-\beta_i^2$ 
(bottom panels),  for two different 
values of the chemical potential: $\mu=0$ (left column) and 
$\mu=300$~MeV (right column) .}
\end{figure}
%%%%%%%%%%%%%%%%%%%%%%%%%%%%%%%%%%%%%%%%%%%%%%%%%%%%%%%%%%%%%%%%%%%%%%%%%%

%%%%%%%%%%%%%%%%%%%%%%%%%%%%%%%%%%%%%%%%%%%%%%%%%%%%%%%%%%%%%%%%%%%%%%
\subsection{Quark Overlap Matrix Elements}
\label{sec_tia} 
%%%%%%%%%%%%%%%%%%%%%%%%%%%%%%%%%%%%%%%%%%%%%%%%%%%%%%%%%%%%%%%%%%%%%%

The instanton formfactors discussed above are designed for momentum
space calculations, in particular for extracting effective interactions
between quarks in the mean-field framework. However,  
in the statistical mechanics treatment of the instanton liquid
partition function presented in sect.~\ref{sec_cocktail}, 
the coordinate space description is the more suitable one.
For that we will need the explicit form of the fermionic overlap 
matrix element $T_{IA}$ which at finite density takes the form  
\be
T_{IA}(z,u;\mu) & = & \ \int d^4x \
\phi_I^\dagger(x-z_I;-\mu) \ (i\Dslash-i\mu\gamma_4) \ \phi_A(x-z_{A};\mu)
\nonumber\\
& = & - \int d^4x \
\phi_I^\dagger(x-z_I;-\mu) \ (i\delslash-i\mu\gamma_4) \
\phi_A(x-z_{A};\mu) \ . 
\label{tiamu}
\ee
The second line is again obtained by virtue of the Dirac equation when 
choosing the sum ansatz for the gauge-field configurations, $A_\mu=A_\mu^I+
A_\mu^A$. $T_{IA}$ plays a crucial role in the fermionic determinant 
of the instanton partition function, where it generates the fermionic 
interaction ('quark hopping amplitude') 
between $I$'s and $A$'s and is therefore responsible for 
correlations in the instanton liquid. In particular, $T_{IA}$ controls 
the probability of forming molecules.

\begin{figure}[tb]
\begin{center}
\epsfig{file=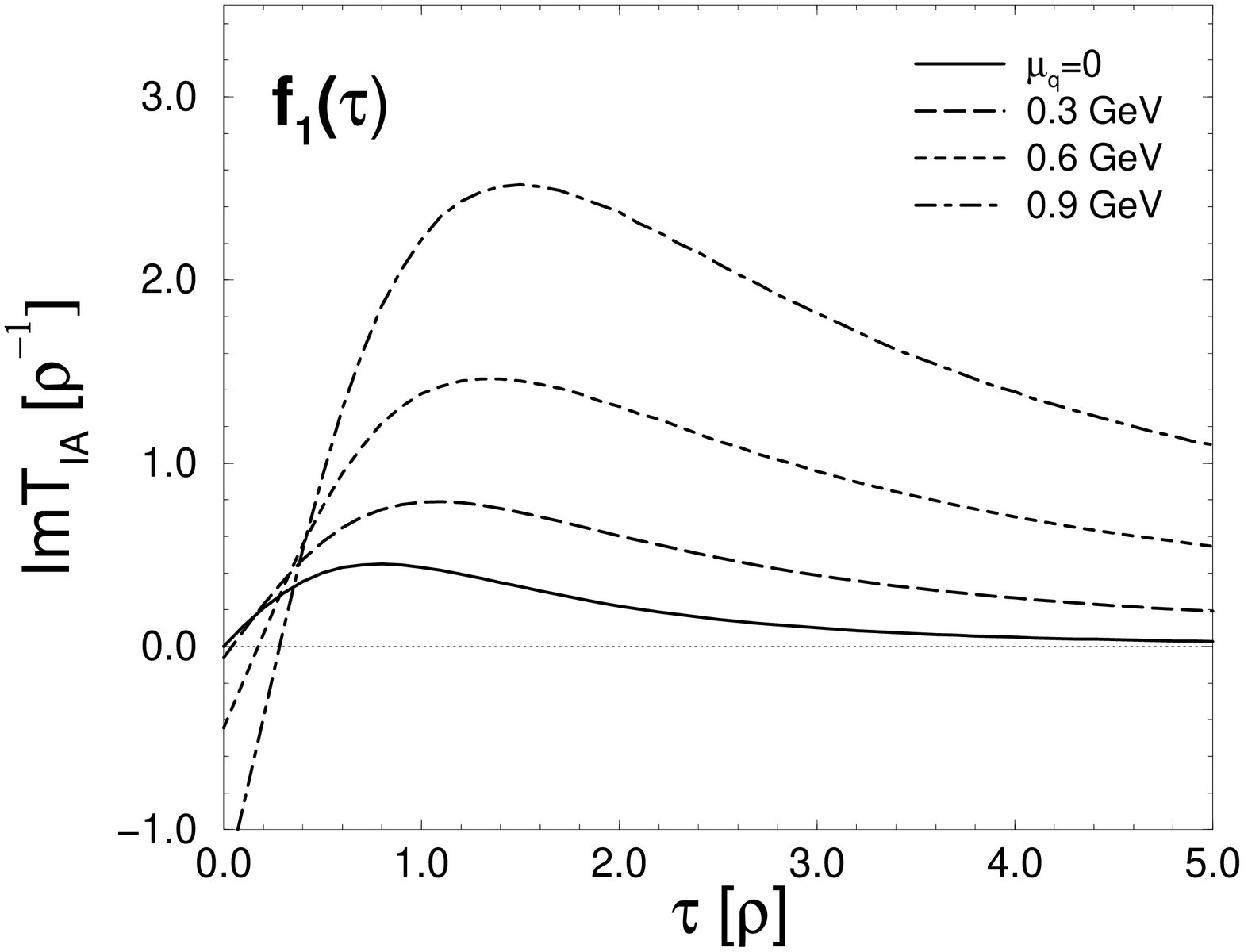,width=6.5cm,angle=-90}
\epsfig{file=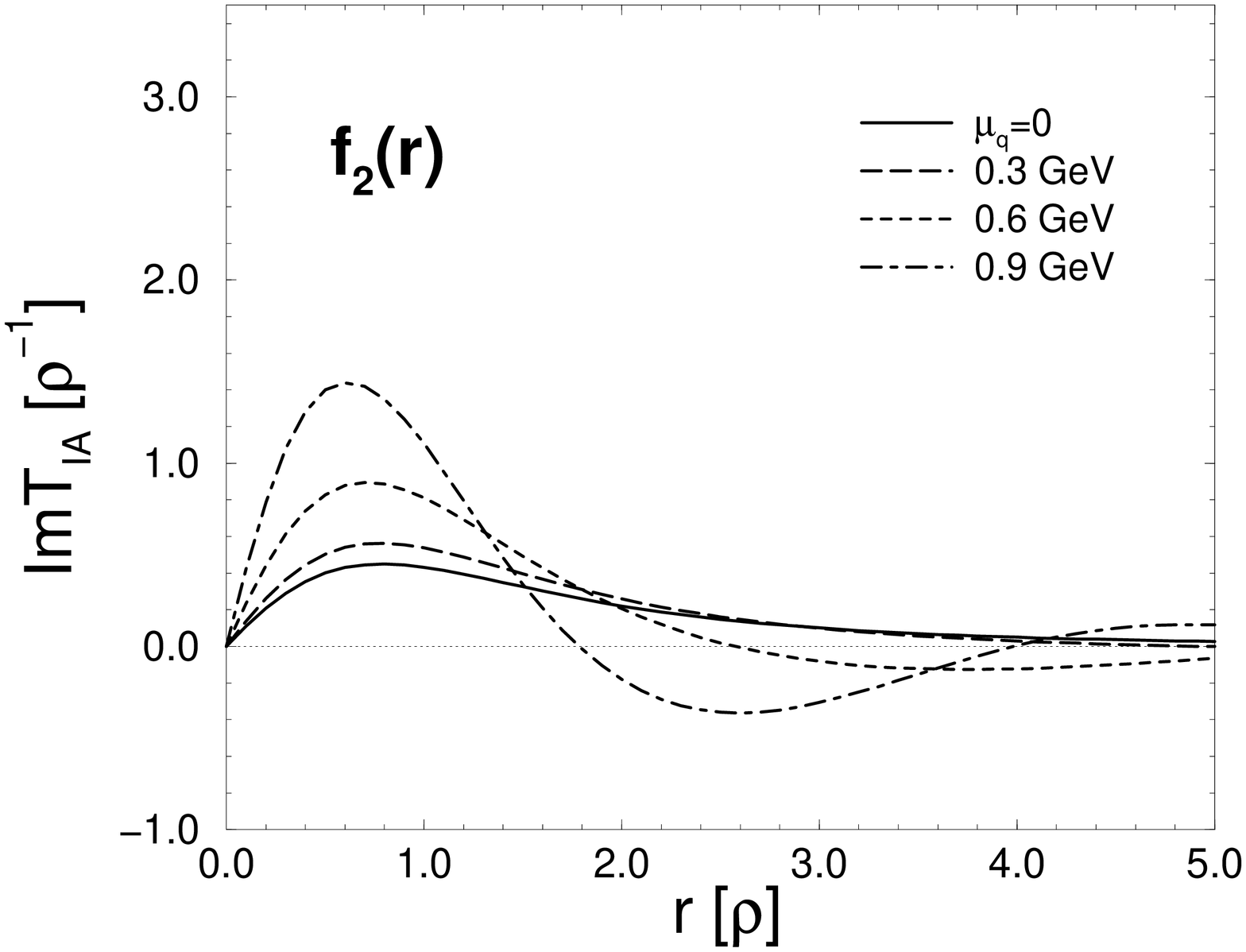,width=6.5cm,angle=-90}
\end{center}
\caption{Quark-induced $I$-$A$-interaction at finite density
for the most attractive color orientation $u_4$=1, $\vec u$=0 as well
as $r$=0 (left panel) and for $u_4$=0, $|\vec u|$=1 and $\tau$=0
(right panel). }
\label{fig_tia}
\end{figure}

 The definite chirality of the zero modes (in the limit of vanishing 
current quark masses) entails  that $I$-$I$ and $A$-$A$ matrix elements 
are zero. In the vacuum Lorentz invariance implies that the overlap matrix 
element can be characterized by a single scalar function~\cite{SV91}, \eg, 
$T_{IA}\equiv i \ u\cdot \hat{z} \ f(z)$. In the medium this is 
no longer true and $T_{IA}$ must be calculated in terms of two independent 
scalar functions $f_1,f_2$ according to 
\be
T_{IA}(z,u;\mu) 
\equiv  i \ u_4 \ f_1(\tau,r;\mu) + i \
\frac{(\vec u \cdot \vec r)}{r} \ f_2(\tau,r;\mu) \ .
\label{tiamu2}
\ee
They are shown in Fig.~\ref{fig_tia}, see also \cite{Rapp_98,Sch_98}. 
Similar to the finite temperature case, we observe a strong 
enhancement with increasing $\mu$ in the temporal direction. 
Moreover, the fermionic interaction becomes very long range, 
$\sim \mu^2/x_4$ (at finite temperature it was limited to 
the Matsubara box of size $1/T$ enforced by periodic boundary 
conditions). In the spatial direction, the exponential damping 
$\exp[-\pi T r]$ in the finite-$T$ case is replaced  
by oscillations $\sim \sin(\mu r)$. 
The latter also effectively suppress the hopping amplitude once 
the $r$-integration in the partition function is performed. 
%%%%%%%%%%%%%%%%%%%%%%%%%%%%%%%%%%%%%%%%%%%%%%%%%%%%%%%%%%%%%%%%%%
%%%%%%%%%%%%%%%%%%%%%%%%%%%%%%%%%%%%%%%%%%%%%%%%%%%%%%%%%%%%%%%%%%
 From the strong  enhancement of $T_{IA}$ with increasing $\mu$ 
one may already  anticipate the relevance of molecule configurations  
at finite densities~\cite{Rapp_98}. This issue will be quantitatively 
investigated  in the 'cocktail' model in sect.~\ref{sec_cocktail}. 

%%%%%%%%%%%%%%%%%%%%%%%%%%%%%%%%%%%%%%%%%%%%%%%%%%%%%%%%%%%%%%%%
\section{The two-flavor problem}
\label{sec_nf2}
%%%%%%%%%%%%%%%%%%%%%%%%%%%%%%%%%%%%%%%%%%%%%%%%%%%%%%%%%%%%%%%% 

  This section will be devoted to study finite density two-flavor 
QCD in the mean-field approximation using the exact momentum 
dependent instanton profile functions as discussed in 
sect.~\ref{sec_formfactors}. As density increases the basic competition 
will be between the chiral condensate $\langle \bar qq\rangle$
and the scalar $ud$ diquark condensate in the $\langle qq\rangle$  
channel, representing the  color superconducting state as discussed 
in Refs.~\cite{RSSV_98,ARW_98}. We will first discuss the corresponding
coupled gap equations formalism (sect.~\ref{sec_nf2A}) and then 
proceed to the numerical results and their interpretation in  
sect.~\ref{sec_nf2B}.   

%%%%%%%%%%%%%%%%%%%%%%%%%%%%%%%%%%%%%%%%%%%%%%%%%%%%%%%%%%%%%%%%%%%
\subsection{Mean-Field Grand Canonical Potential at Finite $\mu$}
\label{sec_nf2A}
%%%%%%%%%%%%%%%%%%%%%%%%%%%%%%%%%%%%%%%%%%%%%%%%%%%%%%%%%%%%%%%%%

\begin{figure}[htb]
\begin{center}
\psfig{file=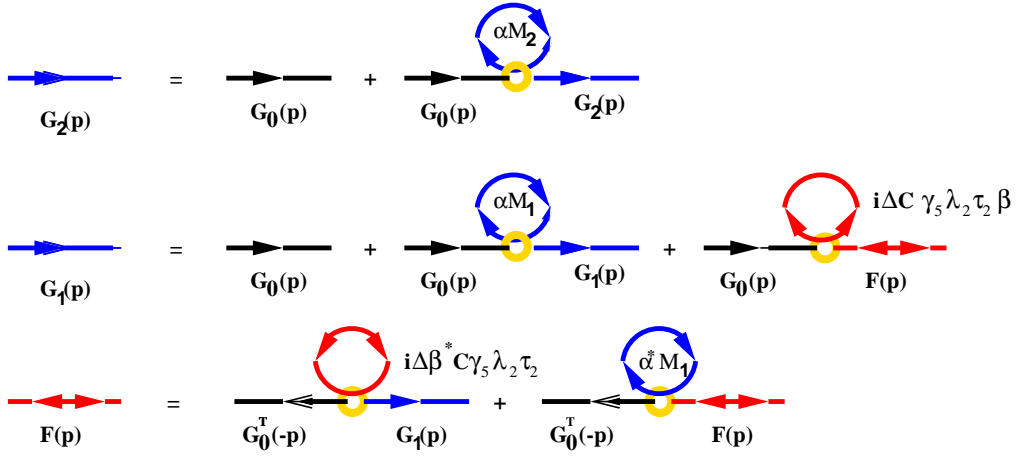,angle=0,height=6 cm}
\end{center}
\caption{Diagrammatic representation of the Gorkov
Eqs.~(\protect\ref{Gorkov}).}
\label{fig_Gorkov}
\end{figure}

For the evaluation of the grand canonical potential we employ the 
Cornwall-Jackiw-Tomboulis (CJT)~\cite{CJT} effective action,
which is elucidated in more detail in Appendix~\ref{app_CJT}.
It involves 3 type of propagators (including their antiparticle
pendants) corresponding to single (anti-) quarks carrying color 
charge that participates in the diquark condensate ($G_1$, $\bar G_1$), 
single (anti-) quarks carrying color charge that is not part of the 
diquark condensate ($G_2$, $\bar G_2$), and (anti-) diquarks
($F$ $\bar F$).  
The minimization of the action with respect to (\wrt) 
these propagators generates the following six gap equations: 
\be
-(G_1-F\bar G_1^{-1}\bar F)^{-1}+G_0^{-1}-M_1 \alpha&=&0
\nonumber\\
-(\bar G_1-\bar F G_1^{-1}F)^{-1}+\bar G_0^{-1}+M_1 \alpha^*&=&0
\nonumber\\
-G_2^{-1}+G_0^{-1}-M_2 \alpha^*&=&0
\nonumber\\ 
-\bar G_2^{-1}+\bar G_0^{-1}+M_2 \alpha^*&=&0
\nonumber\\
\bar G_1^{-1}\bar F(G_1-F\bar G_1^{-1}\bar F)^{-1}+i\Delta \beta&=&0
\nonumber\\
 G_1^{-1}F(\bar G_1-\bar F G_1^{-1}F)^{-1}+i\Delta \beta^*&=&0 \ , 
\label{pre-gap}
\ee
where 
\be
M_1&=&2g\left({1\over 8 N_c^2}(\tr(G_1+G_2)\alpha)+{1\over
\sqrt{2}}{N_c-2\over 16 N_c^2 (N_c^2-1)}(\tr
\lambda_8(G_1+G_2)\alpha)\right),\nonumber \\ 
M_2&=&2g\left({1\over 8 N_c^2}(\tr(G_1+G_2)\alpha)-{2\over
\sqrt{2}}{N_c-2\over 16 N_c^2 (N_c^2-1)}(\tr
\lambda_8(G_1+G_2)\alpha)\right),\nonumber \\ 
\Delta&=&2g{1\over 8 N_c^2 (N_c-1)}\tr(FC \gamma_5 \lambda_2 \tau_2
\beta),
\label{mass_gap}
\ee
are the two chiral masses and the diquark gap, and $G_0$ ($\bar G_0$)
is the bare (anti-) quark propagator defined through Eq.~(\ref{S_source}). 
As before, the traces
involve momentum integrations. The chiral masses are
linear combinations of the $\bar q q$ condensates, $ \tr (G_1 \alpha)$
and $ \tr (G_2 \alpha)$, while the gap $\Delta$ is proportional to the
$qq$ condensate, $\tr (FC \gamma_5 \lambda_2 \tau_2\beta)$.
Eqs.~(\ref{mass_gap}) represent a coupled system of gap equations
in the chiral and diquark masses, $M_1, M_2$ and $\Delta$. 
To determine their solutions, one needs to know the explicit form
of the propagators $G_1, G_2$ and $F$. They are constructed from
the coupled set of Eqs.~(\ref{pre-gap}).    
Using the relation  $\bar G(p)=-G^T(-p)$ and the transposition property 
$C\gamma_5 \gamma_\mu^T= \gamma_\mu C\gamma_5$, 
one can rearrange Eqs.~(\ref{pre-gap}) into Gorkov-type  equations
(note that $G,F,\bar F$ do not commute) as 
\be
G_2(p)&=&G_0(p)+G(p)M_2\alpha(p) G_0(p)\nonumber \\
G_1(p)&=&G_0(p)+G(p)M_1\alpha(p) G_0(p)+F(p)(i\Delta C \gamma_5
\lambda_2 \tau_2 \beta(p)G_0(p)\nonumber \\
F(p)&=&F(p)M_1\alpha^*(p)G_0^T(-p)+G(p)i\Delta\beta^*(p) C \gamma_5
\lambda_2 \tau_2 G_0^T(-p) \ . 
\label{Gorkov}
\ee
A graphic representation of these equations is displayed in 
Fig.~\ref{fig_Gorkov}. 

The Gorkov equations can be solved in algebraic form yielding  
\be
G_2(p)&=&(G_0^{-1}(p)-M_2\alpha(p))^{-1}\nonumber \\
G_1(p)&=&\left(G_0^{-1}(p)-M_1\alpha(p)+\Delta^2\beta^*(p)(G_0^{-1}(-p)-
M_1\alpha^*(p))^{-1}\beta(p) \right)^{-1}\nonumber \\
F(p)&=&i\Delta G_1(p) \beta^*(p)(G_0^{-1}(-p)-M_1\alpha^*(p))^{-1} 
C \gamma_5 \lambda_2 \tau_2\nonumber \\
&=& i\Delta (G_0^{-1}(p)-M_1\alpha(p))^{-1}\beta^*(p)G_1(-p) C \gamma_5
\lambda_2 \tau_2\nonumber \\
&\equiv & \tilde F(p)C \gamma_5
\lambda_2 \tau_2\nonumber \\
\bar F(p)&=&i\Delta C \gamma_5 \lambda_2 \tau_2
(G_0^{-1}(-p)-M_1\alpha^*(p))^{-1} \beta(p) G_1(p)\nonumber \\
&=&i\Delta C \gamma_5 \lambda_2 \tau_2 G_1(-p) \beta(p)(G_0^{-1}(p)-
M_1\alpha(p))^{-1}\nonumber \\
&\equiv & C \gamma_5
\lambda_2 \tau_2 \tilde{\bar F}(p) \ . 
\label{prop}
\ee

%%%%%%%%%%%%%%%%%%%%%%%%%%%%%%%%%%%%%%%%%%%%%%%%%%%%%%%%%%%%%%%
%%%%%%%%%%%%%%%%%%%%%%%%%%%%%%%%%%%%%%%%%%%%%%%%%%%%%%%%%%%%%%%

To obtain  the thermodynamic state variables such as pressure
and energy density we need to know the explicit dependence of the 
thermodynamic potential on the mass parameters. This can be achieved  
by reinserting the explicit solutions of the propagators into the
effective action, Eq.~(\ref{Gamma}), constituing the grand canonical
potential times the 4-volume,  $-V_4\Omega(M_1,M_2,\Delta)$. 
For that purpose we evaluate the propagators more explicitly. 
After some  algebra, one can rewrite $G_1$ from Eq.~(\ref{prop}) 
as
\be
\label{G_1}
G_1(p)&=&1\!\!1_2^{color}\otimes 1\!\!1_2^{flavor}\otimes 
 \left\{ -i\gamma_0 \left[(\omega-i\mu)((\omega+i\mu)^2+k^2+
 \tilde M_1^{* 2})\right.\right. \nonumber \\
& & \left. \left. +(\omega+i\mu)
 \Delta^2 (\beta_r^2+\beta_i^2)-2ik\Delta^2
\beta_r \beta_i \right]-i \vec \gamma \cdot \hat k \right. \nonumber \\
& & \left.
\times \left[k((\omega+i\mu)^2+k^2+\tilde M_1^{* 2})+k 
\Delta^2 (\beta_r^2+\beta_i^2)+2i(\omega+i\mu)\Delta^2
 \right] \right. \nonumber \\ 
& & \left.
-\tilde M_1\left[ ((\omega+i\mu)^2+k^2+\tilde M_1^{* 2})+\tilde
M_1^*\Delta^2(\beta_r^2- \beta_i^2)\right] \right\} {\cal D}^{-1},
\ee
where $\tilde M_1(p)=M_1 \alpha(p)$, and 
\be
\label{Denom}
{\cal D}&=&|(\omega-i\mu)^2+k^2+\tilde M_1^{2})|^2
+ \Delta^4(\beta_r^2- \beta_i^2)^2-8 k\mu \Delta^2 \beta_r \beta_i \nonumber \\
& & +2\Delta^2 (\omega^2+k^2+\mu^2)(\beta_r^2+ \beta_i^2)+2|\tilde
M_1|^2\Delta^2(\beta_r^2- \beta_i^2) .
\ee
Using the relation (\ref{det_trick}) for the Dirac part ($\tr_D$) 
of the trace-log  in  the
kinetic part of Eq.~(\ref{Gamma}) (here the trace does not
include the momentum integration), one finds 
\be
{1 \over 2}\tr_D \ln (-\bar G_1 G_1+\bar G_1 F \bar G_1^{-1}\bar F)
=-4\ln {\cal D},
\ee
and from Eq.~(\ref{G_1}), 
\be
{\rm Re} \left[\tr_D(G_0^{-1}G_1-1)\right]
&=&-4\left[{\rm Re}[((\omega+i\mu)^2+k^2)\tilde
M_1^2]+|\tilde M_1|^4+ \Delta^4(\beta_r^2- \beta_i^2)^2 \right. \nonumber \\ 
& &\left. -4 k\mu \Delta^2 \beta_r \beta_i 
+\Delta^2 (\omega^2+k^2+\mu^2)(\beta_r^2+ \beta_i^2)\right. \nonumber \\ 
& &\left.+2|\tilde M_1|^2\Delta^2(\beta_r^2- \beta_i^2)\right]{\cal D}^{-1},
\\
{\rm Re} \left[\tr_D(G_1 \alpha)\right]
&=& -{4\over M_1}\left[{\rm Re}[((\omega+i\mu)^2+k^2)\tilde
M_1^2]+|\tilde M_1|^4\right. \nonumber \\ 
& &\left.+|\tilde M_1|^2\Delta^2(\beta_r^2-
\beta_i^2)\right]{\cal D}^{-1}, 
\\
{\rm Re} \left[\tr_D(\tilde F \beta)\right]
&=& {4\over \Delta}\left[\Delta^4(\beta_r^2-
\beta_i^2)^2-4 k\mu \Delta^2 \beta_r \beta_i  
+\Delta^2 (\omega^2+k^2+\mu^2)
\right. 
\\
& & \left. \times (\beta_r^2+ \beta_i^2)
+|\tilde M_1|^2\Delta^2(\beta_r^2- \beta_i^2)\right]{\cal D}^{-1} \ .
\ee
The analogous quantities involving  $G_2$ and $M_2$ are 
obtained from the above by substituting $\Delta\to 0$, $M_1 \to M_2$.
 
Using the above relations our final expression for  $\Omega$ becomes 
\be
\label{Omega}
\Omega(M_1,M_2,\Delta)&=& \int_0^{\infty} d\omega {k^2
dk \over 2 \pi^3}\biggl\{ -4 \ln {\cal D}-2 \ln|(\omega-i\mu)^2+k^2+\tilde
M_2^{2})|^2 
\nonumber\\ 
& &  -4{\rm Re}\left[\tr_D(G_0^{-1}G_1-1)\right]
 +8{{\rm Re}[((\omega+i\mu)^2+k^2)\tilde
M_2^2]+|\tilde M_2|^4 \over|(\omega-i\mu)^2+k^2+\tilde
M_2^{2})|^2} \biggr\} 
\nonumber\\ 
& & -{g \over 18}\Big\{\int_0^{\infty} d\omega
{k^2 dk \over 2 \pi^3} {\rm Re}\left[ \tr_D (2 G_1 \alpha +G_2
\alpha)\right] \Big\}^2
 -{g \over 144}\Big\{\int_0^{\infty} d\omega
{k^2 dk \over 2 \pi^3} {\rm Re}\left[ \tr_D ( G_1 \alpha -G_2
\alpha)\right] \Big\}^2
\nonumber\\
& & -{g \over 6}\Big\{\int_0^{\infty} d\omega
{k^2 dk \over 2 \pi^3} {\rm Re}\left[ \tr_D (\tilde F \beta)\right] \Big\}^2
 \ . 
\ee
The global minimum of $\Omega$ \wrt~$M_1,M_2,\Delta$ at each
$\mu$ determines the thermodynamically stable phase and the values of
$M_1,M_2,\Delta$ are the chiral masses and color superconducting gap
in that phase. The extrema of $\Omega$ at each $\mu$ can be found 
by equating the derivatives \wrt~$M_1,M_2,\Delta$ to zero. 
Equivalently, one can verify that after differentiating the expression 
for $\Omega$ \wrt~$M_1,M_2$ and $\Delta$ one recovers Eqs.~(\ref{mass_gap}).
Their solutions correspond to the local extrema of $\Omega$ and represent  
different possible phases. We shall discuss them in the next subsection. Phase
transitions correspond to two distinct minima of $\Omega$ having an equal 
value (first order), or merging together (second order). 

One should recall that in this fomulation the coupling constant $g$ is an
integration variable: an inverse Laplace transformation was used to
exponentiate the instanton vertex. However, in the thermodynamic limit
the saddle-point approximation for the $g$-integration becomes exact 
(since it is multiplied by the 4-volume $V_4$).
Identifying the potential energy of $\Omega$ in Eq.~(\ref{Omega}) as
$-gU$, the integral in question is
\be
{\cal Z} \propto \int dg \ \exp[-gU+\frac{N}{V} \ \ln(g)] \ ,
\ee
where $N/V$ is the total instanton density. The saddle
point is then found to be at
\be
\label{g} 
g_{max}= \frac{N}{V} \ \frac{1}{U} \ .  
\ee
Thus, at the saddle point the new potential energy is
$- N/V \ln(U)$, up to a constant.

The real question is how to determine the 
$\mu$ dependence of $N/V$. This will be addressed in 
sect.~\ref{sec_cocktail} within a statistical mechanics 
treatment taking into account correlations in the instanton ensemble.
It will be shown there that the simplifying 
assumption of a constant total instanton density is indeed reasonably 
justified.
Another approximation concerns the density-dependence of 
the second key parameter, the 
average instanton radius $\rho$, which defines the scale in all 
instanton calculations. Lacking better knowledge, we also assume  
that it does not vary significantly at the chemical potentials
under consideration.   

%%%%%%%%%%%%%%%%%%%%%%%%%%%%%%%%%%%%%%%%%%%%%%%%%%%%%%%%%%%%%%%%%%%
\subsection{$N_f=2$ Phase Diagram}
\label{sec_nf2B}
%%%%%%%%%%%%%%%%%%%%%%%%%%%%%%%%%%%%%%%%%%%%%%%%%%%%%%%%%%%%%%%%%%%
Solutions of the gap equations are extrema of 
$\Omega$, but only  minima represent a thermodynamic phase.
There are 4 types of solutions to (\ref{mass_gap}):
\begin{enumerate}
\item There is a chiral condensate, but no diquark one, \ie, $M_1=M_2=M,
      \Delta=0$,
\item There is a diquark condensate, but no chiral one, \ie, $M_1=M_2=0,
      \Delta\neq 0$,
\item Both condensates are present, \ie, all $M_1,M_2,\Delta\neq 0$,
\item No condensates at all, \ie, $M_1=M_2=\Delta=0$. 
\end{enumerate}
The last case corresponds to free quarks and it is easy to see that, at
$T=0$, having at least one condensate is always more favorable.  
The rather complicated expressions for the formfactors require  
the integrals in Eq.~(\ref{mass_gap}), together with Eq.~(\ref{g}), 
 to be evaluated numerically for each value of  
$\mu$ varying from 0 to 500~MeV. We shall call the three types of 
solutions described above phase 1, 2 and 3, and fix our 
two input numbers as $\rho=1/3$~fm 
and $(N/V)\rho^4=0.0116$, corresponding to $N/V=0.94$~fm$^{-4}$. 

\begin{figure}[htb]
\epsfig{file=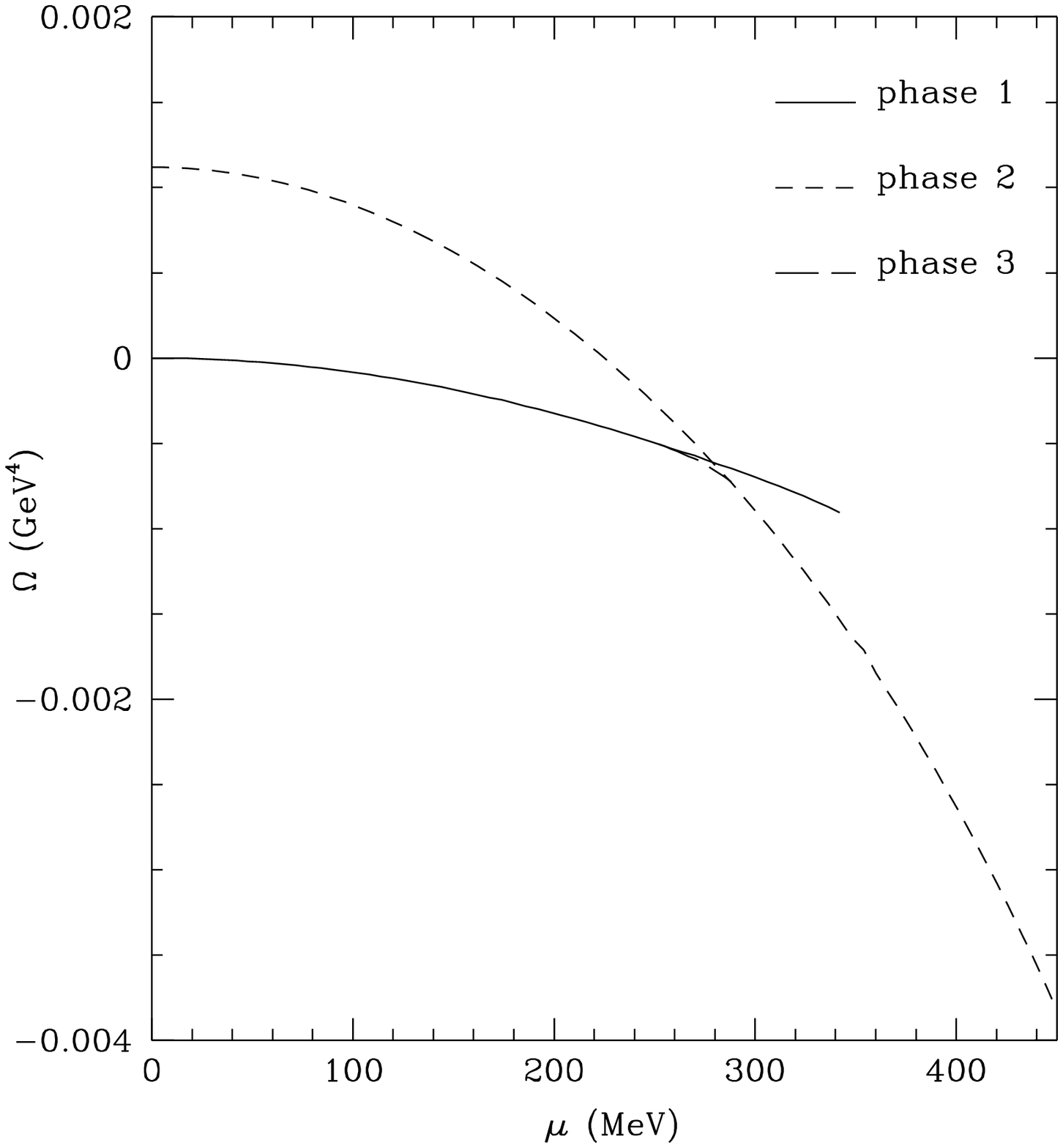,angle=0,width=8cm}
\epsfig{file=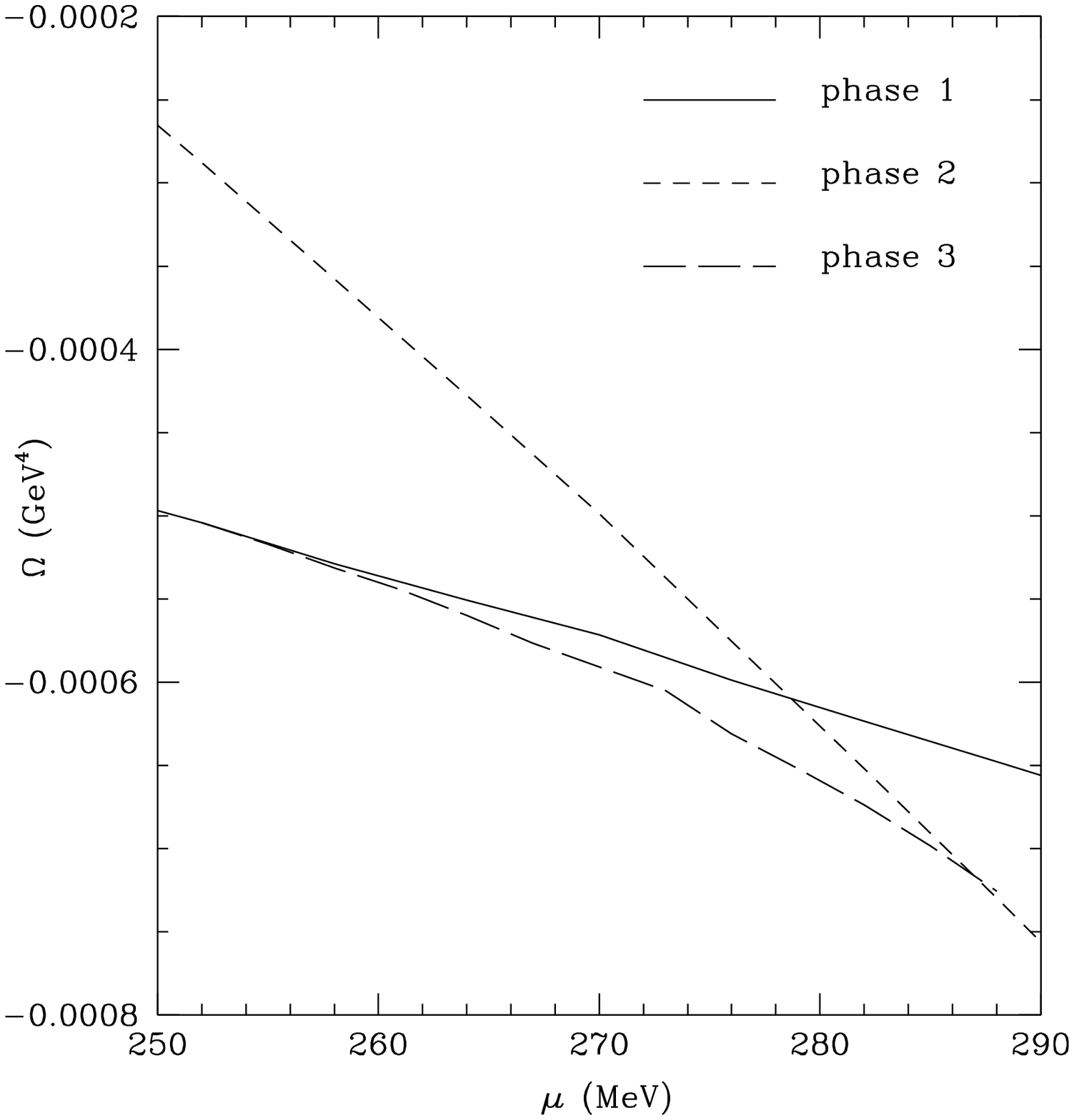,angle=0,width=8cm}
\caption{$\Omega$ for phases 1, 2 and 3. The right panel is a magnification
of the region between 250 and 290~MeV, where phase 3 exists.}
\label{omega_pict}
\end{figure}

\begin{figure}[htbt]
\psfig{file=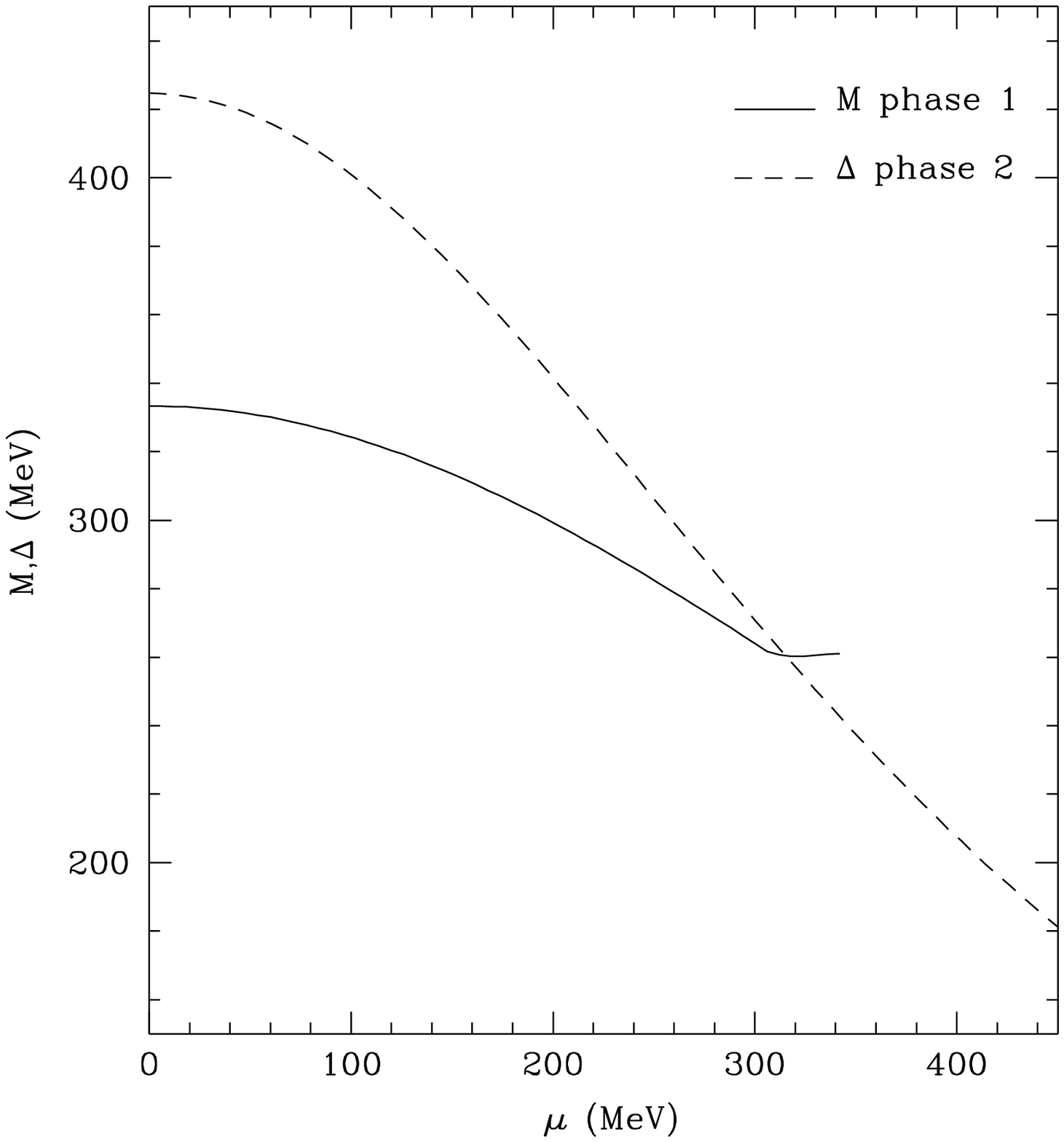,angle=0,width=8cm}
\psfig{file=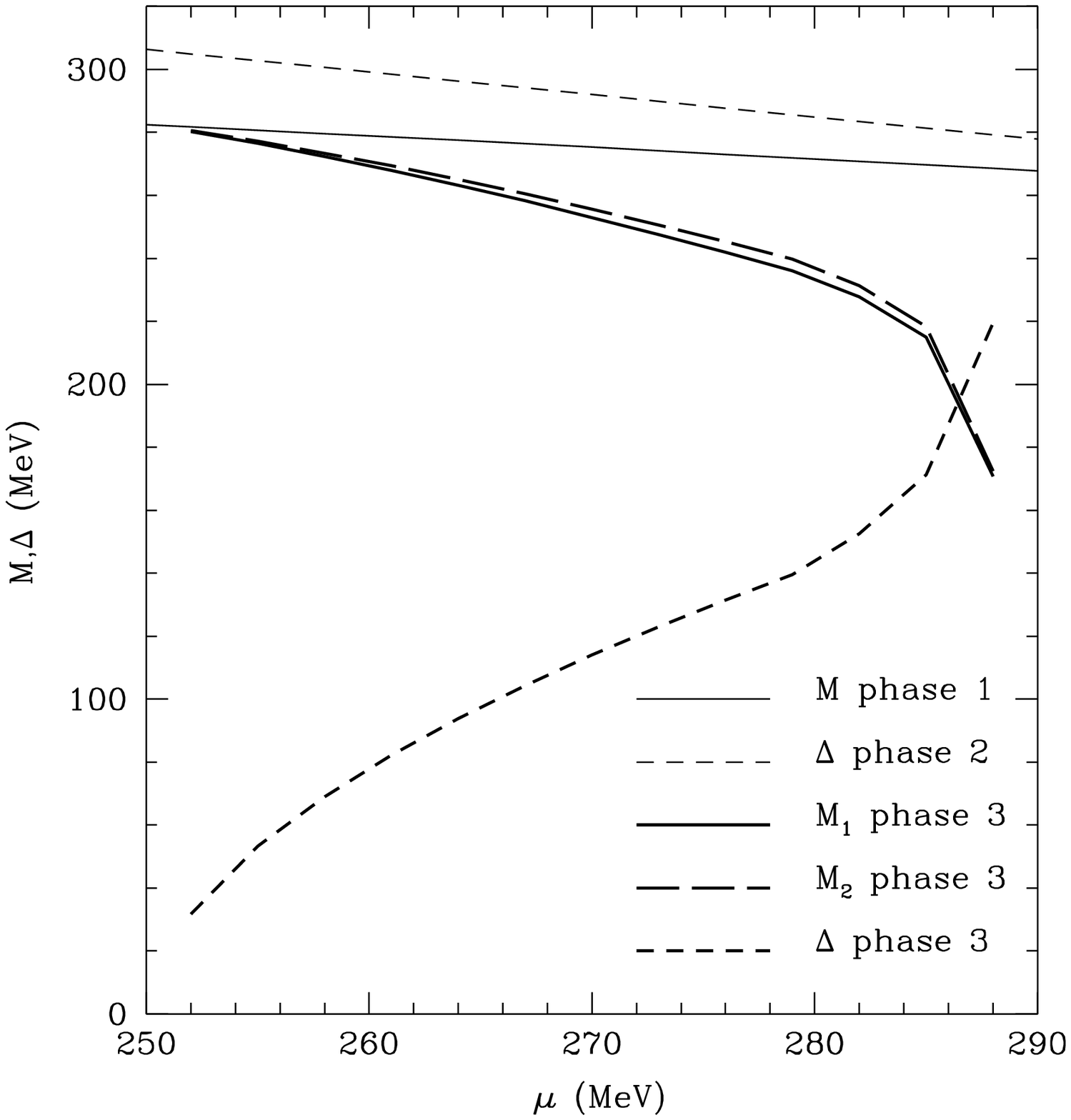,angle=0,height=8cm}
\caption{The left panel shows $M$ for phase 1 and $\Delta$ for phase
2. The right panel is a magnification of the region between 250 and 290
MeV, with the masses and gaps of all three phases.}
\label{mgap_pict}
\end{figure}

With these values we find that at $\mu=0$ the system is in phase 1
with the familiar value of about 330~MeV for the chiral mass M at $\mu=0$. 
We normalize the grand-canonical potential such that the minimum at 
$\mu=0$ has zero pressure, $\Omega=0$. We find minima for phase 1
in the range of $\mu$ from 0 to 360~MeV, for phase 2 for
all values of $\mu$ and for phase 3 only for values of $\mu$
between 250 and 290~MeV.
The values of $\Omega$ for all three phases are shown in 
Fig.~\ref{omega_pict}. 
Phase 1 dominates until 250~MeV where the system makes a transition to  
phase 3, and then, at 288~MeV, there is a transition to phase 2. 
The coupling constant $g$ for all three phases changes little (not shown). 
The mass $M$ for phase 1, the gap $\Delta$ for phase 2 and the two
masses $M_1,M_2$ and the gap $\Delta$ for phase 3 are shown in 
Fig.~\ref{mgap_pict}.

\begin{figure}[htb]
\psfig{file=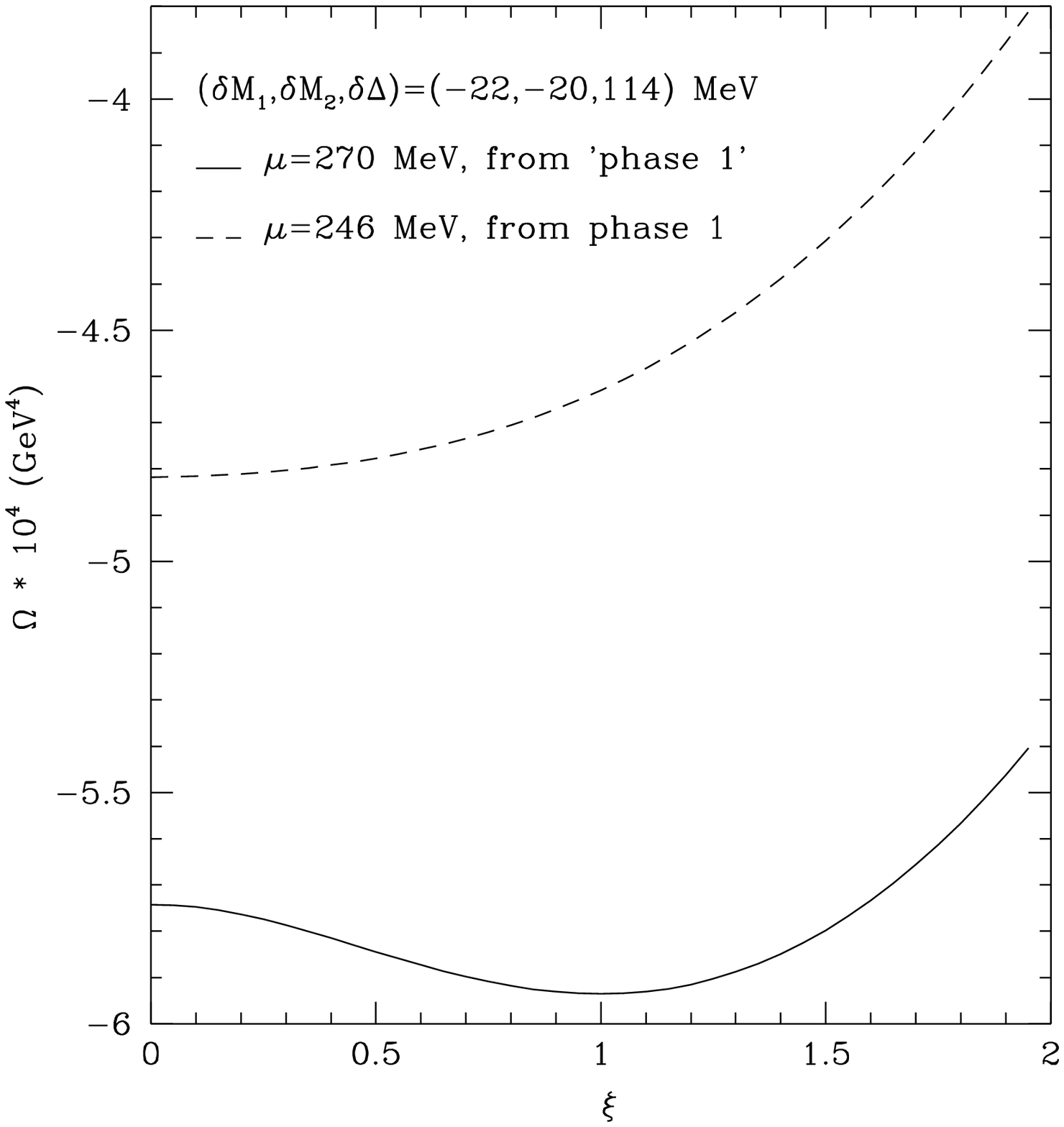,angle=0,width=8cm}
\psfig{file=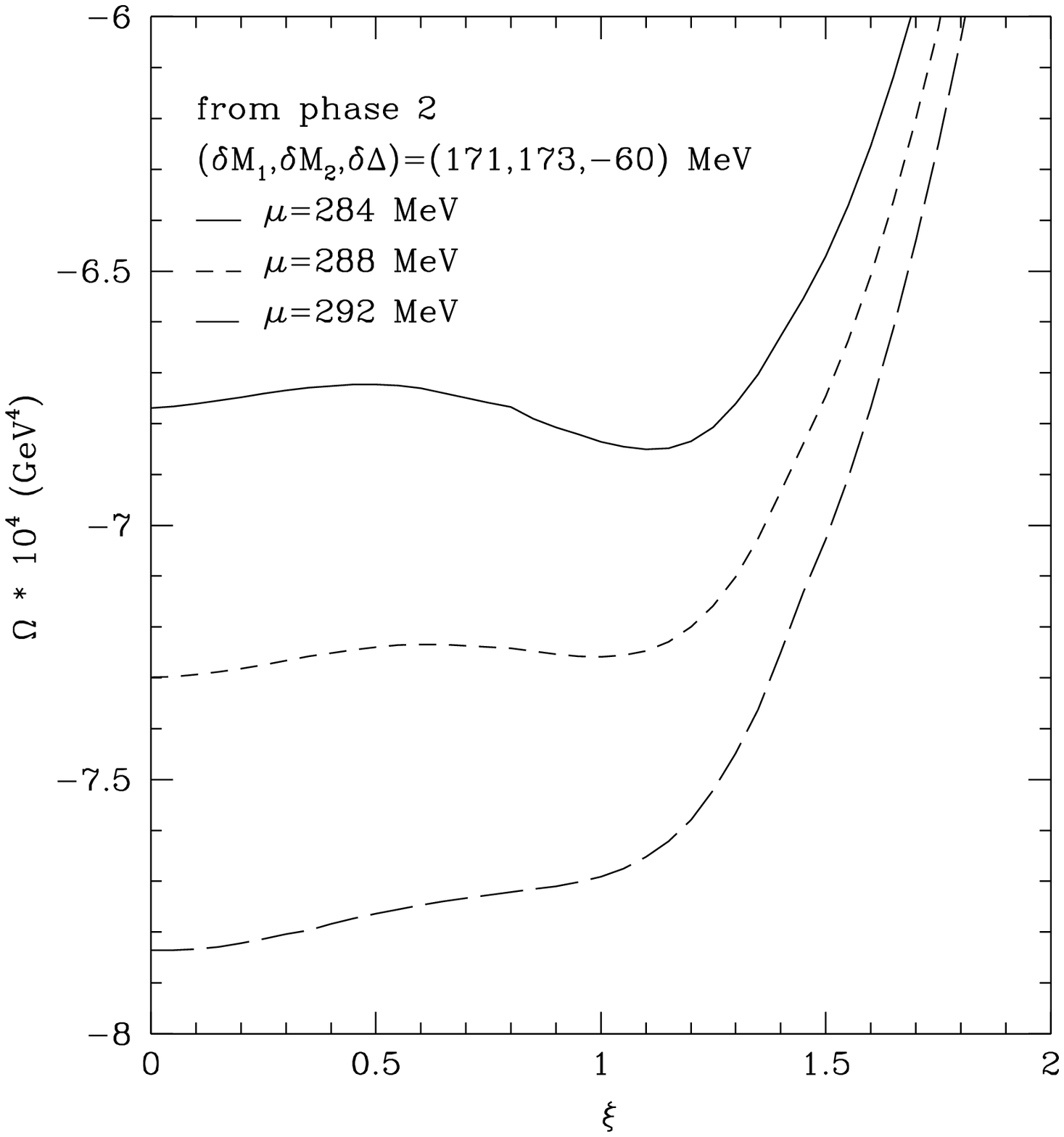,angle=0,width=8cm}
\caption{The left panel shows profiles of $\Omega$ at $\mu=270$ and 
$246$~MeV along the direction that at $\mu=270$~MeV connects the solution 
for phase 1 ($\xi=0$) and phase 3 ($\xi=1$). The right panel shows
$\Omega$ at $\mu=284, 288$ and $292$~MeV along the direction that at 
$\mu=288$~MeV connects the solution for phase 1 ($\xi=0$) and phase 3 
($\xi=1$).} 
\label{hom13}
\end{figure}

To understand the nature of the phase transitions we study the
dependence of $\Omega$ on the three parameters $M_1,M_2$ and $\Delta$.
As seen in the right panel of Fig.~\ref{mgap_pict}, phase 3 starts at
the values of the masses and the gap of phase 1 at $\mu=250$ MeV. This
is an indication of a second order phase transition, as the
corresponding chiral and diquark condensates  (which are proportional
to $M_1,M_2$ and $\Delta$) are the first derivatives of $\Omega$ 
\wrt~these masses and the gap. The emergence of the
phase transition at $\mu=250 MeV$ s exhibited even more clearly in the left  
panel of Fig.~\ref{hom13}. 
At $\mu=270$~MeV, we have plotted the values of $\Omega$
on a straight line in the parameter space $(M_1=M_{1,ph1}+\xi \delta
M_1,M_2=M_{2,ph1}+\xi \delta M_2 ,\Delta=\Delta_{ph1}+\xi \delta \Delta)$,
connecting phase 1 ($\xi=0$) with phase 3 ($\xi=1$). We see that the
solution for phase 1 is not a minimum but a local maximum of
$\Omega$. The second curve shows $\Omega$ just before the transition,
where phase 3 has not yet emerged, but the minimum is quite
flat indicating that the second derivative of $\Omega$ is close to
0. When moving along the same direction $(\delta M_1,\delta
M_2 ,\delta \Delta)$ in the parameter space we see a classic
example of a second order phase transition, when at certain value of
the parameter $\mu$ the symmetry changes (the SU(3) color symmetry is
broken to SU(2)), a diquark condensate appears, the second derivative
of $\Omega$ goes through 0 and the old solution (with the higher
symmetry) turns from a minimum into a maximum. The new absolute minimum
is phase 3. Of course any new extrema can only turn up in pairs, but all
solutions are symmetric \wrt~the sign of $\Delta$ so that 
the two new minima must be at $\pm \Delta$.

 The second phase transition at $\mu=286$~MeV is analyzed in 
the right panel of Fig.~\ref{hom13}.
The middle curve shows the values of $\Omega$ at $\mu=270 MeV$ 
on a straight line in the parameter space $(M_1=M_{1,ph2}+\xi \delta
M_1,M_2=M_{2,ph2}+\xi \delta M_2 ,\Delta=\Delta_{ph2}+\xi \delta \Delta)$,
connecting phase 2 ($\xi=0$) with phase 3 ($\xi=1$). At this value of
$\mu$ phase 2 already dominates over phase 3. The opposite situation is
observed from the upper curve which is for $\mu=284$~MeV with the
cross-section of $\Omega$ in the same direction in the parameter
space. Obviously a first order phase transition occurs for
some $\mu$ between these two values. However, the barrier between the
phases is quite low and the values of the parameters
$(M_1,M_2,\Delta)$ (and the condensates) of the two phases are quite
close, so it is a rather weak first order transition. This is further 
supported by the fact that at $\mu=290$~MeV the minimum for
phase 3 disappears quite rapidly (by going through an inflection point),  
as seen from the right panel of Fig.~\ref{hom13}. For the lowest curve
only phase 2 exists, but the remnant of the inflection point is visible. 

 There are some shortcomings in the mean-field analysis as presented 
here. Below some critical $\mu_{c}$, which marks the onset transition,
no physical quantity should depend on $\mu$. Due to the fact that the 
instanton zero modes explicitly depend on $\mu$ -- however small -- this 
is not respected in our calculation. Nevertheless, the variation of $M$
and $\Omega$ below the phase transition is quite small. The result
might be further improved by taking into account the dependence of 
the instanton density and size on $\mu$. Another problem is that the 
onset transition happens quite early, at $\mu\simeq 250$~MeV, whereas the 
expected onset corresponds to a third of the nucleon mass minus the 
binding energy of nuclear matter $(939-16)/3 {\rm MeV} \simeq 308$~MeV. 
Again, this might be related to various approximations employed. 

  It is interesting to note that we do find all three phases to exist, 
not just the chirally broken and superconducting phases, but
also a phase with chiral symmetry breaking and diquark condensation.
The existence of the latter phase is not a very firm prediction as the
difference in energy density of the three phases in the 
transition regions is rather small. In fact, phase 3 was not observed in 
the NJL calculation of \cite{BR_98} or the instanton calculation
of \cite{Rapp_98} or \cite{CD_99}. The latter work uses slightly 
different techniques to  evaluate the grand canonical potential. 
We will return to the (speculative) phase with simultaneous 
diquark condensation and chiral symmetry breaking in 
the discussion in sect.~\ref{sec_more}.

%But if $\rho$ has shrunk, we have to
%multiply all energies by a scaling factor of
%$M(\mu=0)/M(\mu=250)=1.18$. 
%Then the real $\mu_{crit}$ will be 295 MeV
%and the binding energy per quark will be $M-\mu_{crit}=38 MeV$. This 
%$\mu_{crit}$ is about 16 MeV lower than $m_{Nucleon}/3=312 MeV$ --
%three times the binding of the ordinary nuclear matter which is also
%16 MeV, but per nucleon, not per quark. Still it is quite tempting to
%think that this simple model knows not only about the hadrons, but
%also about the existence of nuclear matter.  

%%%%%%%%%%%%%%%%%%%%%%%%%%%%%%%%%%%%%%%%%%%%%%%%%%%%%%%%%%%%
\section{Three flavor QCD in the chiral limit}
\label{sec_nf3}
%%%%%%%%%%%%%%%%%%%%%%%%%%%%%%%%%%%%%%%%%%%%%%%%%%%%%%%%%%%%

  The situation becomes more involved if one includes the strange 
quark. Since the critical chemical potential $\mu_c\sim 300-350$ 
MeV is larger than the strange quark mass $m_s\simeq 140$ MeV, strange 
quarks have to be included whenever there is time for 
strangeness to equilibrate. There are several qualitatively new 
features in going from $N_f=2$ to $N_f=3$. First, since $N_f=N_c$,
there are new order parameters in which the color and flavor 
orientation of the condensate is locked~\cite{ARW_98b}. Second, 
the instanton-induced interaction is a six-fermion vertex, so it
does not directly induce the BCS instability. 

  In this section we will consider $N_f=3$ flavor QCD in the chiral
limit. In that case, we expect that in the low density phase chiral 
symmetry is broken by a quark condensate $\langle \bar uu\rangle = \langle 
\bar dd\rangle = \langle \bar ss\rangle$. In the high density phase,
quark pairs are condensed. One possible form of ordering is the 
analog of the $N_f=2$ diquark condensate, $\langle q_i^aC\gamma_5
q_j^b\rangle = \Delta^k_c \epsilon_{ijk}\epsilon_{abc}$. This order 
parameter breaks color $SU(3)_C\to SU(2)_C$ and the chiral $SU(3)_L\times
SU(3)_R\to SU(2)_L\times SU(2)_R$. A more attractive possibility is 
provided by the following order parameter~\cite{ARW_98b}
\be
\label{qq_MFA}
\langle q_{i,R}^{a\alpha} q_{j,R}^{b\beta} \rangle &=& 
  \frac{1}{2}(C^\dagger\gamma_5P_R)^{\alpha\beta}
  \left(\Delta_1 \delta_{ia}\delta_{jb} +
       \Delta_2 \delta_{ib}\delta_{ja} \right) .
\ee
Here, $P_R$ is the projector on right-handed quark fields, and there
is an analogous expression for left-handed fields also. 
This order parameters breaks color and chiral symmetry down to the 
diagonal subgroup $SU(3)_{C+L+R}$. Since the color symmetry is completely
broken, there is a gap in the spectrum for all 9 quarks and 8 gluons. 
This already suggests that the phase characterized by Eq.~(\ref{qq_MFA}) 
should be preferred over the $N_f=2$ like phase, in which only 4 quark
states are gaped. We will see this more explicitly in the next section.

  The order parameter (\ref{qq_MFA}) breaks chiral symmetry since 
the residual symmetry couples flavor rotations of right and left 
handed quarks. The diagonal symmetry acts on the quark fields as 
\be 
 q_i^a \to (U^*)_{ij} U^{ab} q_j^b ,
\ee
where $U$ is an element of $SU(3)$. The most general form of the quark 
condensate that is consistent with this symmetry is
\be
\label{qbarq_MFA}
 \langle \bar q_{L,i}^{a\alpha} q_{R,j}^{b\beta} \rangle &=&   
 \frac{1}{2}(P_R)^{\alpha\beta} 
 \left( \left( \Sigma_0- \frac{2}{3} \Sigma_8\right) \delta^{ab}\delta_{ij}
 + 2\Sigma_8 \delta^a_{\;i} \delta^b_{\;j} \right).
\ee
At zero density we expect $\Sigma_8$ to be zero, but in the high 
density phase both $\Sigma_0$ and $\Sigma_8$ will in general be 
non-zero. 

  It is important to note that even though the above argument establishes
that chiral symmetry is broken, it does not show how a chiral condensate
is actually formed. From the superfluid order parameter (\ref{qq_MFA}), we can
directly
form the  chiral order parameter $\langle(\bar q_Lq_R)^2\rangle \sim
\langle \bar q_L\bar q_L\rangle \langle q_R q_R\rangle$, but not the
chiral condensate $\langle\bar q_L q_R\rangle$. This is because (\ref{qq_MFA})
violates right (and left) handed quark number by two units, whereas
$\langle\bar q_L q_R\rangle$ violates right and left handed quark
number by one unit. In other words, the order parameter leaves a
discrete chiral symmetry unbroken,  and this symmetry prevents the
quark condensate from acquiring an expectation value. But this discrete
symmetry is explicitly broken by instantons. In the color-flavor-locked
phase, we can saturate four of the external legs of the  instanton vertex
$(\bar q_L q_R)^3$ with the condensate and obtain an effective interaction 
$\langle \bar q_L\bar q_L\rangle\langle q_R q_R \rangle (\bar q_L q_R)$ 
which leads to the formation of a quark condensate.

 In the case of three massless flavors, the 't Hooft interaction 
is a flavor antisymmetric six-fermion interaction
\cite{SVZ_80b,NVZ_89c,Dia_95}
\be
\label{l_nf3}
{\cal L} &=& G_6 (2\pi\rho)^6 \frac{1}{6N_c(N_c^2-1)}
 \epsilon_{f_1f_2f_3}\epsilon_{g_1g_2g_3}
 \left( \frac{2N_c+1}{2N_c+4}
  (\bar\psi_{L,f_1} \psi_{R,g_1})
  (\bar\psi_{L,f_2} \psi_{R,g_2})
  (\bar\psi_{L,f_3} \psi_{R,g_3}) \right. \\
& & \left. + \frac{3}{8(N_c+2)}
  (\bar\psi_{L,f_1} \psi_{g_1})
  (\bar\psi_{L,f_2} \sigma_{\mu\nu} \psi_{R,g_2})
  (\bar\psi_{L,f_3} \sigma_{\mu\nu}\psi_{R,g_3})
  + ( L \leftrightarrow R ) \right) \nonumber .
\ee
In the following, we will consider the somewhat more general case 
of a $U(1)_A$ violating six-fermion interaction characterized by
two independent coupling constants $G_{6,1}$ and $G_{6,2}$, 
corresponding to the scalar and tensor terms in Eq.~(\ref{l_nf3}).

  In the vicinity of the Fermi surface, six-fermion terms are 
suppressed \wrt four-fermion interactions. This is 
the Cooper phenomenon: Near the Fermi surface, the only interaction
that is not suppressed is $2\to 2$ scattering, where the two particles
are back-to-back. In the more modern language of the renormalization group 
one finds that the strength of the six-fermion interaction is reduced as one
integrates out states away from the Fermi surface~\cite{Sha_95,EHS,SW_98}. 
In the context of the mean-field approximation employed here, we 
will see that the gap equation has a logarithmic enhancement in 
the case of four-fermion interactions, but not for six- (or even higher) 
fermion vertices. 

  For this reason we will have to consider the effect of four-fermion
interactions. We already stressed that instanton-antiinstanton pairs
provide a four-fermion interaction for any number of flavors. In terms
of left- and right-handed fermions, the interaction is 
\be
\label{l_mol}
 {\cal L}_{4}&=& G_4 \left\{
    \frac{2}{N_c^2}\left[
	(\bar\psi_L\gamma_\mu\psi_L)^2+
	(\bar\psi_R\gamma_\mu\psi_R)^2  \right]
   -\frac{1}{N_c(N_c-1)} \left[
	(\bar\psi_L\gamma_\mu\lambda^a\psi_L)^2+
	(\bar\psi_R\gamma_\mu\lambda^a\psi_R)^2 \right] 
   \right. \nonumber \\
  & &  \left.\hspace{0.4cm} -\frac{4}{N_c^2} 
	(\bar\psi_L\gamma_\mu\psi_L)
	(\bar\psi_R\gamma_\mu\psi_R)
   -\frac{2(2N_c-1)}{N_c(N_c^2-1)}
	(\bar\psi_L\gamma_\mu\lambda^a\psi_L)
	(\bar\psi_R\gamma_\mu\lambda^a\psi_R)
	\right\} \ . 
\ee

  In the following, we shall study the condensates (\ref{qq_MFA}) and 
(\ref{qbarq_MFA}) for an interaction given by the sum of the four-fermion 
vertex (\ref{l_mol}) and the six-fermion vertex (\ref{l_nf3}).
In this section, we will consider the coupling constants $G_4$ and 
$G_6$ to be arbitrary parameters, constrained mainly by the known value
of the quark condensate at zero density. In sect.~\ref{sec_coupl}
we shall try to determine these couplings from the partition function 
of the instanton liquid. 

  The system of gap equations for the three-flavor case can be derived
along the same lines as the two-flavor case discussed in the previous 
section. However, the resulting equations are algebraically much more 
involved. In order to keep the presentation reasonably simple, we
will ignore the instanton form-factors and take the interaction to 
be point-like. As we saw in the last section in the case of $N_f=2$,
this approximation does not qualitatively affect the results.

%%%%%%%%%%%%%%%%%%%%%%%%%%%%
\begin{figure}[htt]
\vspace{-2cm}
\centerline{\epsfig{file=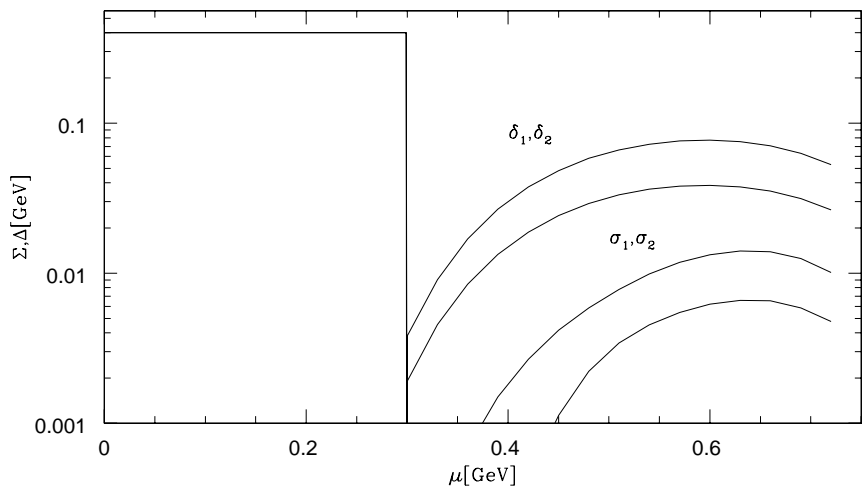,width=120mm}}
\vspace{-2cm}
\caption{Chiral and superconducting gaps $\sigma_{1,2}$ and $\delta_{1,2}$
as a function of the chemical potential for the three-flavor model in the 
chiral limit.}
\label{fig_nf3}
\end{figure}
%%%%%%%%%%%%%%%%%%%%%%%%%%%%

  We shall calculate the thermodynamic potential in the mean-field
approximation. This calculation is somewhat complicated by the fact 
that the color-flavor structure of the propagators is quite involved.
The first step is to determine the quadratic part of the action in
the mean-field approximation. For this purpose, we close off all except 
two legs of the interaction. The result is
\be 
\label{mass_nf3}
 {\cal M} &=& \left( \bar q_{L,i}^a q_{R,j}^b \right)
 \left\{  \left( \delta^{ab}\delta_{ij} \right) 
  \left[ G_4 \left( \frac{16}{3} \Sigma_0 - \frac{2}{9} \Sigma_8 \right) 
   + G_{6,1} \left( 84 \Sigma_0^2 + 8\Sigma_0\Sigma_8
	-\frac{74}{3} \Sigma_8^2 + \frac{9}{2}\Delta_A^2\right) \right. 
 \right.    \nonumber \\
 & & \hspace{6cm}\mbox{} + G_{6,2} \left.\bigg( 144 \Sigma_0^2 
        + 48\Sigma_0\Sigma_8 - 168 \Sigma_8^2 - 30 \Delta_A^2\bigg) \right] 
            \nonumber \\
 & & \hspace{1.5cm}\mbox{} 
  + \left( \delta^a_{\;i} \delta^b_{\;j} \right) 
  \left[ G_4 \frac{2}{3}\Sigma_8 
 + G_{6,1}\left( -24 \Sigma_0\Sigma_8 + 10 \Sigma_8^2
	- \frac{3}{2} \Delta_A^2 \right) \right.
           \nonumber \\ 
 & & \hspace{6cm}\mbox{} + G_{6,2}\left.\left. \bigg( -144 \Sigma_0\Sigma_8 
        + 120 \Sigma_8^2 + 18 \Delta_A^2 \bigg)\right] \right\}  
           \nonumber \\
 & & +\left( q_{R,i}^a C\gamma_5 q_{R,j}^b \right) 
  \left\{ \left(  \delta^a_{\;i} \delta^b_{\;j} -
                      \delta^a_{\;j} \delta^b_{\;i} \right) 
  \Delta_A \left[ \frac{2}{3}G_4 
       + G_{6,1} \left( 3 \Sigma_0 - 2 \Sigma_8 \right) 
       + G_{6,2} 12\left( 3\Sigma_0 - 4 \Sigma_8 \right)  
  \right] \right\} \ .  
\ee 
We note that the interaction is only sensitive to the antisymmetric 
part $\Delta_A=\Delta_1-\Delta_2$ of the $\langle qq\rangle$ order parameter. 
This is different from the OGE interaction considered in \cite{ARW_98b},
but does not make much of a difference in practice, since even in that 
case the solution of the gap equation has $\Delta_S/\Delta_A\ll 1$, 
where $\Delta_S=\Delta_1+\Delta_2$. We also note that there is a
chiral symmetry breaking $\bar qq$ interaction proportional to
$\Delta_A^2$. This is as expected: Color-flavor-locking combined
with instantons leads to chiral symmetry breaking. We also stress 
that both ingredients, instantons and color-flavor-locking, are 
essential.

 The $\langle qq\rangle$ and $\langle \bar qq\rangle$ 
mass terms can be diagonalized simultaneously. 
In general, the mass term can be decomposed as
\be
\label{mass_def}
\left( \bar q_{L,i}^a q_{R,j}^b \right) 
 \left\{ g_0 M_0 + g_1 M_1 \right\}
+\left( q_{R,i}^a C\gamma_5 q_{R,j}^b \right) 
 \left\{ f_1 M_1 + f_2 M_2 \right\} \ ,
\ee
where we introduced the color-flavor matrices 
\be 
\label{mat_def}
 M_0= \delta^{ab}\delta_{ij}, \hspace{1cm}
 M_1= \delta^a_{\;i}\delta^b_{\;j},   \hspace{1cm}
 M_2= \delta^a_{\;j}\delta^b_{\;i} \ .
\ee
The matrices $M_{1},M_2,M_3$ commute. This means that there is 
a color-flavor basis in which (\ref{mass_def}) becomes diagonal.  
We denote the quark fields in this basis by $\phi_\rho$, with
$\rho=1,\ldots,9$. The mass term becomes
\be 
\label{nf3_diag}
\left(\sum_{\rho=1}^{8} 
\left\{ \sigma_1\left(\bar\phi_{\rho,L}\phi_{\rho,R}\right) 
      + \delta_1\left(\phi_{\rho,R}C\gamma_5\phi_{\rho,R}\right) \right\}
 \right)+
\left\{ \sigma_2\left(\bar\phi_{9,L}\phi_{9,R}\right) 
      + \delta_2\left(\phi_{9,R}C\gamma_5\phi_{9,R}\right) \right\} \ , 
\ee
where $\sigma_1=g_0,\, \delta_1=-f_2$ is eightfold degenerate and
$\sigma_2=g_0+3g_1, \,\delta_2=3f_1+f_2$. In our case
\be 
\label{nf3_evals}
\sigma_1 &=& G_4\left(\frac{16}{3}\Sigma_0-\frac{2}{9}\Sigma_8\right)
	    +G_{6,1}\left(84\Sigma_0^2+8\Sigma_0\Sigma_8
	    -\frac{74}{3}\Sigma_8^2+\frac{9}{2}\Delta_A^2\right) 
           \nonumber \\
         & & \hspace{5cm}\mbox{} 
	    +G_{6,2}\bigg( 144\Sigma_0^2+48\Sigma_0\Sigma_8
	    - 168 \Sigma_8^2 - 30 \Delta_A^2\bigg)  \\
\sigma_2 &=& G_4\left(\frac{16}{3}\Sigma_0+\frac{16}{9}\Sigma_8\right) 
	    +G_{6,1}\left(84\Sigma_0^2-64\Sigma_0\Sigma_8
	    +\frac{16}{3}\Sigma_8^2\right)
           \nonumber \\
         & & \hspace{5cm} \mbox{}
	    +G_{6,2}\left(144\Sigma_0^2- 384\Sigma_0\Sigma_8
	    + 192\Sigma_8^2 + 24 \Delta_A^2 \right) \\
\delta_1 &=& \frac{1}{2}\delta_2 \;=\; \Delta_A \left( \frac{2}{3}G_4
	   + G_{6,1} \left(3\Sigma_0-\Sigma_8\right)
           + G_{6,2} \left(-3\Sigma_0+4\Sigma_8 \right)\right) \ . 
\ee
The potential for the mean field is given by closing of all external legs of 
the interaction. This way we get two-loop graphs proportional to $G_4$
and three-loop graphs proportional to $G_6$. In the mean-field 
approximation, we have 
\be
\label{V_nf3}
 V &=& G_4 \left( 2\Delta_A^2 + 12 \Sigma_0^2 + \frac{8}{3} \Sigma_8^2
 \right) + 
  G_{6,1} \left( 504 \Sigma_0^3 - 384 \Sigma_0\Sigma_8^2
  + \frac{320}{3} \Sigma_8^3 + 24\left(3\Sigma_0 - 2\Sigma_8
    \right) \Delta_A^2 \right) \nonumber \\
   & & \hspace{2cm}\mbox{} + G_{6,2}\Big( 864 \Sigma_0^3 
  - 2304 \Sigma_0\Sigma_8^2 + 1280 \Sigma_8^3 
  - 144\left(3\Sigma_0 - 4\Sigma_8 \right) \Delta_A^2 \Big)  \  .
\ee
In the quadratic part of the interaction we can now integrate over
the fermion fields. Since the color-flavor structure is already 
diagonal, we get a sum of nine terms, each (in principle) with 
different gap parameters. We finally obtain the following result
for the free energy
\be
 F &=& -8\epsilon(\sigma_1,\delta_1) -\epsilon(\sigma_2,\delta_2)+V \ .
\ee 
Here, the single particle energy is given by
\be 
\epsilon(\sigma,\delta) = \int \frac{d^3p}{(2\pi)^3}
 \left\{ \sqrt{(E_p-\mu)^2+\delta^2}
	+\sqrt{(E_p+\mu)^2+\delta^2} \right\} \ ,
\ee
and $E_p^2=p^2+\sigma^2$. The mean-field parameters $\Sigma_0,\Sigma_8,
\Delta_A$ are determined by making the free energy stationary
$(\partial F)/(\partial \Sigma_i)=(\partial F)/(\partial \Delta_i) 
= 0$. This gives three coupled gap equations that have to be solved 
numerically. 

 Before we present the results we have to discuss how to fix the 
coupling constants $G_4$ and $G_6$. We take $G_{1,2}$ to have the relative 
size implied by the instanton interaction (\ref{l_nf3}). If we were 
to ignore random instantons, and only had instanton-antiinstanton 
pairs, the four-fermion interaction would break 
chiral symmetry for $G_4>7.5\, {\rm \Lambda}^{-2}$. We consider this 
to be the upper limit on this interaction. In order to see how 
large the gaps in the three-flavor case can possibly be, we take
$G_4$ just below this limit $G_4=7.4 {\rm \Lambda}^{-2}$. $G_6$ is 
then fixed by the requirement that for $m_s=150$ MeV (see next 
section) we get a reasonable constituent $u,d$ mass of 400 MeV. 
This gives $G_6=12.0 {\rm \Lambda}^{-5}$. 

 Results for the various gaps are shown in Fig.~\ref{fig_nf3}.
Note that the superconducting gap is smaller as compared to the
two-flavor case. This is because diquark condensation is now
due to pairs, not individual instantons, and we restricted the 
size of the corresponding coupling such that it does not lead 
to chiral condensation at $\mu=0$. This is similar to the scenario
of Alford et al.~\cite{ARW_98b}, where superconductivity for $N_f=3$ 
is driven by one-gluon exchange. Again, reason dictates that the 
corresponding coupling is below the critical coupling for chiral 
condensation. 

  Instantons lead to chiral condensation in the diquark condensed 
phase. This is immediately clear because if we take the six-fermion
vertex and close off four legs by two diquark insertions, the 
remaining interaction violates chiral symmetry. Color-flavor locking
is nevertheless essential. For the two-flavor superfluid order parameter,
instantons only lead to a non-zero $\langle \bar ss\rangle$. Alford et 
al. realized that chiral symmetry would be broken, but could not calculate 
the size of the effect in their model. Here we find it to be very small.
The maximum constituent mass generated in the diquark condensed 
phase is less than 10 MeV. Qualitatively, this is not hard to 
understand: the constituent mass arises from terms in (\ref{mass_nf3}) 
that are proportional to $\Delta_A^2$. These terms arise as exchange 
terms from the original interaction (\ref{l_nf3}), so they are 
suppressed by degeneracy factors $2N_f N_c$. In addition to that, 
the constituent mass is driven by the superconducting gap squared, 
which is already about an order of magnitude smaller than the zero
density chiral gap. 

  There is one more important direct instanton effect in the high
density phase. For $N_f=3$ all chirally invariant four-fermion 
interactions (molecules, OGE, etc.) are $U(1)_A$ invariant, and 
do not distinguish between scalar and pseudoscalar diquarks. This
means that a parity broken vacuum characterized by the order 
parameter 
\be
\label{qq_odd}
\langle q_i^a C q_j^b\rangle = \bar\Delta_1 \delta_{ia}\delta_{bj}
 + \bar\Delta_2 \delta_{ib}\delta_{ja}
\ee
is degenerate with the parity conserving vacuum considered here. 
The same is true for an arbitrary linear combination of positive
and negative parity condensates. The degeneracy is lifted by the 
six-fermion interaction in conjunction with finite quark masses or 
non-vanishing chiral condensates. This implies that the difference 
in energy density between the parity broken and parity conserving 
vacua is small. This is different from the $N_f=2$ case, where the 
four-fermion interaction distinguishes between (\ref{qq_odd}) and 
(\ref{qq_MFA}), and the energy difference is big. This effect is
also different from the scenario considered by Pisarski and Rischke
\cite{PR_98}, who argued that the parity broken vacuum is degenerate
with the parity conserving one if instanton effects are small. 
In three-flavor QCD in the chiral limit parity broken and parity 
conserving vacua are almost degenerate, even if instanton
effects are not small.

%%%%%%%%%%%%%%%%%%%%%%%%%%%%%%%%%%%%%%%%%%%%%%%%%%%%%%%%%%%%%%%
\section{Flavor symmetry breaking}
\label{sec_fsb}
%%%%%%%%%%%%%%%%%%%%%%%%%%%%%%%%%%%%%%%%%%%%%%%%%%%%%%%%%%%%%%%
 
  The situation is even more complicated if we take flavor symmetry 
breaking into account. For simplicity, we will restrict ourselves
to $m_u=m_d=0$ and $m_s\neq 0$. It is clear that as $m_s\to\infty$, we 
have to recover the two-flavor scenario, with the order parameter 
given by 
\be
\label{del_ud}
\langle q_i^aC\gamma_5q_j^b\rangle 
= \Delta_{ud} \epsilon_{ij3}\epsilon^{ab3} \ .
\ee
Note that in the two-flavor case the color orientation of the 
condensate is arbitrary, but for three flavors the choice (\ref{del_ud})
is preferred because it preserves an $SU(2)$ subgroup of the diagonal
$SU(3)_{C+L+R}$. We might also consider additional gap parameters
that have a different color orientation, but the corresponding gap
equation simply decouples and the solution (except in the limit 
$m_s\to\infty$) is not energetically favored.

  Since flavor symmetry is broken, the structure of the quark 
condensate is also more complicated. The following ansatz generalizes
Eq.~(\ref{qbarq_MFA}):  
\be 
\label{qbarq_fsb}
 \langle \bar q_{L,i}^{a\alpha} q_{R,j}^{b\beta} \rangle =   
 \frac{1}{2}(P_R)^{\alpha\beta} \left(
  \left( \Sigma_0- \frac{2}{3} \Sigma_8\right) \delta^{ab}\delta_{ij}
 + \Sigma_s \delta^{ab}\delta_{i3}\delta_{j3} 
 + 2\Sigma_8 \delta^a_{\;i} \delta^b_{\;j} 
 + \Sigma_{8,1} P_1 + \Sigma_{8,2}P_2  \right) \ ,
\ee
where $P_1=\delta^{a3}\delta_{i3}\delta^{b3}\delta_{j3}$ and 
$P_2=\delta^{a3}\delta_{i3}\delta^b_{\;j}+\delta^a_{\;i} \delta^{b3}
\delta_{j3}$. 

There are a number of complications that occur once flavor symmetry 
is broken, and it is hard to take into account all of these effects 
at the same time. In the following we will concentrate on the dynamical
interplay between a flavor symmetric four-fermion interaction generated
by one-gluon exchange or instanton pairs and the flavor symmetry breaking
four-fermion vertex that comes from the six-fermion 't Hooft interaction
and a strange mass insertion. In addition to that, we have to take into
account that there is no pairing between strange and non-strange quarks
if the mismatch between the Fermi momenta is too big. The BCS instability
arises for pairs with total momentum zero where both of the individual
momenta are on the Fermi surface. This is not possible if the masses
are different and the Fermi surfaces are shifted. In the presence of
pairing the Fermi surface is not sharp, but smeared out over an 
energy range given by the gap. This means that pairing between strange 
and non-strange quarks is suppressed if the mismatch between the Fermi 
momenta exceeds the gap, 
\be  
\label{mismatch}
m_s^2/(4p_F) >\Delta \ .  
\ee 
Moreover, there is the problem that the color-flavor matrix 
characterizing the most general diquark mass term does not commute 
with the color-flavor structure of the $\langle \bar qq\rangle$ 
mass term. This 
means that we cannot simultaneously diagonalize the two mass terms,
and write the free energy in the simple form (\ref{Omega}). Instead it
seems unavoidable to deal with the full (spin, color, flavor, and 
$\langle qq\rangle$ versus $\langle \bar qq\rangle$) 
matrix structure of the quark propagator. 
On the other hand, we found that, except possibly in a small regime,  
quark and diquark condensates do not coexist below the critical chemical 
potential, and that the quark condensate in the high density phase is 
small. In the following we will therefore treat the quark condensate in
the high density phase as a small perturbation.

   The important new ingredient if $m_s\neq 0$ is the presence of
a four-fermion interaction which operates exclusively in the $u,d$
quark sector. 
This interaction arises from the 't Hooft 
interaction, Eq.~(\ref{l_nf3}),  by closing off two external legs 
by a strange mass insertion. The result is
\be
\label{l_nf2_ms}
{\cal L} &=& G_6 (m_s\rho)(2\pi\rho)^4 \frac{1}{2N_c(N_c^2-1)}
 \epsilon_{f_1f_2}\epsilon_{g_1g_2}
 \left\{ \frac{2N_c-1}{2N_c}
  (\psi_{L,f_1}^\dagger \psi_{R,g_1})
  (\psi_{L,f_2}^\dagger \psi_{R,g_2}) \right. 
\nonumber\\ 
 & & \hspace{1cm}\left.  + \frac{1}{8N_c}
  (\psi_{L,f_1}^\dagger \sigma_{\mu\nu} \psi_{R,g_1})
  (\psi_{L,f_2}^\dagger \sigma_{\mu\nu}\psi_{R,g_2})
  + ( L \leftrightarrow R ) \right\} \ ,
\ee
which (of course) has the form of the $N_f=2$ 't Hooft interaction, 
but with a coupling constant controlled by the parameter $3m_s/(4\pi^2
\rho^2)$. So, unlike in the OGE-based works, the value of the strange quark mass
has not just kinematical but also dynamical significance.

 Our model then consists of a flavor symmetric four-fermion interaction, 
the flavor symmetry breaking four-fermion interaction (\ref{l_nf2_ms}),
and the flavor symmetric six-fermion interaction (\ref{l_nf3}). In 
order to compare with work of ARW~\cite{ARW_98}, we take the flavor
symmetric four-fermion interaction to be one-gluon exchange. We could
equally well have used the instanton-antiinstanton induced interaction --
qualitatively this makes very little difference. 

 In this model, the mass term becomes
\be 
\label{mass_fsb}
 {\cal M} &=& \left( \bar q_{L,i}^a q_{R,j}^b \right)
 \left\{  \left( \delta^{ab}\delta_{ij} \right) 
	   \left[ \frac{16}{3}K\Sigma_0 
       + \left(7 G_{4,1} + 6 G_{4,2}\right)\Sigma_0
       + \left(84 G_{6,1} + 144 G_{6,2}\right)
		\left( \Sigma_0^2+\Sigma_0\Sigma_s \right) \right] \right.
	   \nonumber \\
 & & \hspace{1.2cm}\mbox{} + \left.
	  \left( \delta^{ab}\delta_{i3}\delta_{j3} \right)
  \left[ m_s + \frac{16}{3}K\Sigma_s
       - \left(7 G_{4,1} + 6 G_{4,2}\right)\Sigma_0
       - \left(84 G_{6,1} + 144 G_{6,2}\right)
		\Sigma_0\Sigma_s  \right] \right\}
	   \nonumber \\
 & & +\left( q_{R,i}^a C\gamma_5 q_{R,j}^b \right) 
  \left\{ \left(  \delta^a_{\;i} \delta^b_{\;j} \right)
    \left[ \frac{K}{3} \left( 2\Delta_A -\Delta_S\right)\right]
	+ \left(\delta^a_{\;j} \delta^b_{\;i} \right) 
    \left[-\frac{K}{3} \left( 2\Delta_A +\Delta_S\right)\right] \right.
	   \nonumber \\
 & & \hspace{2.5cm}\mbox{} + \left.
    \left(  \epsilon^{3ab} \delta_{3ij} \right)
      \left[ \frac{4}{3}K\Delta_{ud} 
	    + \frac{1}{2} \left( G_{4,1} -12 G_{4,2}\right)
	      \left(\Delta_A+2\Delta_{ud}\right) 
  \right] \right\}  \  . 
\ee 
The quark mass term is already diagonal, while the diquark mass term 
is of the form
\be
\label{dmass_def}
\left( q_{R,i}^a C\gamma_5 q_{R,j}^b \right) 
 \Big\{ f_1 M_1 + f_2 M_2 + f_3 M_3 \Big\} \ 
\ee
with $M_{0,1,2}$ as before and $M_3=\epsilon^{3ab}\epsilon_{3ij}$.
This mass matrix has four eigenvalues, 
\be
 \delta_1 &=& \pm  f_2         \\
 \delta_2 &=& \pm (f_2-f_3)    \\
 \delta_{3,4} &=& \frac{1}{2} \left( 3f_1+2f_2+f_3
   \pm \sqrt{9f_1^2 + 2f_1f_3 + f_3^2} \right) \ , 
\ee 
with degeneracies $d_i=4,3,1,1$. The free energy is given by $F=-\sum_i
d_i\epsilon(\sigma_i,\delta_i)+V$ as before, where the potential is
\be 
 V &=& 16K\Big( 3\Sigma_0^2 + 2 \Sigma_0\Sigma_s + \Sigma_s^2\Big)
     + 6\Big(7G_{4,1}+6G_{4,2}\Big)\Sigma_0^2
     +  \Big(504 G_{6,1} + 864 G_{6,2}\Big) \Sigma_0^2 
	\Big( \Sigma_0+\Sigma_s \Big) \nonumber \\
   & & \mbox{} + 4K\Big(3\Delta_A^2-3\Delta_S^2
	 + 4\Delta_{ud}^2 + 4\Delta_A\Delta_{ud} \Big)
	 + \Big(G_{4,1}-12G_{4,2}\Big) 
	   \Big(\Delta_A+2\Delta_{ud}\Big)^2 \ .
\ee
An important difference as compared to the flavor symmetric case 
is that we have to take into account the kinematic restriction 
discussed above. In channels that involve pairing between 
strange and non-strange quarks we restrict the integration 
to the regime $(E_p-\mu)^2>(m_s^2/(4p_F)-\delta^2)$. For a 
more detailed discussion, we refer the reader to \cite{SW_99}
and \cite{ABR_99}. Again, the gap equation follows from the
requirement that $F$ is stationary \wrt~the gap 
parameters $\Sigma_i$ and $\Delta_i$. 

%%%%%%%%%%%%%%%%%%%%%%%%%%%%
\begin{figure}[t]
\epsfig{file=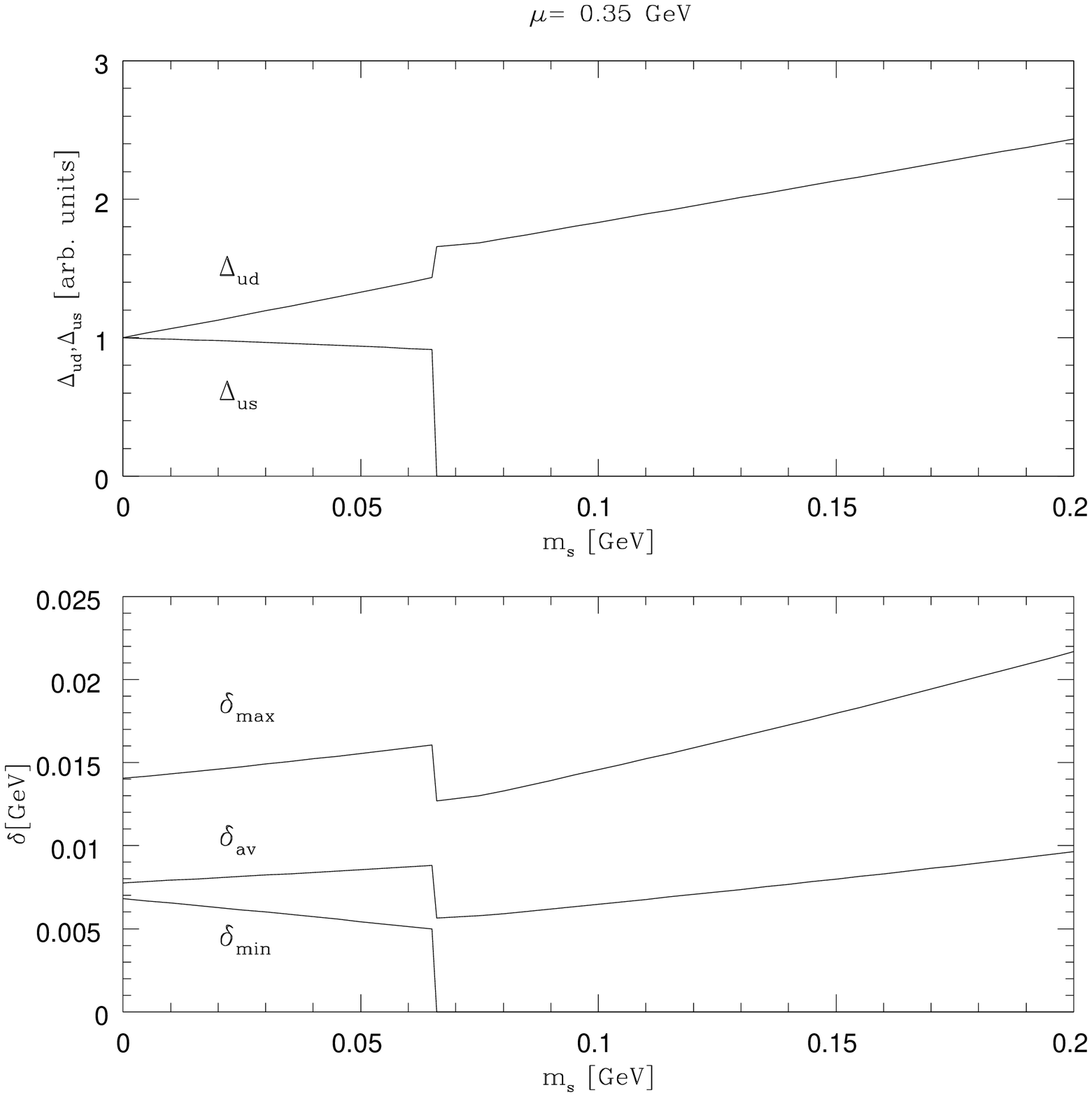,width=8.5cm,angle=0}
\epsfig{file=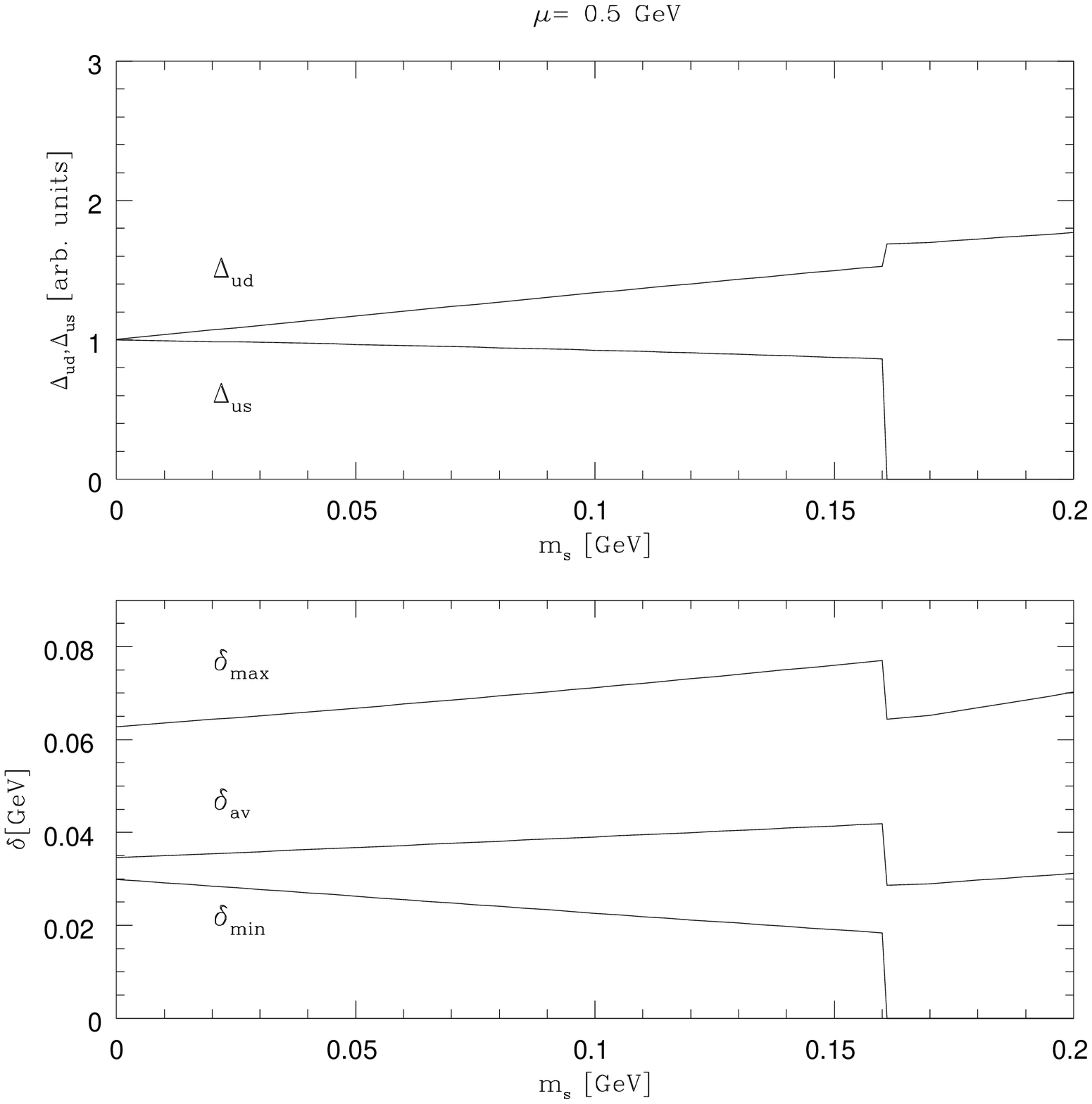,width=8.5cm,angle=0}
\caption{Superfluid order parameters and gaps as a function of
$m_s$ for $\mu=350$~MeV (left panel) and $\mu=500$~MeV (right panel). 
The upper panels show the up-down
and up-strange components $\Delta_{ud}$ and $\Delta_{us}$
of the color-flavor locked state. The lower panels show the
maximum, minimum, and average superconducting gap.}
\label{fig_fsb}
\end{figure}
%%%%%%%%%%%%%%%%%%%%%%%%%%%%
  Numerical results are shown in Fig.~\ref{fig_fsb}. 
The parameters were fixed as described in the last section. In the 
figure we plot $\Delta_{ud}$ and $\Delta_{us}$ as a function of $m_s$ 
for two different chemical 
potentials $\mu=0.35$ and 0.5 GeV. In particular we show the value of 
the largest, the smallest, as well as the average gap. In the two-flavor 
case, the smallest gap is zero and the average gap is $4/9$ of the 
maximum gap. In the three-flavor case, the smallest gap is 
$1/2$ and the average gap $5/9$ of the maximum gap. 

  We observe that there is a sharp transition between the two-flavor 
scenario ($\Delta_{us}=0, \Delta_{ud}\neq 0$) and the three-flavor 
scenario ($\Delta_{ud}=\Delta_{us}\neq 0$)
that takes place around the physical value of the strange quark mass,
$m_s^{crit}=65$ MeV for $\mu=0.35$ GeV, and $m_s^{crit}=160$ MeV
for $\mu=0.5$ GeV. The ratio $\Delta_{ud}/\Delta_{us}$ grows 
roughly linearly already for small $m_s$. This is different from
the results of \cite{SW_99,ABR_99}, and an instanton effect. As
discussed above, instantons induce a four-fermion interaction 
among light quarks which is proportional to $m_s$. We should 
note that we have not included the possibility of a dynamically 
generated contribution to the strange quark mass in the superfluid
phase. In terms of the current mass, this effect will shift the 
critical mass to smaller values. 

%%%%%%%%%%%%%%%%%%%%%%%%%%%%%%%%%%%%%%%%%%%%%%%%%%%%%%%%%%%%%%%%%%
\section{Statistical Mechanics of the Instanton Liquid}
%%%%%%%%%%%%%%%%%%%%%%%%%%%%%%%%%%%%%%%%%%%%%%%%%%%%%%%%%%%%%%%%%
\label{sec_cocktail} 

%%%%%%%%%%%%%%%%%%%%%%%%%%%%%%%%%%%%%%%%%%%%%%%%%%%%%%%%%%%%%%%%%
\subsection{The Cocktail Model at non-zero Density}
%%%%%%%%%%%%%%%%%%%%%%%%%%%%%%%%%%%%%%%%%%%%%%%%%%%%%%%%%%%%%%%%%

  In this section  we take a step beyond the mean-field
approximation in that we allow for possible clustering effects in 
the instanton ensemble. This is achieved by employing a 
somewhat different formalism. In the previous sections we started
from an effective quark interaction obtained by integrating out the 
gluonic (instanton-) fields in the underlying partition function. 
In this section we reverse the strategy and integrate over the fermion 
(quark) fields first. This leads to the following partition function 
for the instanton ensemble at finite density:  
\be
{\cal Z}_{inst}(\mu)=\sum\limits_{N_+,N_-} \frac{1}{N_+! N_-!}
\prod\limits_{I=1}^{N_+,N_-} \int d\Omega_I \ n(\rho_I) \
e^{-S_{int}} \ \rho_I^{N_f} \prod\limits_{f=1}^{N_f}
\det (i \not\!\!D+im_f-i\mu\gamma_4)  \ . 
\label{zinst}
\ee
The original QCD path-integral over all possible gluonic field 
configurations has been converted into an integration over the 
collective coordinates $\Omega_I=\{z_I,\rho_I,u_I \}$ (position, 
size and color orientation) of $N_+$ instantons and $N_-$ antiinstantons. 
The single-instanton amplitude $n(\rho_I)$ contains the semi-classical
tunneling rate (including one-loop quantum corrections), as well as
the Jacobian arising from the introduction of collective coordinates.
The instanton interactions can be divided into a gluonic part $S_{int}$ 
and a fermionic part represented by the determinant of the Dirac operator.
It is usually approximated in the subspace of zero modes, \ie,  
\be
\det(i\not\!\!D-i\mu\gamma_4) \simeq \det
 \left( \begin{array}{cc} 0 & T_{IA}(\mu) \\
   T_{AI}(\mu) & 0
 \end{array} \right) \ , 
\ee
where $T_{IA}$ is the fermionic overlap matrix element discussed 
in sect.~\ref{sec_tia}, see Eq.~(\ref{tiamu}). When restricted to the 
zero mode basis, the fermionic determinant is in fact equivalent to the 
sum of all closed
loop diagrams to all orders in the 't Hooft effective
interaction. The non-hermiticity of the finite-$\mu$ Dirac 
operator is reflected by the fact that  
\be
T_{AI}(\mu) &=& T_{IA}^\dagger(-\mu) 
\nonumber\\
 &\ne& T_{IA}^\dagger(\mu) \ , 
\label{sign} 
\ee
\ie, the fermionic determinant
is complex, entailing the well-known 'sign'-problem in the partition 
function, which will be addressed below. 
 
 In the statistical mechanics treatment chosen here, spontaneous  
chiral symmetry breaking in the vacuum is generated by randomly 
distributed uncorrelated anti-/instantons which allow for a 
delocalization of the associated quark quasi-zero modes  
corresponding to the formation of a nonzero $\langle q\bar q\rangle$ 
condensate state. In other words, quarks can travel arbitrarily 
long distances by randomly jumping from one instanton to another 
(antiinstanton), and thus may carry their chiral
charge to spatial infinity where it effectively becomes `lost'.

In this picture chiral restoration can in principle proceed in two 
ways: either instantons disappear altogether, or they rearrange 
into some {\em finite} clusters which do no longer support any finite 
$\langle q\bar q\rangle$ condensate. In the limit of very large
temperature or density instantons will eventually disappear, 
because Debye screening of the large gluon fields inside the 
instantons leads to a strong suppression of the tunneling rate.
Nevertheless, in the case of finite temperature QCD it was 
argued~\cite{IS_94} that this cannot be the relevant mechanism for
chiral restoration, 
since lattice simulations have observed the transition at rather 
low temperatures of $T_c^\chi\simeq 150$~MeV. This, in turn, led 
to the suggestion that chiral restoration proceeds through the 
formation of  $I$-$A$ molecules. Further support for this idea 
is provided by lattice measurements of the instanton density at finite 
temperatures, showing no depletion below $T_c^\chi$ and a smooth 
onset of the  expected Debye-screening above~\cite{CS_95,Alles}. 
More recent studies~\cite{DeForcrand,Ilg_new} seem to find more 
direct indications for $I$-$A$ molecule formation at $T\approx T_c$, 
but their quantitative role in the transition remains to be clarified. 

  At finite density additional possibilities for clustering of the 
(chirally asymmetric) random instanton liquid into chirally symmetric 
configurations are available. In particular, random instantons 
can be supported by diquark condensates for $N_f=2$ or by a combination 
of diquark and quark-antiquark ones for larger $N_f$. However, as we 
will show in this section, effects of $I$-$A$ molecule formation may 
still be an important element in the finite-$\mu$ chiral restoration 
transition, as was shown
in Ref.~\cite{Rapp_98}. In some sense this is not really surprising:  
when both T and $\mu$ are sufficiently 
large so that the all condensates are absent (\ie, in the true QGP phase), 
$I$-$A$ molecules should constitute the preferred configuration for  
any remaining instanton component in the system. 

  To investigate the interplay between the various components 
in the finite density partition function more quantitatively  
we resort to the 'cocktail-model' introduced
in Refs.~\cite{IS_89,IS_94}. Here, the instanton ensemble is 
decomposed into a mixture of random (``atomic'') and ``molecular'' 
configurations, which yields a grand canonical  partition function 
of the form 
\be
{\cal Z}_{inst}^{a+m}=\sum_{N_a,N_m}
 {(z_a V_4)^{N_a} \over N_a!}  \ {(z_m V_4)^{N_m} \over N_m!} \ . 
\ee
In the thermodynamic limit $V_4\to \infty$, and using the Stirling formula,  
the free energy (thermodynamic potential) becomes  
\be
\Omega_{inst}^{a+m}(n_a,n_m;\mu)=
-\frac{\ln[{\cal Z}^{a+m}_{inst}]}{V_4}
=-n_a \ \ln\left[ \frac{{\rm e} z_a}{n_a} \right]
- n_m \ \ln\left[ \frac{{\rm e} z_m}{n_m} \right] \ . 
\label{Ominst}
\ee
The atomic and molecular 'activities' are~\cite{IS_89,IS_94}
\be
z_a & = & 2 \ C \ \rho^{b-4} \ {\rm e}^{-S_{int}} \
\langle T_{IA}(\mu) T_{AI}(\mu)\rangle^{N_f/2}
\nonumber\\ 
 & = & 2 \ C \ \rho^{b-4} \ {\rm e}^{-S_{int}} \
\left(\frac{n_a}{2} \int d^4z \ du \ [T_{IA}(\mu) T_{AI}(\mu)]
 \ \rho^2 \right)^{N_f/2} 
\nonumber\\
z_m & = & C^2 \ \rho^{2(b-4)} \ {\rm e}^{-2S_{int}} \
\langle [T_{IA}(\mu) T_{AI}(\mu)]^{N_f}\rangle \ 
\nonumber\\
 & = & C^2 \ \rho^{2(b-4)} \ {\rm e}^{-2S_{int}} \ 
 \int d^4z \ du \ \left[ T_{IA}(\mu) T_{AI}(\mu)\right]^{N_f} 
 \ \rho^{2N_f} \ . 
\label{zam}
\ee
The underlying approximation in this approach is that the 
values of the hopping amplitudes $T_{IA}$ in each individual configuration 
are replaced by a product of their mean square values in an uncorrelated 
ensemble. To establish a connection to  the (chiral) quark condensate
and constituent mass, one can use the mean-field approximation to express 
them via the ``atomic'' density $n_a$ as 
\be
\langle \bar qq\rangle &=& -\frac{1}{\pi\rho} \left(\frac{3}{2}n_a\right)^{1/2}
\\
M &=& - C_{M} \ \frac{2}{3} \ (\pi \rho)^2 \langle \bar qq\rangle  
    =  - C_{M} \ (\pi \rho) \ \sqrt{\frac{2}{3}n_a} \ , 
\label{Mq}
\ee
respectively. In the original work of Ref.~\cite{Shu_82}, where these 
relations have been first derived, the coefficient $C_M=1$, leading 
to an effective quark mass of $M^*\simeq 200$~MeV. The latter is to 
be understood as the average mass of a constituent quark at finite 
(euclidean) momentum participating in tunneling processes associated 
with instantons. With increasing four-momentum $M^*$ is appreciably 
reduced and therefore does not correspond to the usual constituent 
quark mass $M$ (defined at zero momentum). We account for this by 
using $C_M=2$, yielding $M=400$~MeV. The minimization of the total 
$\Omega$ over $n_a$ and $n_m$ determines their equilibrium values in 
the  ensemble for given $T,\mu$. The corresponding gap equation for 
the constituent quark mass is the direct analog of the mean-field equation
conjugated to the $\langle\bar q q\rangle $ condensate, but expressed
in terms of different variables. 

 For a refined treatment at finite densities we supplement the cocktail 
model by two additional components. Following the arguments given above, 
we have to account for the possibility that the random instanton component 
can become engaged in diquark chains, first observed in numerical simulations 
of the instanton ensemble in the high-density limit~\cite{Sch_98}.  
The pertinent term in the free energy reads
\be
\Omega_{d}(n_d;\mu)=-n_d \ln\left[ \frac{{\rm e} z_d}{n_d}\right] 
\label{Omegad}
\ee
with the associated activity 
\be
z_d   =  2 \ C \ \rho^{b-4} \ {\rm e}^{-S_{int}} \
\left(\frac{n_a}{2} \int d^4z \ du \ [T_{IA}(\mu) T_{IA}(\mu)]_{\bar 3}
 \ \rho^2 \right)^{N_f/2} \ . 
\label{zd}
\ee
The subscript ``$\bar 3$'' indicates the color projection in some 
predefined direction  characterizing  the color vector of  
the diquark. In the same way that $n_a$ determines the constituent quark
mass $M$, the superconducting gap is related to the density
$n_d$ as 
\beq
\Delta = C_\Delta \frac{\pi\rho}{(N_c-1)} \sqrt{\frac{2}{3} n_d}  
\label{deltaMF}
\eeq
In analogy to Eq.~(\ref{Mq}) we use $C_\Delta=2$, but also perform 
calculations with $C_\Delta=1.5$ to assess the inherent uncertainty of 
this mean-field estimate. 
Notice that in contrary to Eqs.~(\ref{zam}), the fermionic overlap
matrix element enters Eq.~(\ref{zd}) as $(T_{IA})^2$, which, in fact, 
causes the $z$-integration to diverge. This is precisely the 
BCS-singularity of an attractive interaction in the particle-particle
channel, here encountered in coordinate space.  
The standard procedure to treat this singularity is to start from 
a new ground state which a priori has the gap built into the fermion  
propagators, thereby regulating the integrals.  
The net effect of the gap on the overlap integrals is a damping
factor for the intermediate quark propagators, which is delineated in
appendix~\ref{sec_damp}. 

The second refinement consists of including a 
Fermi sphere of quark-'quasi-particles' in the 
free energy, representing the contribution of quark non-zero modes.   
The final expression for the thermodynamic potential then becomes
\be
\Omega(n_a,n_m,n_d;\mu)=\Omega_{inst}^{a+m+d}(n_a,n_m,n_d;\mu) + 
\Omega_{quark}^{QP}(M,\Delta;\mu) \ ,
\label{Omega_cktl}
\ee
where the  quark contribution is simply given by
\be
\Omega_{quark}^{QP}(M,\Delta;\mu)  =  \epsilon_q(M,\Delta;\mu)
-\mu \ n_q(M,\Delta;\mu)
\ee
with
\be
n_q(M;\mu) &=& g_q \int d^3k  \ n_F(\mu-\omega_k)
\nonumber\\
\epsilon_q(M;\mu) &=& g_q \int d^3k \ \omega_k \
 n_F(\mu-\omega_k) \ ,
\ee
$g_q=2 N_c N_f$ and $\omega_k^2=M^2+k^2$ (in the superconducting 
phase the dispersion relations of paired quarks picks up an additional 
dependence   on the gap $\Delta$, see below).

At this point it is instructive to compare the 'cocktail model'
with other approaches.
The (simpler) philosophy employed in  the first part of this paper
introduces the multi-fermion interactions with a  {\em constant} coupling 
$g$, independent of condensates and temperature/density.  
However, since $g$ is itself generated through instantons, density 
variations of the latter will affect the coupling.  
A step towards including this feature was made
by Carter and Diakonov~\cite{CD_99}: their coupling  parameter, 
called  $\lambda$ (which essentially corresponds to  $z_a$ without the 
fermionic determinant part), 
was subjected to minimization. The resulting gap equation relates
the mean instanton density $n_a$ to $\lambda$ and the condensates.
However, instead of calculating $\lambda$ and then finding  $n_a$ (as
done here), they 
stayed in the mean-field approximation in which 
the instanton density remains unchanged. Our results to be discussed
below support the (approximate) validity of this assumption.

To investigate possible mechanisms for chiral symmetry restoration
at finite density in some detail, in particular the competition between
diquark and molecule formation, we will in the following separately discuss 
various versions of the cocktail with increasing complexity, \ie, 
the two-flavor model including (anti-) instantons, molecules and a Fermi
sphere of constituent quarks (sect.~\ref{CKTnf2}), additionally including 
the simplest $ud$-pairing as discussed in Refs.~\cite{ARW_98,RSSV_98} 
(sect.~\ref{CKTnf2+}), and the three-flavor case (sect.~\ref{CKTnf3}). 
In sect.~\ref{sec_coupl} we also give  estimates for the 
density-dependence of effective coupling 
constants for   molecule-induced (anti-)quark-quark interactions, 
which naturally emerge from the formalism employed in this section.

%%%%%%%%%%%%%%%%%%%%%%%%%%%%%%%%%%%%%%%%%%%%%%%%%%%%%%%%%%%%%%%%%%%%%
\subsection{Two Flavor Cocktail Model without Diquark Condensates} 
%%%%%%%%%%%%%%%%%%%%%%%%%%%%%%%%%%%%%%%%%%%%%%%%%%%%%%%%%%%%%%%%%%%%%
\label{CKTnf2}

In this subsection we basically follow the approach of Ref.~\cite{IS_94}, 
generalizing it to finite density~\cite{Rapp_98}.
The issue here is to assess the potential role of $I$-$A$ molecule formation  
in chiral symmetry restoration at finite density, without the additional
complication of superconducting gaps. 

For the actual calculations we now have to face the problem of the
complex fermionic determinant appearing in the various activities, 
Eqs.~(\ref{zam}) and (\ref{zd}). As has been suggested in Ref.~\cite{Rapp_98},
it can be solved under the assumption that  the  gluonic interaction 
does not exhibit a pronounced dependence on the color angles, approximating 
it by an average value (see below). As a result, the color dependence
in the activities only enters through the combinations of $T_{IA}$'s, 
which then can be integrated analytically, rendering the fermionic
determinant real.  For two flavors one obtains
\be
z_a(z_4,r) & \propto & \int du T_{IA}(\mu) T_{IA}^\dagger(-\mu)
= \frac{1}{2N_c} [f_1^+f_1^-+f_2^+f_2^-]
\nonumber\\
z_m (z_4,r) & \propto & \int du 
[T_{IA}(\mu) T_{IA}^\dagger(-\mu)]^{N_f}
=\frac{(2N_c-1) \{f_1^+f_1^-+f_2^+f_2^-\}^2
+\{f_1^+f_2^--f_1^-f_2^+\}^2}{4N_c(N_c^2-1)} \
\label{colav2}
\ee
where $f_i^{\pm}\equiv f_i(\pm\mu)$ are defined through Eqs.~(\ref{tiamu})
and (\ref{tiamu2}). Note that the color averaging of the former complex 
expressions is sufficient to yield real-valued activities. Also note that, 
whereas $z_m(z_4,r)$ is a positive definite quantity, this is not the case 
for $z_a(z_4,r)$ due to the oscillations in $f_{1,2}(r)$. In fact, when 
further integrating $z_a(z_4,r)$ over space-time (the remaining four 
collective coordinates), delicate cancellations occur, which require 
accurate numerical values for the $f_{1,2}^{\pm}$; otherwise one easily 
encounters negative/incorrect results for both activities at finite 
chemical potential. The gluonic interaction entering Eqs.~(\ref{zam}) has 
been approximated by an average repulsion $S_{int}=\kappa \rho^4(n_a+2n_m)$, 
where $\kappa=\beta/2\bar\rho^4 (N/V)$, $\beta=b/2+3N_f/4-2$,
$b=\frac{11}{3}N_c-\frac{2}{3}N_f$. The free parameter $\kappa$
characterizes the diluteness of the ensemble. In the case $N_c=3$
and $N_f=2$ we have chosen $\kappa\simeq 130$ in order to reproduce
the phenomenological value for the diluteness of the instanton vacuum.
The normalization constant $C\propto (\Lambda_{QCD})^b$ can also be fixed
in the vacuum by requiring that the absolute minimum of the 
thermodynamic potential
$\Omega_{inst}(\mu=0;n_a,n_m)$ appears at a total instanton density of
$N/V=n_a+2n_m=1.4$~fm$^{-4}$, being realized for $n_a$=1.34~fm$^{-4}$
and $n_m$=0.03~fm$^{-4}$, which gives $\Lambda_{QCD}\simeq 260$~MeV. The
smallness of the molecular component in the vacuum is a consequence
of the large entropy associated with quantum fluctuations of the color 
angles, randomizing the system. Available lattice data agree that such 
correlations are indeed small at $T=\mu=0$.  
 
%%%%%%%%%%%%%%%%%%%%%%%%%%%%%%%%
\begin{figure}[tp]
{\makebox{\epsfig{file=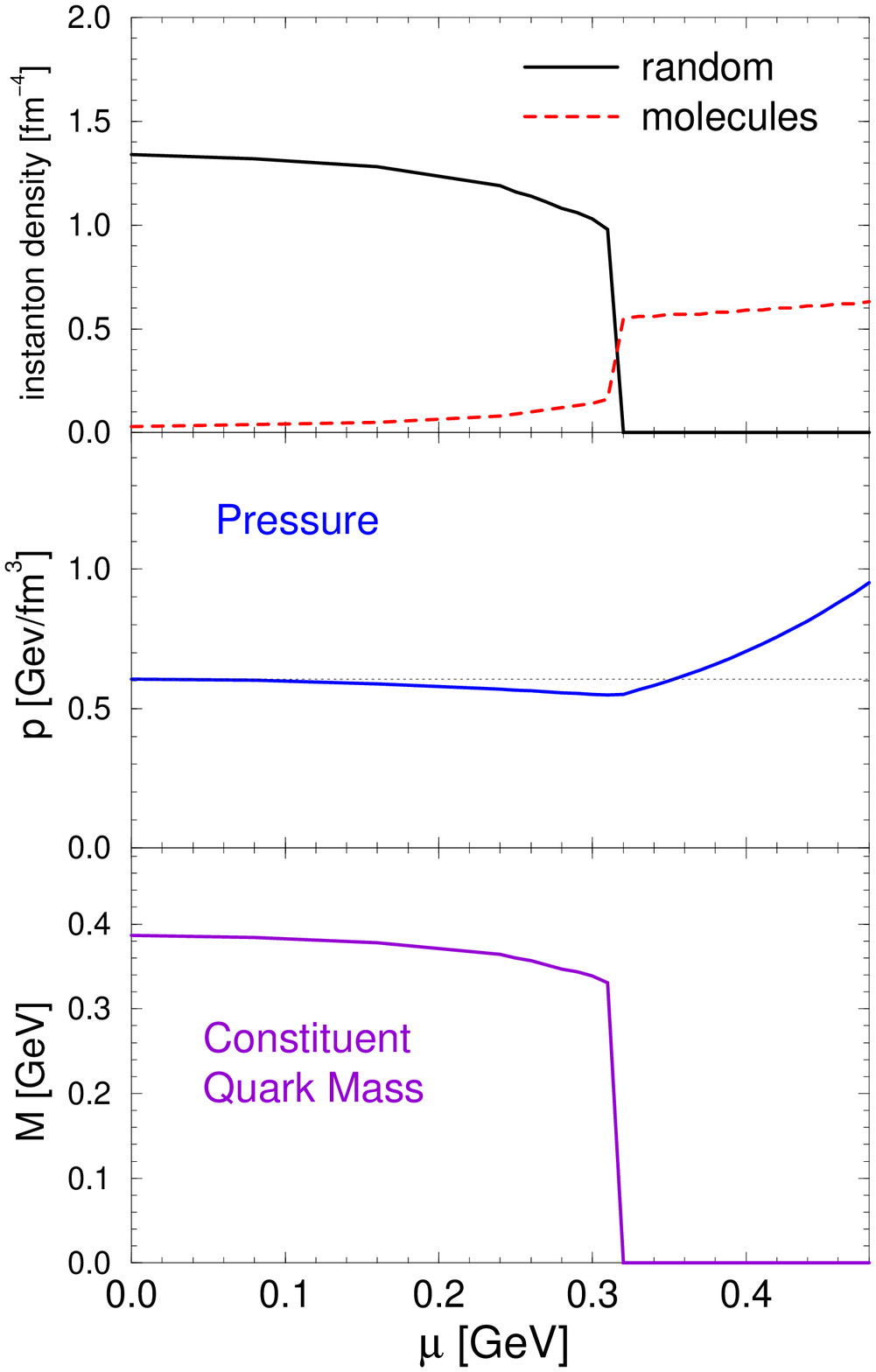,width=90mm,angle=0}}}
\hspace{-0.7cm}
{\makebox{\epsfig{file=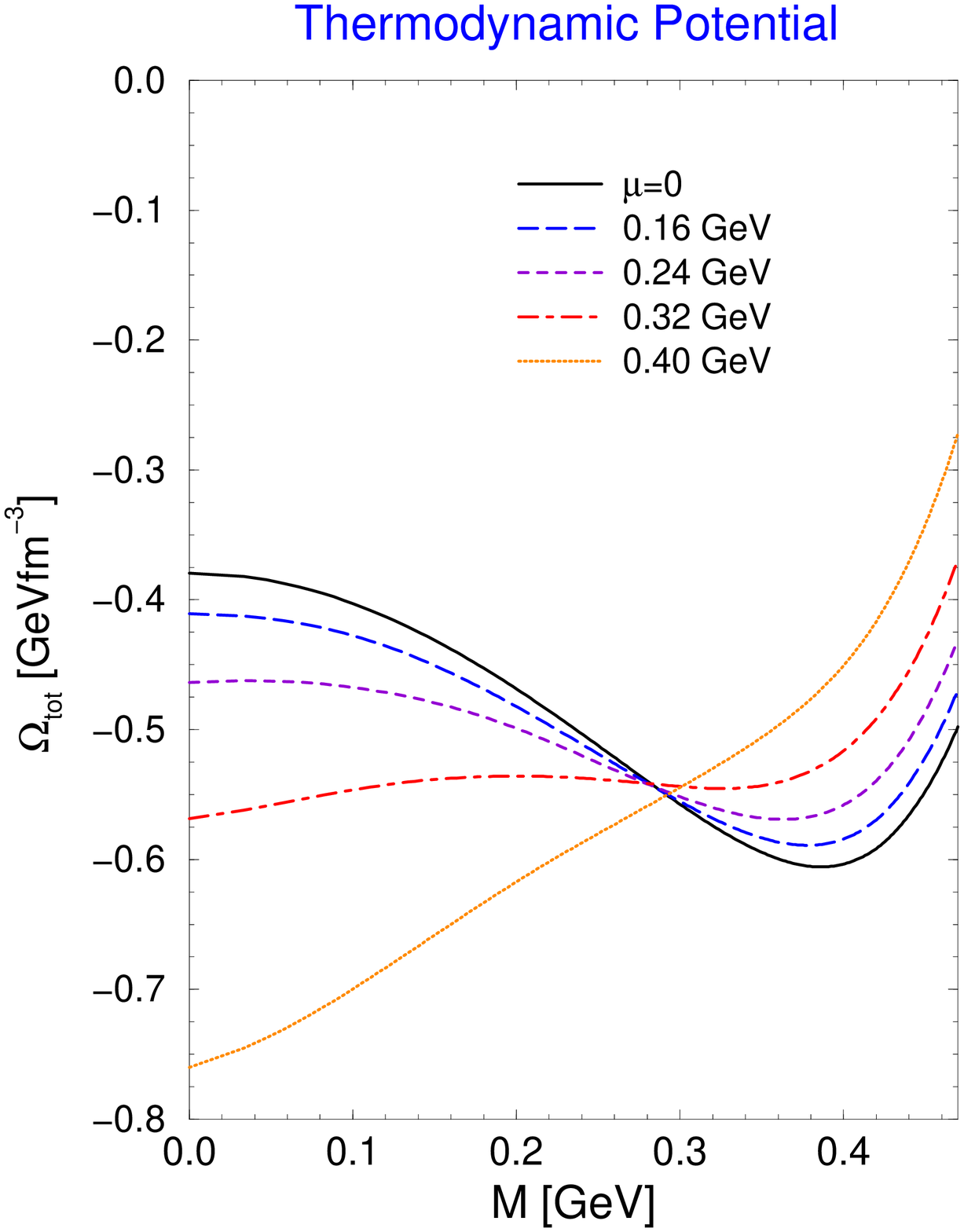,width=75mm,angle=0}}}
\caption{Results for the two-flavor cocktail model {\it without} diquark
pairing;  the left part shows the densities of random instantons ('atomic' 
component) and $I$-$A$ molecules (upper left panel), the pressure $p=-\Omega$ 
(middle left panel) and the constituent quark mass (lower left panel) after 
minimizing the free energy, Eq.~(\ref{Omega_cktl}), \wrt~$n_a$ and 
$n_m$. In the right panel the free energy is displayed as a function of the
constituent quark mass $M \propto n_a^{1/2}$, indicating a first order
transition from the minimum at finite $M$ to the one at $M=0$. }
\label{fig_ct2}
\end{figure}
%%%%%%%%%%%%%%%%%%%%%%%%%%%%%%%%

Our numerical findings for the $N_c=3, N_f=2$ atomic+molecular instanton
cocktail model at finite density are summarized in Fig.~\ref{fig_ct2}. 
At small $\mu$ essentially nothing happens until, at a critical
value $\mu_c\simeq 310$~MeV, the system jumps into the chirally
restored phase, the latter being  characterized by $n_a=0$.
The transition is of first order, as can be seen by inspection of the
$M$-dependence of the free energy (right panel of Fig.~\ref{fig_ct2}).
Below $\mu_c$, the pressure actually decreases slightly with increasing
$\mu$ indicating a mixed phase-type instability, similar to what
has been discussed in Refs.~\cite{ARW_98,Bub_96}.
The total instanton density at the transition (residing in
$I$-$A$-molecules) is  appreciable, $N/V=2n_m\simeq 1.1$~fm$^{-4}$,
providing the major part of the pressure at this point. In other words:
a substantial part of the non-perturbative vacuum pressure persists
in the chirally restored phase. The vacuum pressure of 
$p(\mu=0)=0.6$~GeV~fm$^{-3}$ (indicated by the dotted line in the middle
left panel of Fig.~\ref{fig_ct2}) is not recovered until a chemical
potential of $\mu_0=350$~MeV, corresponding to a free quark number 
density of $n_q^0=1.15$~fm$^{-3}$ (naively, this  translates into a nucleon 
density of $n_N=2.4n_0$ with the normal nuclear matter density 
$n_0=0.16~$fm$^{-3}$). For all chemical potentials in between, 
$0<\mu<\mu_0$, 
the system is mechanically instable, possibly indicating droplet 
formation in the region $M<\mu<\mu_0$ (as suggested in Ref.~\cite{ARW_98}), 
where, with a finite quark density, the pressure $p(\mu)$ is below its 
vacuum value.

A further comment concerning the numerical value
of the critical chemical potential, which is very close
to a third of the nucleon mass, is in order. Given the various 
approximations applied it should be regarded as a coincidence. 
At the same time it most likely provides  
a {\it lower} bound for the true value, as has been the case for
the finite temperature calculations of Ref.~\cite{SS_96} (resulting
in a critical temperature which is about 20\% lower than observed 
on the lattice and might well be related to the fact that the 
instanton model lacks explicit confinement). On the other hand, 
the inclusion of superconducting gaps may further reduce the critical
chemical potential and in fact dominate the transition, as will be 
discussed in the following section.

%%%%%%%%%%%%%%%%%%%%%%%%%%%%%%%%%%%%%%%%%%%%%%%%%%%%%%%%%%%%%%%%%%%%%
\subsection{Two Flavor Cocktail Model Including Pairing}
%%%%%%%%%%%%%%%%%%%%%%%%%%%%%%%%%%%%%%%%%%%%%%%%%%%%%%%%%%%%%%%%%%%%%
\label{CKTnf2+}

  We now address the effects of diquark condensation. For clarity, 
let us first ignore the $I$-$A$ molecule component in the cocktail to
study the competition between chiral and diquark condensates only.  
The additional inclusion of a pairing gap resulting from $ud$ diquark 
formation at the Fermi surface as discussed in Ref.~\cite{RSSV_98} modifies
the quark quasiparticle contribution according to  
\be
\Omega_q^{\Delta}(\mu)=\tr \log\left[D(M,\Delta)\right]  \ . 
\label{Omqdel}
\ee
Here, $D$ denotes the quasiparticle quark-propagator now including the 
BCS gap $\Delta$. All interaction contributions are effectively
accounted for through the instanton part $\Omega_{inst}^d$, 
cf.~Eqs.~(\ref{Omegad}), (\ref{zd}).  
The resulting expression for the free energy then becomes  
\be
\Omega(n_a,n_d;\mu)= \Omega_{inst}^{a+d}(n_a,n_d;\mu)
+\frac{1}{N_c} \left[2\Omega_q^\Delta(M,\Delta;\mu)+(N_c-2)
\Omega_q^{QP}(M;\mu)\right]  \ 
\ee
(the last term accounting for unpaired quarks), which now has to be
minimized \wrt~$n_a$ and $n_d$. The results for two different values
of the vacuum instanton density and the (not precisely determined)
coefficient $C_\Delta$ for calculating the pairing gap 
(cf.~Eq.~(\ref{deltaMF})) are summarized in Tab.~\ref{tab_ad}.
We find that color superconductivity appears at critical chemical 
potentials around  $\mu_c\simeq$~300~MeV, very similar to the 
values of the previous section where only $I$-$A$ molecule formation
was considered. The associated gaps range between 120-180~MeV. 
These results are consistent  
within 20\% with the findings of sect.~\ref{sec_nf2} and those of  
Ref.~\cite{CD_99}. We should also note that the calculated gaps
in our original work~\cite{RSSV_98} are significantly smaller
(below $\sim$~100~MeV) as those were effectively obtained for 2+1
flavors, \ie, in Ref.~\cite{RSSV_98} we included the effect of a reduced
(constituent) strange quark mass $M_s$ in the closed-off 
strange quark loop of the six-fermion 
instanton vertex, which decreases the effective instanton-induced coupling
constant for the four-quark interaction by about 60\% in the chirally 
restored phase. Another feature that emerges here is that the total 
instanton density changes little across
the transition, which a posteriori justifies to assume it as constant
in the mean-field calculations of sect.~\ref{sec_nf2}. 
 
As in the previous section, there is an intermediate constituent 
quark-diquark phase, similar to what
was found in sect.~\ref{sec_nf2}, but here it again has 
small negative pressure which makes it mechanically unstable 
against the formation of a mixed phase. 

\begin{figure}[htb]
\begin{center}
{\makebox{\epsfig{file=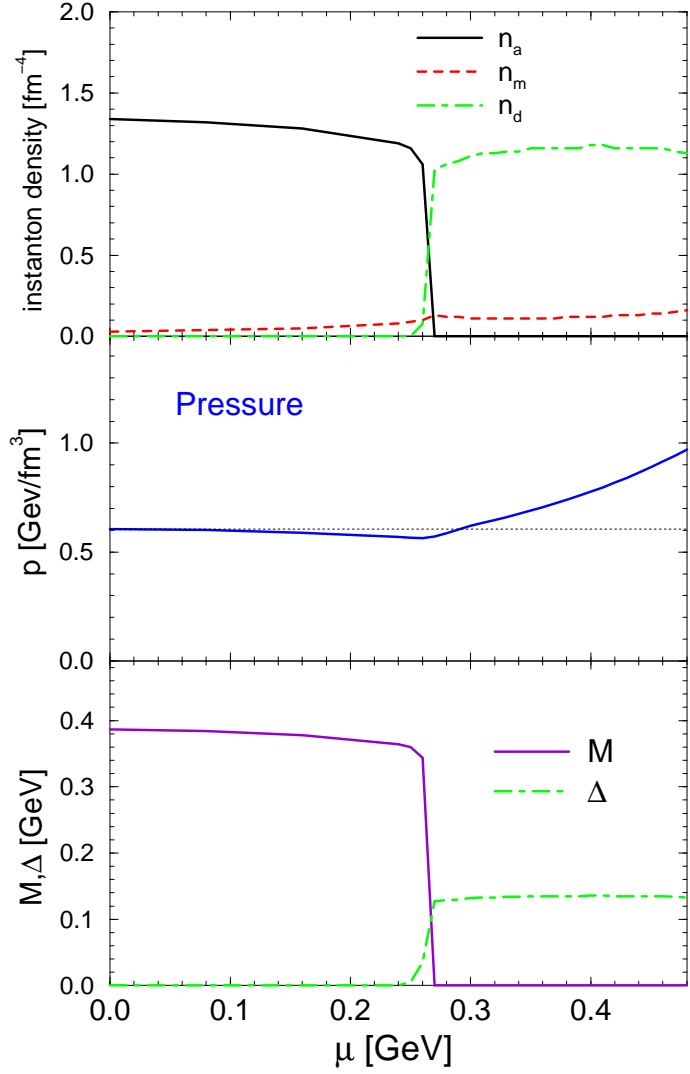,width=120mm,angle=0}}}
\end{center} 
\caption{The full two-flavor cocktail model including $I$-$A$ molecules 
and diquark pairing;  
upper panel:  densities of random instantons ('atomic'
component, full line), $I$-$A$ molecules (dashed line) and instantons
engaged in 'diquark chains' (dashed-dotted line);  middle panel:  
pressure $p(\mu)=-\Omega(\mu)$, maximized over $n_a, n_m$ and $n_d$  
at each value of $\mu$; 
lower panel: constituent quark mass (full line, using $C_M=2$ in  
Eq.~(\protect\ref{Mq}) to give
$M(\mu=0)=400$~MeV) and diquark gap
(dashed-dotted line, using $C_\Delta=1.5$ in Eq.~(\protect\ref{deltaMF}),
as discussed in the text).} 
\label{fig_ct2+}
\end{figure}

Let us now turn to the full two-flavor cocktail model with simultaneous
account for the chiral and diquark condensates as well as $I$-$A$ molecules.
The total free energy
\be
\Omega(n_a,n_d,n_m;\mu)= \Omega_{inst}^{a+d+m}(n_a,n_d,n_m;\mu)
+\frac{1}{N_c} \left[2\Omega_q^\Delta(M,\Delta;\mu)+(N_c-2)
\Omega_q^{QP}(M;\mu)\right] , \
\ee
is to be minimized \wrt~$n_a$, $n_m$ and $n_d$. The results
are shown in Fig.~\ref{fig_ct2+}, using the coefficient of $C_\Delta=1.5$
in Eq.~(\ref{deltaMF}) (which most closely resembles the results of
the calculations in sect.~\ref{sec_nf2}).

%%%%%%%%%%%%%%%%%%%%%%%%%%%%%%%%%%%%%%%%%%%%%%%%%%%%%%%%%%%%%%%%%%%%%%
\begin{table}[htb]
\begin{tabular}{ccccc}
 $n_a(\mu=0)$ [fm$^{-4}$] & $C_\Delta$ & $\mu_c$ [MeV] &
$\mu_0$ [MeV] & $\Delta(\mu_c)$ [MeV] \\ \hline
 1   & 2   & 300 & 355 & 158 \\
 1   & 1.5 & 265 & 290 & 120 \\
 1.4 & 2   & 310 & 375 & 188 \\
 1.4 & 1.5 & 270 & 305 & 142 \\
\end{tabular}
\caption{Parameter dependence of the critical chemical potential for chiral
restoration in a cocktail model with chiral and diquark condensates
(no $I$-$A$ molecules included).  }
\label{tab_ad}
\end{table}
%%%%%%%%%%%%%%%%%%%%%%%%%%%%%%%%%%%%%%%%%%%%%%%%%%%%%%%%%%%%%%%%%%%%

%%%%%%%%%%%%%%%%%%%%%%%%%%%%%%%%
%%%%%%%%%%%%%%%%%%%%%%%%%%%%%%%% 
As to be expected from the previous analysis, there is a delicate
competition for chiral restoration between random instantons engaged 
in diquarks and $I$-$A$ molecule formation. The former do, in fact, 
induce the chiral transition  for all parameter ranges considered. 

Towards higher densities it may happen that a second transition occurs
within the chirally restored phase characterized by a substantial jump 
in the molecule density and an accompanied drop of the diquark gap,
(its location is somewhat sensitive to parameter
choices). The singularity in $z_d$, on the other hand, guarantees that
there is always a finite $\langle qq\rangle$ condensate present, 
albeit possibly strongly reduced in molecule-dominated phases. 
In Fig.~\ref{fig_ct2+}, the transition occurs at $\mu_c=270$~MeV
into a chirally broken diquark phase, which, again, is mechanically
unstable. The vacuum pressure is only recovered at $\mu_0=295$~MeV, \ie, 
the combined effect of diquarks and molecules further lowers the 
$\mu_0$-values found when including only molecules (Fig.~\ref{fig_ct2},
where $\mu_0=350$~MeV) or only diquarks (Tab.~\ref{tab_ad}, where
$\mu_0=305$~MeV). 

Another important point to note is that the total instanton density, 
$N/V=n_a+n_d+2n_m$, is indeed essentially  independent of the chemical
potential within $\sim$15\% or so.  
{\it E.g.}, in  Fig.~\ref{fig_ct2},  the upward jump in the density of
molecules is close to half the drop in density of the random
instanton component. A similar continuity is seen in Fig.~\ref{fig_ct2+}
for the more realistic case of the transition
between the two kinds of random instanton liquids, associated with 
the two condensates $\langle\bar qq\rangle$ and $\langle qq\rangle$. 
After all this is not too surprising since, at least for chemical 
potentials $\mu\le 0.4$~GeV,  the by far dominant fraction
of the total energy in the system is carried by  
the gluonic component residing in instantons, and the  
free energy itself must of course be continuous across the transition.

Finally, a comment concerning the finite temperature behavior is
in order. On the zero-density and finite-$T$ axis it has been shown 
that molecules drive the chiral restoration. This implies
that at finite density, starting from low $T$, the role of molecules
should become more and more relevant; at the same time, 
color-superconducting  gaps are well-known  to be suppressed. Thus
we expect that for {\em any} value of $\mu$ where there is 
diquark condensed phase for $T=0$, a rearrangement into a molecule-dominated
phase should occur when raising the temperature. As indicated by our
zero temperature results, for chemical potentials $\mu\ge 300$~MeV or so, 
the corresponding $T_c(\mu)$ line in the phase   
diagram might actually reach down to  fairly low temperatures.

%%%%%%%%%%%%%%%%%%%%%%%%%%%%%%%%%%%%%%%%%%%%%%%%%%%%%%%%%%%%%%%%%%%%%
\subsection{Chiral Restoration for more Flavors}
%%%%%%%%%%%%%%%%%%%%%%%%%%%%%%%%%%%%%%%%%%%%%%%%%%%%%%%%%%%%%%%%%%%%%
\label{CKTnf3}

As was already mentioned in 
the introduction, in vacuum the tendency towards chiral 
restoration in the instanton model strongly  increases with the 
number of flavors, leading to a chirally symmetric vacuum state 
for $N_f$ as low as $\sim$~5. The  reason for that has been discussed 
in sect.~\ref{molint}: the increased number of quark lines 
enhances the 
interactions between instantons and antiinstantons, making the random
liquid less favorable. More specifically, the
integral over color orientations within the  
molecule, being proportional to $\langle (\cos \theta)^{2N_f} \rangle$,
increases more strongly than  that for  the averaged 'random'-instanton
configurations raised to the $N_f$-th power, 
$\langle (\cos \theta)^2\rangle^{N_f}$. The former integral is strongly
peaked at $\theta=0$, creating a ``locking'' of the color orientation
within a molecule. 

 From continuity one may thus expect that   
the critical chemical potential for chiral restoration should be further 
reduced when moving from two to more massless flavors. 
Here we would like to 
pursue the question in how far a $third$ massless flavor impacts
the results of the two-flavor case. For simplicity, we now ignore
color superconductivity and consider an interplay between random and
molecular components only. As discovered in \cite{ARW_98b} and further
elaborated in the first part our article, starting from $N_f=3$
a color-flavor locking phenomenon sets in, leading to
a complicated set of $\langle qq\rangle$ and $\langle \bar q q\rangle$ 
condensates. However, they
are relatively small in magnitude, and can therefore be neglected
for the present purpose.
 
Compared to the  $N_f$=2 case, the color integration    for $N_f$=3 is 
substantially more involved. Using the appropriate relations for the 
integration over a string of six SU(3)-color matrices 
(see, \eg, Ref.~\cite{Now_91}), one obtains 
\be
z_m(z_4,r) & \propto & \int du [T_{IA}(\mu) T_{IA}^\dagger(-\mu)]^{N_f}
\nonumber\\
 & = & \frac{3 \left( f_1^+ f_1^- + f_2^+ f_2^- \right)}
 {16N_c(N_c+2_)(N_c^2-1)} 
\left( 3N_c \left[ (f_1^+)^2 (f_1^-)^2 + (f_2^+)^2 (f_2^-)^2 \right] + 
\left[ 4-N_c \right] \left[ f_1^+f_2^- - f_1^- f_2^ +\right]^2 \right) \ .  
\label{colav3} 
\ee 
Note that $z_m(z_4,r)$ in the three-flavor case has no definite sign 
before integration over space-time (similar to the $N_f$=1 case, 
\ie, $z_a(z_4,r)$ from Eq.~(\ref{colav2})), which 
inevitably entails partial cancellations.  
Apparently, only for an even number of flavors the space-time integrand of 
the finite density activities is positive definite. 
Performing  again the minimization procedure in $n_a$ and $n_m$ 
for the thermodynamic
potential  we find that the critical chemical potential 
is indeed further reduced,  by about 10\% to $\mu_c\simeq$~270~MeV
(as compared to 310~MeV for $N_f$=2). 
This indeed complies with the above mentioned expectation 
that at a sufficiently large number of flavors a purely ``molecular''
vacuum is more preferable than a ``random'' one, and thus chiral symmetry
would be unbroken even in vacuum. 
As this phenomenon reflects itself also along the  $\mu$-axis, we expect 
$\mu_c$ to be further reduced at $N_f$=4, possibly crossing zero 
at $N_f$=5.

%%%%%%%%%%%%%%%%%%%%%%%%%%%%%%%%%%%%%%%%%%%%%%%%%%%%%%%%%%%%%%%%%%%%%
\subsection{Molecule-Induced Effective Couplings}
%%%%%%%%%%%%%%%%%%%%%%%%%%%%%%%%%%%%%%%%%%%%%%%%%%%%%%%%%%%%%%%%%%%%%
\label{sec_coupl}

In this section we  apply the cocktail model to a microscopic 
estimate of the density-dependence in the effective coupling constants 
for (anti-)quark-quark interactions. In the mean-field framework 
employed in sects.~\ref{sec_nf2},~\ref{sec_nf3},~\ref{sec_fsb} such 
density-dependencies were not accounted for. 

To begin with let us recall the expression for the 
effective coupling constant for single-instanton induced interactions 
as, \eg, used in Ref.~\cite{RSSV_98}:  
\be
G_{inst} = \int d\rho \ n(\rho,\mu) \ \rho^{N_f} \ (2\pi\rho)^4, 
\label{Ginst}
\ee
with 
\be 
n(\rho,\mu) = C_{N_c} \ \left(8\pi^2/g^2\right)^6 \
 \exp\left[-8\pi^2/g(\rho)^2\right]\ \rho^{-5}
\label{nrho}   
\ee
denoting the single-instanton distribution, \ie, the semi-classical 
tunneling amplitude. Replacing the $\rho$-integration by an 
average value for $\bar \rho$, one has 
\be
G_{inst} = n^{\pm} \ (\bar \rho)^{N_f} \ (2\pi\bar\rho)^4 
\ee
with the anti-/instanton density $n^{\pm}$. Notice that (for $N_f=2$) 
six additional 
powers of $\rho$ appear in the integral of Eq.~(\ref{Ginst}) as 
compared to the usual expression for the density, 
$n^\pm=\int d\rho \ n(\rho)$. This  entails 
a significantly larger average value $\bar\rho$ than the typical instanton 
size of about $\rho\simeq 1/3$~fm. In order to be consistent 
the standard value of the zero-density constituent quark mass 
$M \simeq 400$~MeV (which requires $G_{inst}\simeq 19.1$~fm$^2$), 
one should have the effective size saturating this integral
to be $\bar\rho=0.51$~fm. 

The effective coupling constant for molecule-induced interactions 
has been first derived in Ref.~\cite{SSV_95}, where it is written 
as  
\be
G_{mol}= \int n(\rho_1,\rho_2) \ d\rho_1 \ d\rho_2 \ 
\frac{1}{T_{IA}^2} \ (2\pi\rho_1)^2 \ (2\pi\rho_2)^2   
\label{Gmol} 
\ee
with the total molecule amplitude 
\be
n(\rho_1,\rho_2)=\int du \ d^4z \ n(\rho_1) \ n(\rho_2) \ 
T_{IA}(u,z)^{2N_f} \ \rho_1^{N_f} \ \rho_2^{N_f}  \ ,  
\label{nmol} 
\ee
which is nothing but the molecular activity given in Eq.~(\ref{zam}). 
The graphical interpretation of Eq.~(\ref{Gmol}) is quite transparent: 
starting from the molecule amplitude, where {\em all} 2$N_f$ quark legs
are closed within the molecule, one 'cuts' open two of them (corresponding
to the division by $T_{IA}^2$) which provides an effective 
interaction  between four external (anti-) quarks. 
Inserting (\ref{nmol}) into (\ref{Gmol}), and replacing again the size  
integrals by using average values for $\bar\rho$ we obtain 
\be
G_{mol}=\frac{(2\pi\bar\rho)^4 {\bar\rho}^{2N_f} n^+ n^-}{8} 
\int du \ d^4z \ T_{IA}(u,z)^{2N_f-2} \ . 
\label{Gmol2}
\ee 
We see that for $N_f=2$ the individual $\rho$ integrations carry additional
powers of $\rho^4$, while  it is $\rho^5$ for $N_f=3$. Those powers
should be compared to $\rho^6$ in Eq.~(\ref{Ginst}) and 
$\rho^0$ for the usual instanton density. As these last two integrals 
lead  to $\bar\rho\simeq$ 0.51 and  1/3 fm, respectively, we 
interpolate between them (lacking a more accurate determination of the size 
distribution), \ie,  estimate the appropriate average size values entering 
Eq.~(\ref{Gmol2}) to be $\bar\rho\simeq 0.43$~fm for $N_f=2$ and 
$\bar\rho\simeq 0.47$~fm for $N_f=3$. The corresponding zero-density
values for the coupling constants are (including a factor of 16 
accounting for the difference in color coefficients between 
Eqs.~(\ref{l_mes}) and (\ref{l_mesmol})): 
\be
G_{inst}(\mu=0,\bar\rho=0.51{\rm fm}) & = & 19.1~{\rm fm}^2 
\nonumber\\ 
16~G_{mol}^{N_f=2}(\mu=0,\bar\rho=0.47{\rm fm}) & = & 0.86~{\rm fm}^2 
\nonumber\\
16~G_{mol}^{N_f=3}(\mu=0,\bar\rho=0.43{\rm fm}) & = & 0.044~{\rm fm}^2 \ .
\ee  
Obviously, $G_{inst}\simeq 20\times(16~G_{mol}^{N_f=2})$, and 
$(16~G_{mol}^{N_f=2})\simeq20\times(16~G_{mol}^{N_f=3})$. 
Recalling that the fermionic matrix element, averaged over positions
and sizes of an $I$-$A$-pair, is given by~\cite{SS_98} 
\be
\rho^2\langle |T_{IA}|^2\rangle &=& \frac{2\pi^2}{3N_c} \frac{N\rho^4}{V} 
\simeq {1\over 40}
\ee  
(for $\rho=1/3$~fm),
one readily understands the decreasing magnitudes of the coupling constants
in terms of the additional powers of dimensionless diluteness of the ensemble 
entering Eq.~(\ref{Gmol2}). Note that the effect of color integrations
favors molecules, and the corresponding factors compensate, to some degree,
powers of the small diluteness parameter.

\begin{figure}[htb]
\begin{center}
\epsfig{file=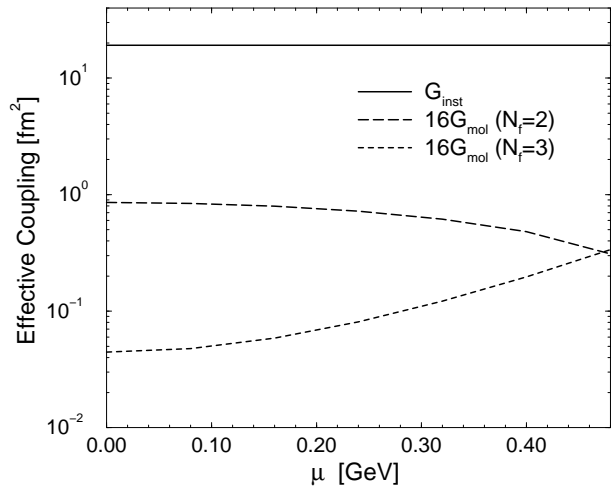,width=75mm,angle=-90}
\end{center}
\caption{Instanton-/molecule-induced effective 4-quark coupling 
constants as a function of chemical potential.}
\label{fig_Gmu}
\end{figure}

We are now in position to evaluate the density-dependence of 
the effective coupling constants. Using the fermionic 
matrix elements at finite $\mu$, Eq.~(\ref{tiamu}), and performing
the color integrations as discussed in sects.~\ref{CKTnf2},~\ref{CKTnf3} 
we obtain the results shown in Fig.~\ref{fig_Gmu}; we recall that  
throughout this paper possible effects from the Debye-screening of 
instantons at high densities have been ignored. Therefore  
$G_{inst}$ is in fact a constant as it does not depend on $T_{IA}(\mu)$. 
On the other hand, the behavior of the molecule-induced 
couplings depends on the number of flavors: whereas in the $N_f=2$ case
it decreases with $\mu$, the opposite is found for $N_f=3$. This 
is directly related to the $\mu$-dependence of the activities
calculated in sect.~\ref{CKTnf2}, since 
$G_{mol}^{N_f=2}(\mu)\propto T_{IA}^2$ (corresponding to $z_a$) 
and $G_{mol}^{N_f=3}(\mu)\propto T_{IA}^4$ (corresponding
to $z_m^{N_f=2}$).      

  It is instructive to compare the values of the coupling constants 
with the perturbative OGE interaction. For large-angle 
scattering\footnote{For small-angle scattering there appears
an additional logarithmic enhancement, which becomes relevant
for color superconductivity at asymptotically high $\mu$~\cite{Son_98}.} 
the coupling is
\be 
G_{OGE}={4\pi\alpha_s \over  p^2_{eff}} \ ,  
\ee 
where $p_{eff}$ is some effective momentum transfer averaged over 
the Fermi sphere. With $p_{eff}=0.5-1$~GeV and $\alpha_s(p_{eff})
\approx 0.3$ we have $G_{OGE}=0.15-0.7$~fm$^2$. This is comparable
to the effect of molecules for $N_f=2$, and significantly larger than 
that for $N_f=3$ at low $\mu$. One should note, however, that the structure 
of the interactions is different. For example, OGE is flavor independent
whereas instanton-induced interactions are flavor antisymmetric.

%%%%%%%%%%%%%%%%%%%%%%%%%%%%
%%%%%%%%%%%%%%%%%%%%%%%%%%%%

%%%%%%%%%%%%%%%%%%%%%%%%%%%%%%%%%%%%%%%%%%%%%%%%%%%%%%%%%%%%%%%%%%
\section{More Phases, Outlook  and Experimental Consequences}
\label{sec_phdia}
%%%%%%%%%%%%%%%%%%%%%%%%%%%%%%%%%%%%%%%%%%%%%%%%%%%%%%%%%%%%%%%%%%%%%

%%%%%%%%%%%%%%%%%%%%%%%%%%%%%%%%%%%%%%%%%%%%%%%%%%%%%%%%%%%%%%%%%%%%
\subsection{The phase diagram}
\label{sec_phd}
%%%%%%%%%%%%%%%%%%%%%%%%%%%%%%%%%%%%%%%%%%%%%%%%%%%%%%%%%%%%%%%%%

 In this section we would like to put the main results obtained in 
this work into perspective and discuss the emerging picture of the QCD 
phase diagram.

  The main feature of the phase diagram for two-flavor QCD in the chiral 
limit 
and at $T=0$ is that at a critical chemical potential $\mu_c$ 
the system undergoes a transition from the chirally broken phase to the 
superconducting phase~\cite{ARW_98,BR_98,CD_99}. 
With instanton-induced formfactors we
also found a small window in which a chirally broken `constituent
quark' plus  diquark condensate
phase may exist. Unfortunately, this is not a robust prediction of 
the model, since the free energy differences between the phases are 
small (in fact, in the cocktail model analysis of sect.~\ref{sec_cocktail}
this phase is mechanically unstable). 
 We will discuss this issue from a slightly different 
perspective in the next section. 

In any case, the mean-field approach predicts a strong first order phase 
transition, either from the vacuum phase to superfluid quark matter 
phase, or from the chirally broken diquark phase to the superfluid
phase. This implies the existence of an inhomogenous phase at 
intermediate density, with dense quark matter bubbles immersed in the 
chirally broken phase. Of course, a more refined treatment should 
reproduce the fact that matter clusters into nucleons and nuclei. 
In the high density phase chiral symmetry 
is restored, but color-$SU(3)$ is broken to $SU(2)$. We have not 
explored the possibility of further breaking $SU(2)$ via color-6 
condensates. These condensates seem to generate very small gaps 
for the up and down quarks of the third color~\cite{ARW_98}.

  For three massless flavors we also find a first order phase transition 
from the chirally broken vacuum phase to the superconducting phase.
In spite of the fact that instanton-induced dynamics are very different
from OGE considered in Ref.~\cite{ARW_98b}, we also find 
 that the preferred order parameter 
in the superconducting phase exhibits color-flavor-locking. Both
color-$SU(3)$ and chiral $SU(3)_L\times SU(3)_R$ are broken, while
the diagonal $SU(3)_{C+L+R}$ is preserved. But even though 
color-flavor locking in general implies that chiral symmetry is broken,
instantons are crucial for generating a non-zero value of $\langle
\bar qq\rangle$. In practice, the value of the chiral condensate
turns out to be small.

Furthermore, we have considered the more general case of QCD with 2+1  
flavors allowing for a massive strange quark. Of course, the two 
cases $N_f=2,3$ discussed 
above emerge as limiting cases for $m_s\to 0$ and $m_s\to\infty$. For
$m_s\simeq 2\sqrt{\Delta\mu}$ pairing between light and strange
quarks becomes impossible, and there is a phase transition between
the color-flavor locked phase and the two-flavor superconductor. 
Again, a significant difference between the instanton model and 
schematic interactions abstracted from one-gluon exchange appears.
In the case of OGE, the strange quark mass has a 
purely kinematical effect. Instantons induce four-fermion interactions 
of the type $m_s(ud)(\bar u\bar d)$, which can generate large 
asymmetries between the the $\langle ud\rangle$ and $\langle us\rangle
=\langle ds\rangle$ components of the color-flavor locked state
even for $m_s<m_s^{crit}$. Similar effects are well known from hadronic
spectroscopy. 

\begin{figure}[htb]
\begin{center}
\epsfig{file=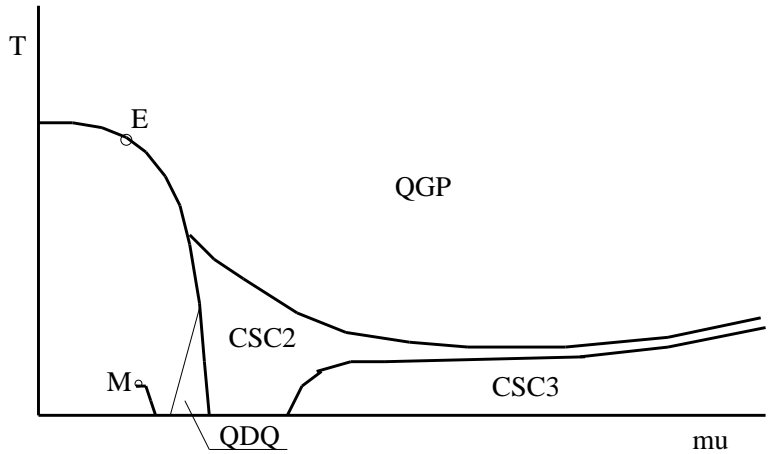,width=6cm,angle=270}
\end{center}
\vspace{1cm}
\caption{Schematic QCD phase diagram in the chemical potential
$\mu$ - temperature $T$ plane. The small $T$-/$\mu$-region
corresponds to ordinary hadronic matter, with broken chiral
symmetry. The point M (from ``multifragmentation'')
is the endpoint of the nuclear liquid-gas phase transition. The point E
indicates where the first order line either terminates in a second order
endpoint (for $m_u,m_d=0$) or disappears (for finite light quark masses).
CSC2 and CSC3 label the $N_f=2$ and  $N_f=3$-type superconducting phases.
The hypothetical intermediate quark-diquark phase is indicated by QDQ.}
\label{fig_phases}
\end{figure}
 In the present work we have not addressed the effects of finite 
temperature. One would expect that the first order chiral phase transition 
at $\mu\neq 0$ will persist for some range in temperature, until the 
transition becomes second order at a tricritical point~\cite{HJS*_98,BR_98}. 
In BCS theory one can also estimate that superfluidity disappears at 
a critical temperature $T_c(\mu)\simeq 0.6\Delta(\mu,T=0)$
(however, as discussed in sect.~\ref{sec_cocktail} the impact of
instanton-antiinstanton molecules presumably reduces the BCS-coefficient 
appreciably).
As explained above, the boundary between the two types of superconductors 
should be approximately determined by the condition (\ref{mismatch}).
 Since, asymptotically, the gap is expected to
slowly grow as a function of chemical potential, the critical 
temperature will also grow. As a function of chemical 
potential the system will eventually reach the color-flavor locked state
for any value of the strange quark mass. 
On the other hand, 
as a function of temperature (at given $\mu$) one might expect that,
since the gap in the two-flavor superfluid is somewhat larger
than in the color-flavor locked phase, the system first makes a transition
to the two-flavor superconductor, and then to the quark-gluon plasma
phase.
Combinig these conjectures leads to the schematic phase diagram
dsiplayed in Fig.~\ref{fig_phases} (where we have also indicated
the nuclear liquid-gas transition line).

%%%%%%%%%%%%%%%%%%%%%%%%%%%%%%%%%%%%%%%%%%%%%%%%%%%%%%%%%%%%%%%%%%% 
\subsection{More Phases?}
\label{sec_more}
%%%%%%%%%%%%%%%%%%%%%%%%%%%%%%%%%%%%%%%%%%%%%%%%%%%%%%%%%%%%%%%%%%% 

  Finally, we would like to make a few comments on the range of
applicability and the limitations of our approach. First, we have 
restricted ourselves to an instanton model. Even though many of 
our conclusions apply to any phenomenologically successful effective
interaction, many of our numbers are indeed model-dependent.

   Furthermore, we have made extensive use of the mean-field
approximation (MFA). The MFA is valid provided the condensates are 
sufficiently smooth, and their fluctuations can be neglected. This 
assumption becomes better at large density, because in this limit 
the coupling is weak and Cooper pairs are large compared to the 
inter-quark distances (as is the case in ordinary superconductors).
 In the opposite limit of small density the simplest phase one can
generate in the mean-field approach is a chirally asymmetric Fermi 
gas of constituent quarks with masses on the order of 
400~MeV. Remarkably enough, our approach predicts that 
this phase is unstable against separation into a mixed phase, with 
high density quark droplets separated by pure vacuum~\cite{Bub_96,ARW_98}.
While this result appears to be very suggestive, the MFA cannot predict 
the actual composition of the clusters.

Of course, we know that the ``correct''  clusters are nucleons. Below nuclear
matter saturation density, nucleons themselves will form a mixed phase
of clusters (nuclei), but for larger density one has homogeneous nuclear 
matter. In the vicinity of nuclear saturation density 
($n_N=n_0=0.16$~fm$^{-3}$), the 
equation of state is known experimentally. The behavior of the 
energy density as a function of density is commonly parameterized
as
\be
\epsilon= \left(m_N-\delta m_N\right)n_N + \frac{K n_0}{18}
 \left(1-{n_N \over n_0}\right)^2,
\ee
where $\delta m_N\approx 16$~MeV is the binding energy (per nucleon)
of nuclear matter
and the compression modulus is on the order of $K^{-1}=200-300$~MeV.  
This means that nuclear matter is rather stiff, that is, the pressure
grows very fast as a function of density. The physical reason for the
steep rise in pressure is not just Fermi motion, but, more importantly, 
a strong repulsive core in the nucleon-nucleon interaction. 
The standard lore is that the steep rise in the pressure 
with density above saturation density will eventually stop, due to the
transition to some other phase of matter, either
previously considered scenarios such as pion and kaon 
condensation~\cite{mesonic_cond}, or superconducting quark matter.  
In neutron stars, $K^-$ condensation is additionally favored because of
a large electron chemical potential.

Although lacking explicit confinement, the instanton model
does provide the interaction to bind quarks into nucleons~\cite{SSV_94}. 
The  question whether the model provides the repulsive core necessary 
to prevent nucleons from collapsing into 6 and more quark clusters  
remains open. Nevertheless, it seems plausible that a sufficiently
sophisticated treatment could lead to nuclear matter as the correct 
ground state at small density. 

 Can there be other phases, in addition to nuclear matter and superfluid 
quark matter? In the remainder of this section we will discuss a number
of possibilities connected with diquark fluids or Bose condensates.
These diquarks would not be Cooper pairs, but tightly bound states. 
As discussed in sect.~\ref{sec_diq}, the instanton model seems to 
predict such states as bound scalar $ud$ diquarks. The corresponding energy 
per baryon $E/B= 3M_{dq} /2 \sim 800-900$ MeV is lower than the nucleon 
mass, so it seems natural to look for a diquark phase.

 Naively, one would expect diquarks at $T=0$ to be Bose-condensed in 
the zero momentum state, because in addition to the gain in binding 
energy, there is a gain over nuclear or quark matter
because no Fermi motion is required. However, since diquarks are
colored, this phase could not be color neutral. This means that 
we have to consider either (i) significant motion of diquarks or 
(ii) add color-neutralizing quarks. 
 
  Let us start with the first idea, with only diquarks at $T=0$. 
A good starting point is the question why -- if the scalar diquark
is bound -- a two-baryon state, such as the deuteron, does not decay 
into a more tightly bound three-diquark state. The simplest color 
singlet combination of three $(ud)_i=\phi_i$ scalar diquark fields 
is $\phi_{i1} \phi_{i2}\phi_{i3}\epsilon_{i1,i2,i3}$, but this 
wave function is antisymmetric, violating Bose statistics. 
This means that we have to consider p-wave diquark states $\phi_i^m$ 
(\eg, in a  bag), where $m$ is the third component of angular
momentum, in a symmetric combination:    
\be 
\psi=\phi_{i1}^{m1}
\phi_{i2}^{m2}\phi_{i3}^{m3}\epsilon^{i1,i2,i3}\epsilon_{m1,m2,m3}.
\ee
Simple estimates show that such a state is no longer more economical.

Similarly, one can construct a wave function for infinite diquark 
matter starting from a set of plane wave states. The ground state would 
then be a quite peculiar  ``color crystal'', where the diquark momentum 
is determined from the balance of kinetic and potential energy, 
the latter resulting from the color-electric fields. We have not
attempted to calculate the energy in this phase, but it seems
much more natural to consider a quark-diquark phase.

 The quark-diquark (QDQ) phase could occur as an
intermediate phase between nuclear matter and the color 
superconducting phase. Let us  
focus on a flavor composition corresponding to neutron
matter (relevant for dense stars): $ud$ diquarks plus an equal 
amount of $d$ quarks. In this case, the total color and electric 
charge are zero. However, if the density is very low, color has 
to be neutralized over large distances, and confinement prohibits 
a phase like this. It is amusing to note that even without 
confinement, there is no low density QDQ phase. One reason is
that the nucleon is bound \wrt~a diquark and a quark. 
Another reason is that at very low density Fermi motion in the 
QDQ phase is more costly than in the nuclear phase. For example, 
let us set the threshold for the QDQ phase,  $M_{dqq}=M_{dq}+M$  
equal to the nucleon mass $m_N$. Then the density of color-compensating 
$d$ quarks is equal to the density of neutrons in the nuclear 
phase at the same baryon density. Furthermore, since both have 
the same degeneracy factor ($g_s$=2 due to spin, the color 
being fixed by the color of the diquark condensate), both the $d$ 
quark and the neutron have the same Fermi momentum $p_f$. The 
kinetic energy $p_F^2/(2 M)$ is then smaller for the
neutron because of its larger mass.

  Nevertheless, it is not obvious which phase is preferred at
densities a few times nuclear matter density. The QDQ phase is 
very different from both nuclear matter and color superconducting
($N_f=2$) quark matter, in particular both chiral symmetry and
color are broken. This suggests that if such a phase exits, it
is probably an isolated minimum, separated by first 
order transitions on both sides. Let us try to estimate if 
such a window may exist. We assume that $\langle\bar qq\rangle$ 
is large in this phase, still providing an effective quark mass 
$\cal O$(400 MeV), and therefore this phase should have no strange 
component, in contrast to the CSC3. The interactions in nuclear matter 
crucially combine long range attraction with short range repulsion.  
A similar description may be approximately valid for diquarks 
as well. Let us use a simple model, with only  repulsive interactions
represented by the scattering length\footnote{The radius of the 
nucleon repulsive core is about 0.4 fm, but diquarks (and constituent 
quarks) are smaller objects. If they are instanton-generated, their 
core should be of the order of the typical instanton radius $\rho
\approx 1/3$~fm~\cite{Shu_82}, which we took as a representative value.} 
$a$. The energy per baryon of the diquark Bose gas is
\be
{\epsilon_{dq} \over n_B}={12 \pi a n_{dq} \over m} 
 \left( 1+ {128 \over 15\sqrt{\pi}}(a^3n_{dq})^{1/2}\right) \ , 
\ee
where $n_{dq}$ denotes the diquark density. 
The first term is just the mean-field interaction of the 
condensed diquarks, the second term stems from non-condensed 
bosons, as follows from the classic Lee-Yang paper~\cite{LY}. In this 
approximation the unphysical behavior of the ideal Bose gas is overcome,
the chemical potential grows with the density, and at some density
it becomes favorable to split some diquarks into quarks. We are then 
led to a mixture of a Bose gas of diquarks and a Fermi gas of quarks 
(now of the same color), with chemical potentials related by the 
equilibrium condition $\mu_{dq} = 2 \mu$. Such a description leads 
to a  more natural transition to quark matter with Cooper pairs at 
high density.

%%%%%%%%%%%%%%%%%%%%%%%%%%%%
\begin{figure}[t]
%\vskip -0.4in
\epsfxsize=4.0in
\centerline{\epsffile{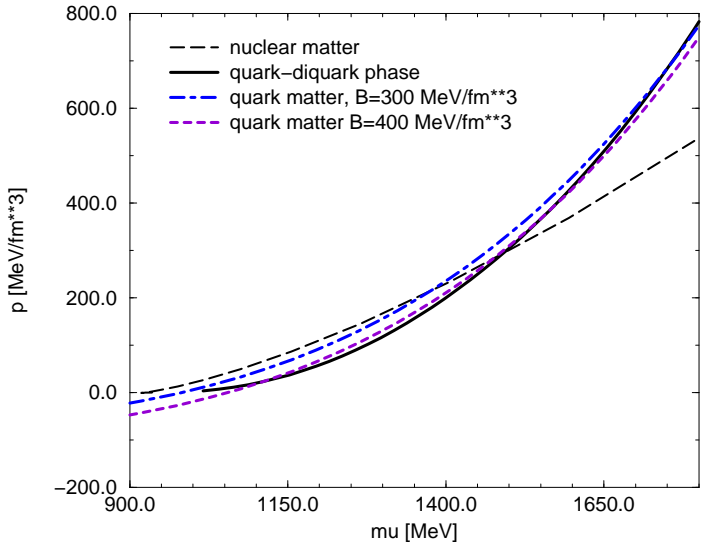}}
\caption{ Pressure versus baryonic chemical potential (3 times $\mu$ 
for quarks as used above) for 3 phases: nuclear matter, the quark-diquark 
phase made of a Bose gas of interacting diquarks and a Fermi gas of 
constituent quarks (no bag constant), and a Fermi gas of $u,d,s$ 
quarks with current masses (for two different bag constants).  }
\label{fig1_fb}
\end{figure}
%%%%%%%%%%%%%%%%%%%%%%%%%%%%

   So far we have ignored the role of confinement. One possibility is
that the transition to the quark-diquark phase leads to deconfinement. 
If not, one has to include the energy associated with separated color 
charges. It is well known that the confining potential in vacuum is 
linear $V(r)=Kr$, with a string tension $K\simeq 1$~GeV/fm. However,  
for small $r$ this relation only holds for heavy, point-like quarks, 
not for light ones. Various 
non-relativistic models of hadronic structure use some 
effective $K$, reduced by a significant factor, in order to obtain a  
good description of hadronic masses. 
Indeed, it was was found on the lattice -- by measuring 
the string tension after smoothing the gauge fields\footnote{For 
example, the results of \cite{FIMT_98} can be approximated as 
$V=K |R-R_0| \theta(R-R_0)$ with the standard string tension 
$K\simeq 1$~GeV/fm but a rather large $R_0\simeq 0.7$~fm. } -- 
that the effective potential between constituent
quarks is very small at small $r$, but approaches $V(r)=Kr$ at
large distances. The 
motivation for smoothening is the due to the extended nature of 
constituent quarks, as opposed to essentially point-like heavy flavors
($c,b$).  
To put it differently: color-strings only form if constituent quarks
do not overlap. For such  a potential the average energy of 
colored strings seems to be  negligible for the relevant densities
considered above.

%%%%%%%%%%%%%%%%%%%%%%%%%%%%%%%%%%%%%%%%%%%%%%%%%%%%%%%%%%%%%%%%%%%%%%%%
\subsection{Observable signatures} 
%%%%%%%%%%%%%%%%%%%%%%%%%%%%%%%%%%%%%%%%%%%%%%%%%%%%%%%%%%%%%%%%%%%%%

In this section we would like to discuss potential experimental signatures
of the quark-diquark and quark superconducting phases, in particular
with regard to heavy-ion experiments and neutron stars. 
Let us start with the quark-diquark 
phase. It is easy to estimate the critical temperature for the 
quark-diquark mixture by applying  Einstein's ideal gas expression for
Bose condensation 
\be 
T_c=3.31 n_{dq}^{2/3}/M_{dq} \ . 
\ee
Assuming $n_{dq}=3*n_0$ and $M_{dq}=0.6$~GeV we find $T_c=120$~MeV, 
which is expected to be further reduced by the short range repulsive
core. Heavy-ion collisions at SIS/BEVALAC energies (1-2~AGeV), 
where comparable compression is
reached, lead to a heating of the system of up to $T\approx 100$~MeV, and 
so we conclude that even if this phase exists and our estimates are valid, 
it can be involved only peripherally. The same is even more true
for the color superconducting phase. Because of the high density 
and low critical temperature of the color superconductor, 
it is not likely to be produced in heavy-ion collisions. 

  Even if the gap is still larger than expected so that  matter in 
the color superconducting phase would be  produced in heavy-ion 
collisions, its presence will be hard to establish. The two 
most spectacular manifestations of superconductivity, perfect
conductivity and the Meissner effect, are very difficult to detect for
a short-lived sample. The transition to the superconducting
state has an effect on the equation of state, but the 
condensation energy $\epsilon\sim\mu \Delta^2 p_F/(2\pi^2)$ is
small compared to the energy density of a Fermi gas. In this
context, the effect of quark-diquark phase is likely to be
larger. Finally, the superconducting phase, in particular
color-flavor-locking, may have an effect on the flavor 
composition. But this effect is not very specific since  
strangeness enhancement is a general consequence of 
quark matter formation.

  For these reasons, compact stars and stellar explosions 
are probably a  more appropriate  place to search for observational
consequences of quark superconductivity. For a recent review
on the structure of neutron stars we refer the reader to 
Refs.~\cite{HHJ_99,HPS_94}. Owing to theoretical uncertainties in the 
nuclear equation of state, the central density of neutron
stars is not very well known. The main experimental constraint
comes from the fact that neutron stars with masses 
$M_{NS}=1.45M_{Sun}$ have definitely been observed. This still 
allows for central densities as low as $3n_0$ or as high as $10n_0$.
Ultimately, a better handle on the central density will come
from measurements of neutron star radii. 

 In neutron star structure calculations quark matter is 
almost always treated as a simple Fermi gas, confined by a 
bag constant $B$. Let us only note here that there are two 
distinct scenarios: (i) For sufficiently small values of $B$ 
strange quark matter is absolutely stable, and the entire star 
is in this phase, while (ii) for large $B$ the outer part 
of the star consists of nuclear matter, while the interior
contains various mixed phases (quark matter sheets, rods 
or clusters). These funny arrangements owe their existence
to the possibility to move charge from the quark phase 
to the hadronic phase. The shape is then determined by the
interplay of the equation of state and long range Coulomb 
forces. Quark superconductivity will again have an influence on the
equation of state, but as before we expect this to be a correction 
on the order of ${\cal O}((\Delta/\mu)^2)$, small compared to the 
uncertainties in the bag constant. 

  In neutron stars, a more direct measure of the gap is 
provided by the cooling history of the star. Without a gap,
neutron stars can efficiently cool by $\beta$-decay of thermally
excited $u,d$ quarks. This process seems to lead to unacceptably
large cooling rates~\cite{Iwa_82}. Such a problem emerged  
already for neutrino emission from nuclear matter. In 
that case it is solved due to nuclear superfluidity. Since both
neutrons and protons are gapped, there are no single-particle
states in the vicinity of the Fermi surface. For  
two-flavor quark matter, a similar problem arises since the quarks
of the third color have no or only very small gaps. It is 
absent in the color-flavor-locked phase, because all quarks
acquire a gap. For realistic values of the strange quark mass, 
the system is likely to be in the two-flavor phase at moderate  
densities, so that the ``cooling-problem'' for  neutron 
stars might still persist.
 
Another feature of the CSC-phases is that 
Cooper pairs are electrically charged, so one might expect the
color superconductor to be electrically superconducting as well. 
This is not quite true. For both the $N_f=2$ and color-flavor
locked phase, there is a modified charge operator that is not
broken, so there is a linear combination of the photon and 
the diagonal gluons that remains massless. Nevertheless, since
the photon inside the superconducting phase is different from 
the photon outside, magnetic flux will be $partially$ expelled. 
Magnetic fields in pulsars are very large, up to $10^{12}$ Gauss.
Fields on the order of $10^{15}$ Gauss were recently suggested 
to drive the so called ``magnetars'', and $10^{18}$ gauss
is the absolute upper limit allowed by star stability (virial theorem). 
However, even such fields are not yet large enough to significantly
influence the CSC phase. On the other hand, simple scaling
considerations show that in order to have the field completely 
expelled from the quark matter core would cost energy of order 
${\cal O}(R^3)$ 
(where $R$ is the star radius), while transferring the field lines
through some channels into the superconductor only costs ${\cal O}(R)$. 
Inside these channels the field should be at the critical value 
$B_c\sim 10^{19}$ Gauss. They can be either  macroscopic 
(if the superconductor is of the first kind) or microscopic 
Abrikosov vortices  (if it is of the second kind).

%%%%%%%%%%%%%%%%%%%%%%%%%%%%%%%%%%%%%%%%%%%%%%%%%%%%%%%%%%%%%%%%%%%
\section{Summary and Conclusions}
\label{sec_concl}
%%%%%%%%%%%%%%%%%%%%%%%%%%%%%%%%%%%%%%%%%%%%%%%%%%%%%%%%%%%%%%%%%%

 We have studied the interplay of instantons, superfluidity/-conductivity
and chiral symmetry breaking in QCD at finite density. Unlike many 
schematic models based on short-range interactions abstracted from 
one gluon exchange, the instanton model has the virtue of providing 
a realistic phenomenology of the zero-density ground state, including
such features as spontaneous chiral symmetry breaking and a correct
description of (light) hadron spectroscopy. 
This gives us some confidence for a semi-quantitative 
investigation of cold quark matter at moderate densities  
as might be encountered, \eg, in the core of neutron stars.

  We started out by reviewing some properties of 
the instanton-induced quark-quark interaction at $\mu=0$: 
it originates from the same effective interaction that leads to chiral 
condensation and a light pion in the QCD vacuum, predominantly acting  
in the scalar-isoscalar, color antitriplet $qq$ channel.  
We found these correlations to be sufficiently strong to 
generate a bound-state pole in the corresponding  diquark propagator.

 In a second step we studied the density dependence of the 
instanton-induced (`t Hooft) interaction. It 
arises through the modification of the quark zero modes as
we introduce a chemical potential. 
%E   In particular we showed
%E  that the modified zero modes automatically incorporate 
%E  the BCS instability.
Both the density-dependence of the instanton form factors, governing 
the effective quark-(anti-)quark interactions,  and of the 
instanton-antiinstanton overlap matrix elements, relevant 
for a statistical treatment of the instanton liquid, have been assessed. 
The main difference to earlier calculations (using local
approximations or schematic formfactors)  manifests itself in 
quantitatively somewhat larger superconducting gaps ($\sim$~150-200~MeV).  

In the first main part of this article we have  performed a systematic 
study of the finite-$\mu$, $N_c=3$-QCD phase structure for 
2, 3 and 2+1 flavors within the mean-field approximation: 
\begin{itemize}
\item[(i)]
For two massless flavors
we confirmed the usual transition to a $ud$ superconductor 
at $\mu_c\simeq 300$~MeV with gaps $\Delta\simeq 200$ MeV. 
Apart from some quantitative deviations  
%\footnote{In the subsequent MFA calculations for $N_f>2$ 
%(items (ii), (iii)) we have ignored the moderate quantitative deviations 
%caused by the  complicated non-local formfactors.}
the use of microscopic instanton formfactors entailed the 
possible existence of a novel intermediate phase, namely 
chirally broken quark-diquark matter (characterized by simultaneous 
chiral symmetry breaking and diquark condensation).
\item[(ii)] For three massless flavors we 
returned to a simplified treatment based on a idealized sharp 
formfactor (amenable to the more involved calculations with somewhat
less accuracy). 
We found that, like schematic one-gluon exchange interactions, 
instantons lead to color-flavor locking, despite their very  
different color-flavor vertex structure. In addition,  
there emerged an interesting new feature: in contrast to OGE,  
instantons generate 
a non-zero (albeit small) chiral condensate in the superfluid phase.
\item[(iii)] The consequences of finite (current) strange quark 
masses $m_s$ have been investigated. 
With increasing  $m_s$ a sharp phase transition
from the color-flavor locked phase to the two-flavor
superconductor occurs. At small chemical potential, the critical strange 
quark mass is smaller than the physical mass. As the chemical
potential grows, the critical mass is also expected to grow. 
We find that the color-flavor locked phase will appear only 
at $\mu>450-500$~MeV. This  implies that chiral symmetry would 
be restored initially and then broken again at higher density, 
but with much smaller condensates.
\end{itemize} 

In the second main part of our article a statistical mechanics treatment
of the instanton liquid has been employed. This enabled us to 
incorporate effects that go beyond the standard 
mean-field approximation, in particular those associated with 
instanton-antiinstanton molecules. We demonstrated how to handle a 
complex fermion determinant in this context (using a gluonic 
interaction that has been averaged over the relative color orientation). 
Without molecules, the results were shown to be in reasonable
agreement with the mean-field analysis in the first part.
Including correlations, molecule formation constitutes a  $\sim$10\% 
effect in the free energy/critical chemical potential 
of the  $T$=0-, $N_f=2$-transition. 
However, with rising temperature $I$-$A$ molecules 
are expected to play an increasingly important  role in the 
transition between the superconducting and the plasma phases, 
corresponding to the critical temperature for superconductivity. 
Furthermore, the statistical mechanics approach allows to assess the 
$\mu$-dependence of the total instanton density.
It turns out that the latter is indeed approximately constant for all
$\mu$-values under consideration,  
which is not really  surprising as the gluonic energy within the
instanton component carries the dominant fraction of the free energy
of the system. This supports the respective 
assumption made in earlier works as well as in the first part of 
this article. 
  
 Based on our results, we conjectured a qualitative picture of
the QCD phase diagram in the $T$-$\mu$ plane.  We argued that
for realistic values of the strange quark mass, there are both
finite-$\mu$ and finite-$T$ transitions between the color-flavor
locked (CSC3) and two-flavor superconducting (CSC2) states.
With increasing density and at small $T$, chiral symmetry is first 
restored and then (weakly) broken again.
We also elucidated on the 
possibility of a new quark-diquark phase, in which nucleons are dissolved 
into a Bose gas of $ud$-diquarks and a Fermi gas of unpaired quarks, 
characterized by simultaneously broken color and chiral symmetry. 
It represents a natural possibility for an intermediate phase
between the hadronic and superconducting phase,
although numerical estimates for its existence are rather uncertain.

Concerning experimental consequences, we conclude that 
heavy-ion reactions are unlikely to reach into the rich
high-density/low-temperature phase structure discussed here. 
In this respect neutron stars are much more promising. The 
pairing gaps should leave their traces in  cooling rates
and even in the equation of state provided the pairing gaps are 
large enough. Decomposition of the
``external'' into ``internal'' magnetic fields presumably imply   
complicated configurations inside the stars.  

As an outlook, we expect that it should be rather straightforward 
to generalize our approach to a simultaneous account of finite 
$\mu$ and $T$. Another major   
challenge is to go beyond the mean-field approximation in the 
quark sector in order to address clustering of quarks into
nucleons, and, even more difficult, the nucleon-nucleon 
interaction. The instanton liquid
model does provide the correct properties on the one-nucleon 
level. Turning the instanton liquid into a realistic description of 
nuclear matter, however,  requires a long way to go. 
These and related issues certainly  provide exciting 
opportunities for future research in the field of finite-density QCD.

\vspace{1cm}

{\bf ACKNOWLEDGEMENTS} \\
R. R. acknowledges support from the A.-v.-Humboldt foundation
as a Feodor-Lynen fellow. T. S. is supported by NSF-PHY-513835.
This work is supported in part by US-DOE grants
DE-FG02-88ER40388, DE-FG06-90ER40561 and DE-FG02-91-ER40682.

%--------------------------------------------------------------------

\newpage
%%%%%%%%%  APPENDIX %%%%%%%%%%%%%%%%%%%%%%%%%%%%%%%%%%%%%%%%%%%%%%%%%%%%%%%%
\begin{appendix}
%%%%%%%%%%%%%%%%%%%%%%%%%%%%%%%%%%%%%%%%%%%%%%%%%%%%%%%%%%%%%%%%%%%%%%%%%%%

%%%%%%%%%%%%%%%%%%%%%%%%%%%%%%%%%%%%%%%%%%%%%%%%%%%%%%%%%%%%
\section{Fierz Transformations}
\label{fierz}
%%%%%%%%%%%%%%%%%%%%%%%%%%%%%%%%%%%%%%%%%%%%%%%%%%%%%%%%%%%%

Let us denote a general four-fermion interaction of the type
(\ref{l_nf2}) by $(\bar \psi_a {\cal O} \psi_b)(\bar \psi_c 
{\cal O} \psi_d)$. An effective ``mesonic'' interaction 
corresponds to the sum of the direct and exchange ($s$+$u$) 
channels. The corresponding kernel is the sum of the 
original interaction and its Fierz transform:
\be
"s+u"=(\bar \psi_a {\cal O} \psi_b)(\bar \psi_c {\cal O} \psi_d)+
(\bar \psi_a {\cal O}_1 \psi_d)(\bar \psi_c {\cal O}_1 \psi_d) \ .
\ee
An effective diquark ($t$-channel) interaction is obtained by 
first transposing and then performing the Fierz transformation
\be
T=(\bar \psi_a {\cal O} \psi_b)(\psi_d^T {\cal O}^T \bar \psi_c^T)=
\epsilon(\bar \psi_a {\cal O} \psi_b)(\psi_d^T C {\cal O} C \bar
\psi_c^T)=
\epsilon(\bar \psi_a {\cal O}_2  C \bar \psi_c^T)(\psi_d^T C {\cal
O}_2 \psi_b) \ . 
\ee
We use Euclidean gamma matrices in the standard or chiral 
representations, $C=-i \gamma_2\gamma_0, C^2=C^\dagger C=1\!\!1$.
If  the isospin and color parts of ${\cal O}$ are $1\!\!1$, $\epsilon=-1$ 
if the Dirac part of ${\cal O}$ is $\gamma_\mu, \sigma_{\mu \nu}$, and
$\epsilon=1$ when it is $1\!\!1_D, \gamma_5,\gamma_\mu \gamma_5$.
For the flavor part, $\tau_2^T=- \tau_2$ and $\tau_{1,3}^T= \tau_{1,3}$.

For the Fierz transformation the following completeness relations are
used:
\be
\delta_{ii'}\delta_{jj'}={1\over N_c}\delta_{ij'}\delta_{ji'}+2t^n_{ij'}
t^n_{ji'} 
\ee
for color, 
\be
\delta_{AA'}\delta_{BB'}&=&{1\over 2}\delta_{AB'}\delta_{BA'}+{1\over 2}
\tau^a_{AB'}\tau^a_{BA'}  
\\
\tau^a_{AA'}\tau^a_{BB'}&=&{3\over 2}\delta_{AB'}\delta_{BA'}-{1\over 2}
\tau^a_{AB'}\tau^a_{BA'}  
\ee
for isospin, and, defining $S=1\!\!1\otimes 1\!\!1$,$V=\gamma_\mu 
\otimes \gamma_\mu$, $P=\gamma_5 \otimes 
\gamma_5$, $A=\gamma_5\gamma_\mu \otimes\gamma_\mu$,$T=\sigma_{\mu \nu}
\otimes\sigma_{\mu \nu}$, with $\sigma_{\mu \nu}=1/2[\gamma_\mu,
\gamma_\nu]$, 
\be
\pmatrix{S\cr V\cr T \cr A\cr P\cr}'=\pmatrix{1/4&1/4&-1/8&-1/4&1/4\cr
1&-1/2&0&-1/2&-1\cr -3&0&-1/2&0&-3\cr-1&-1/2&0&-1/2&1\cr
1/4&-1/4&-1/8&1/4&1/4\cr}\pmatrix{S\cr V\cr T \cr A\cr P\cr} \ 
\ee
for the Dirac structures. Due to the tensor product structure of 
the matrices ${\cal O}$, one has to Fierz transform separately the Dirac, 
color and isospin parts by using the above relations and multiply them 
accounting for the fact that fermion fields anticommute. The latter  
generates an additional (-1) in the $u$-channel, and for the $t$-channel it 
excludes any symmetric parts of the ${\cal O}'$ matrices. The 
final results are given in the main text, Eqs.~(\ref{l_mes}) and
(\ref{l_diq}).

%%%%%%%%%%%%%%%%%%%%%%%%%%%%%%%%%%%%%%%%%%%%%%%%%%%%%%%%%%%%%%%%%%%%%%
\section{Instanton Form Factors at Finite $\mu$}
\label{app_ff}
%%%%%%%%%%%%%%%%%%%%%%%%%%%%%%%%%%%%%%%%%%%%%%%%%%%%%%%%%%%%%%%%%%%%%%

The two structures in the form factor are
\be
B&=&4\rho i \int_0^\infty dR R^3 \int_0^\pi d\eta\,
  \sin^2 \eta{1\over R\sqrt{R^2+\rho^2}}
 \left[\left(1-2{t^2\over R^2+\rho^2}-\mu t\right)
 {\sin(\mu r)\over r}\right.\cr
& &\hspace{0.5cm}\left.
 -\left(2{t\over R^2+\rho^2}+\mu\right)
  \cos(\mu r)\right]{\sin(kr) \over kr}e^{-i\omega t}
\ee
\be
A&=&-4\rho\int_0^\infty dR R^3 \int_0^\pi d\eta\,
 \sin^2 \eta{1\over R\sqrt{R^2+\rho^2}}
 \left[-\left(\mu+{t\over r^2}+{2t\over R^2+\rho^2}\right)
\sin(\mu r)\right.\cr 
& & \hspace{0.5cm}\left.
 +\left({\mu t \over r}-{2 r \over R^2+\rho^2}\right)
 \cos(\mu r)\right]
 \left({\cos(kr)\over k r}-{\sin(k r) \over k^2 r^2}\right)
 e^{-i\omega t}.
\ee
Let us separate the real and imaginary parts of $A$ and $B$:
$A=A_c+i A_s, B=B_s+i B_c,$ the subscript referring to  
$\cos(\omega t)$ or $\sin(\omega t)$. We also introduce $k^\pm=k\pm\mu$.
Then 
\be
A_c&=&{2 \rho\over k}\int_0^\infty dR \int_0^\pi d\eta {R \sin \eta 
\cos(\omega R \cos \eta) \over \sqrt{
R^2+\rho^2}}\cr
& & \left[ \left(
{2 R \sin \eta \over R^2+\rho^2}+{\mu \over k R \sin \eta}\right)
\cos(k^+ R \sin \eta)\right.\cr
 & &\left.+\left(\mu -{2\over  k( R^2+\rho^2)}\right)\sin 
(k^+ R \sin \eta)\right] + (\mu \to -\mu)
\ee
\be
A_s&=&{-i 2 \rho\over k}\int_0^\infty dR \int_0^\pi d\eta {R\cos \eta 
\sin(\omega R \cos \eta) \over \sqrt{
R^2+\rho^2}}\cr
& & \left[ \left(
{1\over k R^2 \sin^2 \eta}+{2 \over k (R^2+\rho^2)}-\mu\right)\cos(k^+
R \sin \eta)  \right.\cr
 & &\left .
+\left({k^+\over kR \sin \eta}+{2 R \sin \eta \over 
R^2+\rho^2}\right)\sin 
(k^+ R \sin \eta)\right] - (\mu \to -\mu)
\ee
\be
B_c&=&{-i 2 \rho\over k}\int_0^\infty dR \int_0^\pi d\eta {\cos(\omega R \cos 
\eta)\over \sqrt{
R^2+\rho^2}}
 \left[ \left(
1-{2 R^2 \cos^2 \eta \over R^2+\rho^2}\right)\right.\cr
 & &\left . \cos
(k^+ R \sin \eta)+\mu R \sin \eta\sin 
(k^+ R \sin \eta)\right] - (\mu \to -\mu) 
\ee
\be
B_s&=&{2 \rho\over k}\int_0^\infty dR \int_0^\pi d\eta {R \cos \eta
\sin(\omega R \cos \eta)\over \sqrt{R^2+\rho^2}}\cr & &
 \left[\mu\cos
(k^+ R \sin \eta)-{2 R \sin \eta \over R^2+\rho^2}\sin 
(k^+ R \sin \eta)\right] + (\mu \to -\mu) \ .
\ee
Using the following basic integrals,
\be
\int_0^\pi d\eta \cos( \alpha \cos \eta)\cos(\beta \sin \eta)= \pi J_0(\sqrt{
\alpha^2+\beta^2})
\ee
and, denoting $\omega^\pm=\sqrt{\omega^2+(k^\pm)^2}$, 
\be
\int_0^\infty dR{1\over \sqrt{R^2+\rho^2}} J_0(R \omega^\pm)=I_0({\rho 
\omega^\pm \over 2})K_0({\rho 
\omega^\pm \over 2}) \equiv I_0^\pm K_0^\pm ,
\ee
one can rewrite the above expressions as derivatives of 
$I_0^\pm K_0^\pm $ according to 
\be
A_c&=&{2 \pi \rho \over k}\left[ {2 \over \rho}{d\over d \rho} 
\left({d^2 \over d k^2}-{1 \over k}{d \over d k} \right) + \mu\left({1
\over k}- {d \over d k} 
\right)\right]I_0^+ K_0^+ +(\mu \to -\mu),
\\
A_s&=&{i 2 \pi \rho \over k}\left[ {d\over d \omega} \left({2 \over
\rho}{d\over  
d \rho}\left({d \over d k}-{1 \over k}\right)-\mu\right)I_0^+ K_0^+ +
\int_0^\mu d \mu' {k^+\over k}{d\over d \omega}I_0^+ K_0^+\right]\cr & &
-(\mu \to -\mu) \ . 
\ee
Using ${d \over d \omega}={\omega \over \omega^+}{d \over d \omega^+}$,  
${d \over d \mu}={k^+ \over \omega^+}{d \over d \omega^+}$, 
the integral over 
$\mu'$ is just ${\omega \over k}I_0^+ K_0^+$. Furthermore, 
\be
B_c&=&{i 2 \pi \rho \over k}\left(1-{2 \over \rho}{d\over d \rho}{d^2
\over d \omega^2}-\mu {d \over dk}\right)I_0^+ K_0^+ -
(\mu \to -\mu),
\\
B_s&=&{- 2 \pi \rho \over k}\left(\mu-{2 \over \rho}{d\over d \rho}{d \over dk}
\right) {d\over d \omega}I_0^+ K_0^+  +(\mu \to -\mu).
\ee
Denoting ${d\over dz}I_0^+ K_0^+=I_1^+ K_0^+-I_0^+ K_1^+\equiv \Delta^+$, we get
\be
A_c&=&{2 \pi \rho \over k}\left[ \left({\mu \over k}+{2 (k^+)^2 \over 
(\omega^+)^2}\right)I_1^+ K_1+ {\rho k^+ \over 2 \omega^+}(2k+\mu)
\Delta^+\right]+(\mu \to -\mu),
\\
A_s&=&{i2 \pi \rho \over k}\left[ \left({\omega \over k}+{2\omega k^+\over 
(\omega^+)^2}\right)I_1^+ K_1+ {\rho \omega  \over 2  \omega^+}(2k+\mu)
\Delta^+\right]-(\mu \to -\mu),
\\
B_c&=&{i 2 \pi \rho \over k}\left[{\omega^2- (k^+)^2 \over 
(\omega^+)^2}I_1^+ K_1+{\rho  \over 2 \omega^+}(2\omega^2+\mu k^+)
\Delta^+\right]-(\mu \to -\mu),
\\
B_s&=&{2 \pi \rho \over k}\left[{2\omega k^+\over 
(\omega^+)^2}I_1^+ K_1+ {\rho \omega  \over 2  \omega^+}(2k+\mu)
\Delta^+\right]+(\mu \to -\mu).
\ee
These Fourier transforms of the fermion zero modes at finite $\mu$ agree
with the results obtained in \cite{CD_99}.

%%%%%%%%%%%%%%%%%%%%%%%%%%%%%%%%%%%%%%%%%%%%%%%%%%%%%%%%%%%%
\section{Grand Canonical Potential in the Cornwall-Jackiw-Tomboulis (CJT) 
Formalism}
\label{app_CJT}
%%%%%%%%%%%%%%%%%%%%%%%%%%%%%%%%%%%%%%%%%%%%%%%%%%%%%%%%%%%%
To derive the grand canonical potential in MFA we start from 
a generating functional with bilocal meson and diquark sources, 
\be
\exp(W[J,\bar J,K,\bar K])&=&\int {\cal D}\psi{\cal D}\bar \psi
\exp\left\{(S_0+U+\int\bar\psi J\psi
+\int\psi^T \bar J
\bar\psi^T+\int\bar\psi \bar K\bar\psi^T + \int\psi^T  K\psi)\right\} \ ,
\label{Z_source}
\ee
where $S_0$ is the free fermion action and $U=\int({\cal L}_{mes}+{\cal
L}_{diq})$  with ${\cal L}_{mes}, {\cal L}_{diq}$ given in
Eqs.~(\ref{l_mes},\ref{l_diq}). Using a matrix representation, 
we write the free part of the action in momentum space as
\be
S_0&=&\int {d^4p\over (2 \pi)^4}\left[ \bar\psi(p)(J(p)+{1 \over
2}G_0^{-1}(p))\psi(p) +\psi^T(-p)(\bar J(p) 
\right.
\nonumber\\
&& \qquad \qquad \left. +{1 \over 2}G_0^{-1T}(-p))\bar\psi^T(-p)+ 
\bar\psi(p)\bar K(p)\bar\psi^T(-p) +\psi^T(-p) K\psi(p)\right]
\nonumber\\
 &=& \int {d^4p\over (2 \pi)^4}
\pmatrix{\bar\psi(p),&\psi^T(-p)\cr}\pmatrix{ \bar K(p) &J(p)+{1 \over
2}G_0^{-1}(p) \cr \bar J(p)-{1 \over
2}G_0^{-1T}(-p) & K(p)\cr}\pmatrix{\bar\psi^T(-p) \cr \psi(p)\cr},
\label{S_source}
\ee 
where the sources $J,\bar J, K, \bar K $ have the following properties:
\be
\label{sour_sym}
-\bar J^T(-p)=J(p)\nonumber \\
-\bar K^T(-p)=\bar K(p)\nonumber \\
- K^T(-p)= K(p) \ . 
\ee
The path integral over the fermion fields is
\be
e^{W_0}&=&\int {\cal D}\psi{\cal D}\bar \psi e^S={\det} ^{1\over2}\pmatrix{ \bar K
&J+{1\over 2}G_0^{-1}\cr \bar J- {1\over 2}G_0^{-1T}& K\cr}\equiv \det
{\cal M} \nonumber \\
W_0&=&{1 \over 2} \tr \ln {\cal M} \ ,
\ee
where the trace includes the momentum integration, and the
products between the fields (sources) include convolutions in momentum
as well as all other indices. Using the identities  
\be
\pmatrix{ \bar A & B \cr \bar B & A\cr}&=&\pmatrix{ 0 & B \cr \bar B & 0\cr}
\pmatrix{ 1\!\!1 &\bar B^{-1} A \cr  B^{-1}\bar A & 1\!\!1\cr},
\nonumber\\
\ln \det \pmatrix{ \bar A & B \cr \bar B & A\cr}&=&\ln \det(-B\bar
B)+\tr \ln(1+\pmatrix{ 0 &\bar B^{-1} A \cr  B^{-1}\bar A & 0\cr}
\nonumber\\
&=&\ln \det(-B\bar B)+\tr \sum_{n=1}^{\infty}{(-1)^{n-1}\over n
}\pmatrix{ 0 &\bar B^{-1} A \cr  B^{-1}\bar A & 0\cr}^n 
\nonumber\\
&=&\ln \det(-B\bar B)+\tr \sum_{n=1}^{\infty}-{1\over 2n}(\bar B^{-1}
AB^{-1}\bar A +B^{-1}\bar A\bar B^{-1} A)
\nonumber\\
&=& \ln \det(-B\bar B)+\tr \ln(1-\bar B^{-1}
AB^{-1}\bar A)
\nonumber\\ 
&=&\tr \ln(-B\bar B+BAB^{-1}\bar A) \ ,
\label{det_trick}
\ee
we obtain  
\be
W_0={1\over 2}\tr \ln[-(J+{1\over 2}G_0^{-1})(\bar J- {1\over 2}G_0^{-1T})+
(J+{1\over 2}G_0^{-1})K(J+{1\over 2}G_0^{-1})^{-1}\bar K] \ .
\ee
Next, one introduces the classical two-point fields
$\bar F= {\delta W \over \delta \bar K}$, $F= {\delta
W \over \delta  K}$, $G= {\delta W \over \delta
J}$, $\bar G= {\delta W \over \delta \bar J}$ and performs a  
Legendre transformation
\be
\Gamma[G, \bar G, F,\bar F] =W[J,\bar J,K,\bar K] -\tr(GJ+\bar G \bar
J +KF+\bar K\bar F) 
\ee
with $J,\bar J,K,\bar K$ expressed as functionals of $G, \bar G, F,\bar
F$.  From the non-interacting part $W_0$ we have  
\be
G&=&{1\over2}( J+{1\over 2}G_0^{-1} -\bar K(\bar J-{1\over
2}G_0^{-1T})^{-1} K)^{-1 }
\nonumber\\ 
\bar G&=&{1\over2}( \bar J-{1\over 2}G_0^{-1T} - K(J+{1\over
2}G_0^{-1})^{-1} \bar K)^{-1 }
\nonumber\\
F&=&- (J+{1\over 2}G_0^{-1})^{-1}\bar K \bar G
\nonumber\\
\bar F&=&-(\bar J-{1\over 2}G_0^{-1T})^{-1} K)^{-1 }KG \ ,
\ee
and
\be
\label{Gamma_0}
\Gamma_0=-{1 \over 2}\tr \ln (-\bar G G+\bar G F \bar G^{-1}\bar F)+
{1 \over 2}\tr(G_0^{-1} G - G_0^{-1T} \bar G-2) \ .
\ee
Following the CJT approach~\cite{CJT} for the four-fermion interaction,
one can show that
\be
\label{gamma_expansion}
\Gamma=\Gamma_0+{1\over 4!}\hbar^2
\left(\tr(G{\delta^2 \over \delta
\psi \delta \bar \psi}) S_{int} \tr({\backdel\over \delta
\psi \delta \bar \psi}\bar G) +
\tr(F{\delta^2 \over \delta
\bar\psi^T \delta \bar \psi}) S_{int}\tr({\backdel \over \delta
\psi \delta  \psi^T}\bar F)\right)
+ \Gamma_4 \ ,
\ee
where $S_{int}$ is the original four-fermion interaction and we have
explicitly indicated the dependence on $\hbar$. Now it becomes clear that, in 
order to perform the functional derivatives, it is convenient to
Fierz-rearrange $S_{int}$ into the 3 channels $s$, $t$ and $u$ as outlined 
above. The sum of the four derivatives in Eq.~(\ref{gamma_expansion}) 
naturally suggests  the use of $U$ in Eq.~(\ref{Z_source}) 
with two-by-two derivatives \wrt~the fermion fields
 that are contracted with the same matrix ${\cal O}^i$.

The Hartree-Fock scheme is equivalent to (i) neglecting $\Gamma_4$,
which is the sum of all four-particle irreducible diagrams and is of
order $O(\hbar^4)$,  and (ii)  considering only translationally
invariant solutions (\eg, $G(p,p')=G(p)\delta^4(p-p')$, and the same for
$F, \bar F$). The mesonic and the diquark terms in the above 
lowest-order interaction term have the form $V_4({\bf M}_i)({\bf\bar
M}_i)$, $V_4 ({\bf D}_i)({\bf\bar D}_i)$, where 
\be
{\bf M}_i&=&\int
{d^4p\over (2 \pi)^4} \tr (G {g^{mes}_i}{\cal F}^\dagger  {\cal O}^i
{\cal F})\nonumber \\
{\bf D}_i&=&\int {d^4p\over (2 \pi)^4}\tr( F^\dagger
{g^{diq}_i}{\cal F}^T  {\cal O}^i C {\cal F}^*) \ ,
\ee
\ie,  they are products of mesonic and diquark condensates. 
A priori one might consider condensation in any channel, and then study 
if it is energetically favored. It is reasonable to start with the channels 
that are most attractive (\eg, those which generate bound states in vacuum).   
As we have seen in sects.~\ref{sec_L_eff} and \ref{sec_diq}, 
these are the scalar color singlet 
mesonic channel and the diquark color-${\bf \bar 3}$ one. For the $N_f=2$ case 
there is only one pattern of diquark condensation, which necessarily breaks 
the color symmetry down from $SU(3)_C\to SU(2)_C$. The diquark condensate is 
proportional to a unit vector in $SU(3)_C$. Moreover, because the flavor part 
of the vertex is antisymmetric (due to $\tau_2$), 
the Pauli principle requires that 
the above unit vector belongs to the antisymmetric part of $SU(3)_C$. Without 
loss of generality
 we can choose this vector to be the Gell-Man matrix $\lambda_2$ 
(recall that all Gell-Mann matrices in this paper are normalized to 3, 
\ie, $\tr \lambda_i^2 =3$). We can identify the unbroken 
$SU(2)_C$ with the upper  $2\times 2$ corner of the $SU(3)_C$ group. 
If both $qq$ and $\bar qq$ condensates
are present, the latter should consist of two parts -- one that involves the
two colors from the unbroken $SU(2)_C$ and the second one which involves
quarks and antiquarks from the third color. One should note that there are 
two terms in ${\cal L}_{mes}$ contributing to the $\bar qq$ condensates: the 
isoscalar color singlet one and the isoscalar $SU(3)_C$ octet one proportional 
to $\lambda_8$. Both have projections onto the $SU(2)_C$ singlet as well as 
onto the subgroup represented by third color of (unpaired) quarks. Owing to 
the two different chiral condensates, it is convenient to split the propagators 
$G, \bar G$ into a $SU(2)_C$ part $G_1, \bar G_1$ and a part  involving the 
third color only, $G_2, \bar G_2$. The $F$ and $\bar F$ propagators are 
proportional to the diquark condensate and hence to $\lambda_2$. Putting 
everything together, Eq.~(\ref{gamma_expansion}) becomes  
\be
\label{Gamma}
\Gamma&=&-{1 \over 2}\tr \ln (-\bar G_1 G_1+\bar G_1 F \bar G_1^{-1}\bar F)+
{1 \over 2}\tr(G_0^{-1} G_1 - G_0^{-1 T} \bar G_1-2)\nonumber \\ 
& & -{1 \over 2}\tr
\ln (-\bar G_2 G)+ {1 \over 2}\tr(G_0^{-1} G_2 - G_0^{-1 T} \bar
G_2-2)\nonumber \\
& &+g{1\over 8 N_c^2}|\tr(G_1+G_2)\alpha|^2+g
{N_c-2\over 16 N_c^2 (N_c^2-1)}|\tr \lambda_8(G_1+G_2)\alpha|^2\nonumber \\
& &+g
{1\over 8 N_c^2 (N_c-1)}\tr(FC \gamma_5 \lambda_2 \tau_2
\beta)\tr(\beta^*C \gamma_5 \lambda_2 \tau_2 \bar F),
\ee
where $\alpha$ and $\beta$ are the formfactors defined in
Eqs.~(\ref{alpha}),(\ref{beta}). 

Since $G,\bar G, F, \bar F$ are classical fields, $\Gamma$ should vanish 
under the respective variations.

%%%%%%%%%%%%%%%%%%%%%%%%%%%%%%%%%%%%%%%%%%%%%%%%%%%%%%%%%%%%%%%%%%%%%%%%%%%
\section{Gap-Induced Damping of Single-Quark Propagators}
\label{sec_damp}
%%%%%%%%%%%%%%%%%%%%%%%%%%%%%%%%%%%%%%%%%%%%%%%%%%%%%%%%%%%%%%%%%%%%%%%%%%%
In the superconducting phases, single-quark propagation in the vicinity
of the Fermi surface is damped in the temporal
direction due to the presence of the finite energy gap.
This effect has to be included in the evaluation of the $I$-$A$ overlap
matrix elements, Eq.~(\ref{tiamu}),  at finite density (this is
particularly crucial for the
activity $z_d$ of Eq.~(\ref{zd}), which otherwise would diverge).
Rewriting the overlap matrix elements as
\be
T_{IA} = - \int d^4x \
\bigl[\phi_I^\dagger(x-z_I;-\mu)(i\not\!\partial-i\mu\gamma_4)\bigr]
 \ (i\not\!\partial-i\mu\gamma_4)^{-1} \
\bigl[(i\not\!\partial-i\mu\gamma_4)
\Psi_{0,A}(x-z_{A};\mu)\bigr] \ ,
\ee
they are readily interpreted as a quark hopping amplitude, represented
by two amputated (anti-) instanton vertices and an intermediate (free)
quark propagator  $(i\not\!\partial-i\mu\gamma_4)^{-1}$.
In the following we will evaluate an approximate damping factor for this
propagator. Starting from the
expression of the (massless) momentum space propagator at finite
$\mu$ in the superconducting phase (ignoring the Dirac structure)
\be
G(p,\mu,\Delta) & = & \frac{p_0+\xi}{p_0^2-\xi^2-\Delta^2+i\eta}
\nonumber\\
 & = & \frac{u_p^2}{p_0-\epsilon(p)+i\eta} +
  \frac{v_p^2}{p_0+\epsilon(p)-i\eta}
\ee
($\xi=\omega_p-\mu, \epsilon(p)^2=\xi^2+\Delta^2,
u_p^2=\frac{1}{2}[1+\xi/\epsilon(p)],
v_p^2=\frac{1}{2}[1-\xi/\epsilon(p)]$), 
we compute its Fourier transform as
\be
G(z;\mu,\Delta) & = & \int \frac{d^4p}{(2\pi)^4} e^{-ipz}
  G(p,\mu,\Delta)\nonumber\\
 & = & \frac{1}{2\pi^2r} \int\limits_{p_F}^{\infty} p \ dp \ \sin(pr) \
e^{-\epsilon(p) z_4} \ u_p^2 \ .
\ee
For $\Delta\to 0$ one recovers the result~\cite{Sch_98}
\be
G(z;\mu)=\frac{1}{2\pi^2 z^4} \left[ (2z_4+\mu z^2) \cos(\mu r)
+(z_4^2-r^2+\mu z_4 z^2) \frac{\sin(\mu r)}{r} \right]  \ .
\ee
An approximate correction to $T_{IA}(z;\mu)$ is thus obtained
by supplying it with the ratio
\be
R(z_4;\mu,\Delta)\equiv \frac{G(z_4;\mu,\Delta)}{G(z_4;\mu)} \ , 
\label{Rdelta}
\ee
where we have restricted the {\it l.h.s}~of Eq.~(\ref{Rdelta}) to $r$=0 to 
avoid artificial singularities caused by oscillations in $G(z_4,r;\mu)$.
However, since (for $N_f=2$) only two out of three quarks at the Fermi
surface can
participate in the diquark condensate, the correction (\ref{Rdelta})
enters on average with a smaller power, {\it i.e.},
\be
T_{IA}(z;\mu,\Delta)\simeq R(z_4;\mu,\Delta)^{2/3} \ T_{IA}(z;\mu) \ .
\ee
The net effect of $R(z_4;\mu,\Delta)$ is a damping of the zero mode
propagation, which, in fact, is much less pronounced than the naive
expectation, $\propto e^{-\Delta z_4}$, would suggest.
On the other hand,  $\Omega_{inst}$ {\it disfavors} finite values for 
$\Delta$.
%%%%%%%%%%%%%%%%%%%%%%%%%%%%%%%%%%%%%%%%%%%%%%%%%%%%%%%%%%%%%%%%%%%%%
\end{appendix}
%%%%%%%%%%%%%%%%%%%%%%%%%%%%%%%%%%%%%%%%%%%%%%%%%%%%%%%%%%%%%%%%%%%%%

\end{document}